# *A Hierarchical Multilevel Inference Framework for Structural Cardiovascular Risk Modeling: County-Scale Analysis of Cardiovascular Mortality in Ohio and Pennsylvania (1999–2020)*

**Christopher Blaszczak-Boxe,[1,2] Jose Perez-Chavez,[1,2] Sium Gebremeriam,[1,2] King-Fai Li,[3] Yuk L. Yung,[4] and Mohammad Russel[5]**

[1] Graduate School, Howard University, Washington, DC 20059, USA

[2]Department of Earth, Environment, and Equity, Howard University, Washington, DC 20059, USA

[2]Environmental and Climate Sciences Department, Brookhaven National Laboratory, Upton, NY 11973, USA

[3]Environmental Science Department, UC Riverside, Riverside, CA 92521

[4]Geology and Planetary Science Division, California Institute of Technology, Pasadena, CA 91125, USA

[5]*School of Ocean Science and Technology, Dalian University of Technology, China, Liaoning, Panjin 124221, People's Republic of China*

***Corresponding Author:*** **Chirstopher Blaszczak-Boxe**

**Email: christopher.boxe@howard.edu**

**Abstract**

*Cardiovascular mortality is shaped by interacting demographic, environmental, and structural processes operating across multiple spatial scales. Conventional epidemiologic analyses often rely on aggregate summaries or single-model formulations that obscure hierarchical variation and contextual heterogeneity. We present a reproducible multilevel statistical inference framework integrating* Normal *(age-adjusted), Poisson (count-based), and population-offset* Poisson *models to quantify cardiovascular mortality across nested geographic units while separating demographic effects from structural variation.*

*The framework was applied to county-level mortality data from* Ohio *and* Pennsylvania *(1999–2020) using MLwiN hierarchical models for seven cardiovascular disease* (CVD) *subtypes. Fixed effects included year, sex, race,* PM2.5, *and* O3, *while county-level random intercepts captured spatial heterogeneity. Complete model equations are provided in the* Supplementary Material.

*The framework reveals complementary perspectives on cardiovascular risk unavailable from a single model. Age-adjusted mortality declined more rapidly in* Pennsylvania *than* Ohio, *whereas Poisson models identified post-2010 stagnation or reversal for several* CVD *subtypes. Black populations experienced elevated mortality risks, males exhibited higher mortality than females, and* PM2.5 *showed stronger associations with ischemic and hypertensive mortality in* Pennsylvania. *Population-offset models reduced unexplained variance while preserving county-level structural disparities.*

*Beyond cardiovascular epidemiology, this work introduces a generalizable hierarchical statistical framework for structurally nested health systems. The methodology provides a scalable foundation for disease surveillance, environmental health assessment, health equity research, reproducible statistical analysis, and AI-assisted scientific inference.*

***Computational Contributions***

*This work introduces a reusable computational framework for hierarchical epidemiological inference through:*

- A *unified multilevel modeling architecture integrating* Normal, Poisson, *and population-offset* Poisson *regression.*
- *Explicit decomposition of demographic effects and latent structural heterogeneity using hierarchical random-effects modeling.*

- *A reproducible MLwiN-based statistical workflow with complete equation-level documentation.*
- *Comparative inference across multiple statistical distributions applied to identical disease endpoints.*
- *A scalable framework applicable to other diseases, geographic regions, and environmental health datasets.*
- *Transparent supplementary equations supporting reproducibility, extension, and methodological benchmarking.*

## 1. Introduction

Over the last four decades, the United States has achieved a notable decline in cardiovascular mortality, driven by major medical, public health, and environmental advances. National age-adjusted death rates for ischemic heart disease (IHD) and related cardiovascular disease (CVD) subtypes fell due to expanded use of statins, antihypertensives, and revascularization procedures, alongside tobacco control, dietary changes, and improved screening for hypertension and diabetes.[1,2] Environmental policies like the Clean Air Act also contributed, reducing exposure to fine particulate matter ($PM_{2.5}$) and ozone ($O_3$), both of which are now well-established cardiovascular risk factors.[3,4]

Yet beneath these national gains lies an uneven geography of mortality. Since 2011, CVD mortality declines have slowed or reversed in some subgroups—particularly among younger adults, racial and ethnic minorities, rural populations, and residents of economically depressed regions.[5,6] The COVID-19 pandemic further exposed this fragility: heart disease deaths increased in 2020–2021 amid care disruptions and amplified preexisting inequities.[7]

The industrial Midwest—especially Ohio and Pennsylvania—exemplifies these emerging challenges. Both states have histories of heavy industry, coal combustion, and automotive emissions that contributed to persistent $PM_{2.5}$ and $O_3$ exposure.[8,9] Combined with racial segregation, aging populations, income inequality, and healthcare access disparities, this creates a complex web of risk that standard aggregate mortality metrics often fail to resolve.[10,11]

Most national surveillance systems, including the CDC WONDER platform, report aggregate CVD mortality. Such aggregation masks important differences across disease subtypes—including hypertensive heart disease, heart failure, atherosclerosis, and acute myocardial infarction—which have distinct etiologies, trajectories, and environmental sensitivities. Disaggregated analysis is thus essential for detecting where mortality burdens are growing, and for understanding how structural, demographic, and environmental drivers interact over time.

To address these gaps, we apply a multilevel modeling approach (MLA) using MLwiN to analyze mortality across seven major CVD subtypes (total CVD, IHD, acute myocardial infarction, atherosclerosis, heart failure, hypertensive heart disease, and other heart diseases) from 1999 to 2020 in all counties of Ohio and Pennsylvania. By nesting individuals within counties and modeling both Poisson (absolute deaths) and Normal (age-adjusted rates) distributions—with and without log(population) offsets—we examine how county-level structural factors shape both the magnitude and distribution of CVD mortality.

The Poisson models, particularly when adjusted with log(population) offsets, allow for interpretable per-capita death counts but remain heteroscedastic—reflecting greater variance in high-population or structurally underserved counties. In contrast, the Normal models provide homoscedastic estimates, better isolating relative disparities. Together, these models reveal how mortality burden (where deaths concentrate) and mortality disparity (how unequally risk is distributed) evolve under pressure from environmental injustice, healthcare inequity, and demographic shifts.

Our analysis shows that while age-adjusted mortality continues to decline—more steeply in Pennsylvania than Ohio—absolute deaths for several subtypes (especially hypertensive and heart failure mortality) have plateaued or reversed post-2010, with sharper deterioration during the COVID-19 era. Black populations experienced disproportionately high relative risks (up to +233.5%), and men consistently experienced higher death rates than women. $PM_{2.5}$ was a strong predictor of ischemic and hypertensive mortality in Pennsylvania, while associations in Ohio were weaker or null.

This study introduces an integrated modeling and visualization framework for uncovering place-based cardiovascular risk and mortality. It advances the field by unifying disaggregated disease classification, multilevel statistical modeling, and structural variance interpretation—illuminating a path toward targeted, equity-focused intervention in the U.S. Rust Belt and beyond. This work complements and extends the Global Burden of Disease (GBD) approach by shifting from global, top-down estimates to localized, bottom-up analyses of cardiovascular mortality at the state-county level. Unlike GBD's generalized models, our method integrates raw mortality data with context-specific predictors—such as $PM_{2.5}$, ozone, race, gender, and healthcare access—informed MLwiN multilevel analysis to reveal structural disparities often masked in national averages. By incorporating both Normal (for age-adjusted rates) and Poisson models (for raw counts, with and without log(population) offsets), the study captures complementary dimensions of mortality risk, enhancing inferential validity and enabling rate-based and count-based interpretations. This dual-model design not only enables policy-sensitive hotspot detection and stratified risk profiling but also lays a scalable foundation for equity-driven health planning, how synthetic counterfactual scenarios, and even planetary health analogues.

## 2 Methodology

### 2.1 Study Design Overview

This study applied an integrative, multilevel, and machine learning–enhanced modeling framework to investigate all CVD subtypes (and total) mortality death trends across Pennsylvania and Ohio between 1999 and 2020. Recognizing the nested structure of health outcomes across individual and county levels, we employed multilevel normal and Poisson regression models to quantify the relative contributions of demographic, socioeconomic, and environmental determinants. These approaches allowed simultaneous estimation of individual-level predictors—such as age, race, and gender—alongside county-level exposures, including fine particulate matter ($PM_{2.5}$) and ground-level ozone concentrations.

Mortality data were obtained from the CDC WONDER database, supplemented with state health department records and air quality measurements from the U.S. Environmental Protection Agency's Air Quality System (AQS). Covariates were selected based on prior evidence of cardiovascular disease risk factors, including structural indicators of healthcare access, income inequality, and racial composition.[1,9,10,12,13]

### 2.2 Data Sources and Preprocessing

Mortality data were retrieved from the CDC WONDER database, augmented with health department records from Pennsylvania and Ohio, and aligned with air quality data from the U.S. Environmental Protection Agency's Air Quality System (AQS0.[14,15] Mortality records included stratifications by age group, gender, race, and year of death. Environmental exposure variables included county-level annual average concentrations of fine particulate matter ($PM_{2.5}$) and ground-level ozone ($O_3$).[9,16] Built environment covariates, such as proximity to highways, healthcare accessibility, and urbanization levels, were incorporated into supplementary analyses.

Sociodemographic predictors included race (White, Black, American Indian/Alaskan Native, Asian/Pacific Islander, and Other), gender (male, female), and age groups (0–4 years to 100+ years). Socioeconomic indicators encompassed county-level median household income, educational attainment, poverty rates, and healthcare provider density.[17-20]

To ensure data quality and analytic robustness, preprocessing steps included the following:

- Outlier detection and winsorization of extreme mortality values.
- Imputation of missing air quality values using county-level temporal interpolation.
- Standardization (z-scoring) of continuous variables to facilitate convergence in machine learning models.

**2.3 Multilevel Statistical Modeling**

**Rational for using multilevel analysis (MLA) modeling?** Multilevel modeling (MLM) is a critical analytic approach when evaluating outcomes measured at the individual level within hierarchically structured data, such as individuals nested within geographic, institutional, or social contexts.[21] Traditional regression models assume the independence of observations—an assumption that is violated when individuals share contextual exposures (e.g., neighborhood socioeconomic status or environmental pollutants). Assigning the same contextual value to all individuals within a group introduces intraclass correlation, leading to misestimated regression coefficients and standard errors. This overstates the effective sample size and increases the likelihood of false-positive associations.

In public health research, contextual factors are not merely statistical nuisances but define the structural constraints and opportunities that shape individual behavior. Communities, workplaces, and other macro-level units influence access to healthcare, environmental exposures, and behavioral norms—factors central to health outcomes such as ischemic heart disease. Determining which contextual levels are analytically relevant requires examining where constraints and resources are patterned and how they structure health-relevant behaviors.

Multilevel models allow for the explicit testing of four key hypothesis domains: (1) the magnitude and patterning of outcome variation across levels; (2) associations between individual-level predictors and outcomes; (3) effects of higher-level contextual variables; and (4) cross-level interactions, whereby the influence of individual-level factors may vary by context. This capacity makes MLM uniquely suited for investigating the multiscale determinants of health and generating actionable insights for both population-level policy and targeted interventions.

MLwiN is a purpose-built software developed by the Centre for Multilevel Modelling at the University of Bristol to address the complex requirements of multilevel (hierarchical) data analysis. Its design is particularly well-suited to scenarios involving nested structures—such as individuals within schools, patients within hospitals, or repeated measurements within subjects. MLwiN's intuitive graphical user interface (GUI) supports a range of models, including linear, logistic, multinomial, and Poisson, across two or more levels of data. The software also supports intricate model structures like cross-classified and multiple membership models, which are essential in educational, epidemiological, and social science research. MLwiN offers robust Bayesian estimation through MCMC (Markov Chain Monte Carlo) techniques and provides visually intuitive diagnostics, such as residual plots, to assess model fit and variability across levels.

A multilevel regression framework, via MLwiN, was employed to capture hierarchical structures within the data. Normal and Poisson models were used to assess variations in mortality rates. Multilevel models were constructed in MLwiN software,[22] with supplementary analyses conducted in R (using *lme4* and *brms*) and Python (*statsmodels*, *PyMC*) for cross-validation. Two types of models were fit:

- **Normal Distribution Models**: Examined linear associations between mortality rates and independent variables. Modeled mortality rates as continuous outcomes, estimating linear associations between predictors and mortality.
- **Poisson Distribution Models**: Estimated log-linear relationships, interpreting coefficients as percentage changes in mortality risk. Modeled count data (number of deaths) with log-linear associations, interpreting coefficients as percentage changes in mortality risk.
- **Poisson Distribution Models w/ Offset**: In multilevel Poisson regression, an offset is a variable included with a fixed coefficient of 1. It adjusts for exposure time or population at risk, allowing you to model rates (e.g., deaths per person-year or deaths per 100,000 people) rather than counts alone.
- Random and Fixed Effects: Random intercepts accounted for county-level effects, while fixed effects included year, air pollution levels, built environment factors, and demographic factors to control for unobserved heterogeneity.

- Interaction Terms: To assess potential moderating effects, interaction terms between demographic and environmental factors were included in select models.

These models quantify the relationship between cardiovascular mortality and environmental, demographic, and temporal predictors. The log(population) offset in the final Poisson formulation adjusts for county population size, aligning the analysis with rate-based rather than count-based comparisons.

Random intercepts were specified at the county level to capture unobserved heterogeneity, while fixed effects included year, age, race, gender, $PM_{2.5}$, ozone, and socioeconomic status.

Interaction terms between environmental exposures and sociodemographic variables were tested to identify potential effect modification.

The general form of the multilevel **Normal Distribution Model** was:

$$Mortality_{ij} = \beta_{0ij} + \sum_{k=1}^{K} (\beta_k X_{kij}) + u_{0j} + e_{0ij}$$

$$Mortality_{ij} = \beta_0 + \beta_1 \cdot Year_{ij} + \beta_2 \cdot PM_{2.5,ij} + \beta_3 \cdot O_{3,ij} + \beta_4 \cdot AgeGroup_{ij} + \beta_5 \cdot Gender_{ij} + \beta_6 \cdot Race_{ij} + U_{0j} + \varepsilon_{0ij}$$

- $\beta_{0ij}$ is the intercept term that can vary across counties.
- $\beta_k$ represent the coefficients for different predictor variables $X_k$ such as year, race, gender, age, ozone levels, PM2.5 exposure.
- $u_{0j}$ is the county-level random effect, which accounts for unobserved variability between counties.
- $e_{0ij}$ is the individual-level residual error, which captures unexplained variability within counties.

The random effects ($u_{0j}$ and $e_{0ij}$) follow normal distributions with respective variances.

**Poisson Distribution Models:**

$$Deaths_{ij} \sim \text{Poisson}(\pi_{ij})$$

$$\log(\pi_{ij}) = \beta_0 + \beta_1 \cdot Year_{ij} + \beta_2 \cdot PM_{2.5,ij} + \beta_3 \cdot O_{3,ij} + \beta_4 \cdot AgeGroup_{ij} + \beta_5 \cdot Gender_{ij} + \beta_6 \cdot Race_{ij} + U_j$$

and w/ log-offset:

This makes the model:

$$\log(\pi_{ij}) = \beta_0 + \beta_1 X_{ij} + \ldots + \log(Pop_{ij})$$

Where:

- $\pi_{ij}$ = expected count of events
- $X_{ij}$ = covariates (race, gender, $PM_{2.5}$, ozone, etc.)
- $\log(Pop_{ij})$ = **the offset**, i.e., population of the county/year/race/gender group

$$\log(\pi_{ij}) = \log(pop_{ij}) + \beta_0 + \beta_1 \cdot Year_{ij} + \beta_2 \cdot PM_{2.5,ij} + \beta_3 \cdot O_{3,ij} + \beta_4 \cdot AgeGroup_{ij} + \beta_5 \cdot Gender_{ij} + \beta_6 \cdot Race_{ij} + U_j$$

- $\beta_1$: Change in log mortality rate for each one-year increase.
- $\beta_2$: Change in log mortality rate for each unit increase in ozone concentration.
- $\beta_3$: Change in log mortality rate for each unit increase in PM2.5 concentration.
- $\beta_4$: Difference in log mortality rate across different age groups.
- $\beta_5$: Difference in log mortality rate across different genders.
- $\beta_6$: Difference in log mortality rate across different races.
- $U_j$: Random intercept for county j, accounting for unobserved county-level factors.
- $\pi_{ij}$: is the expected mortality rate.

### Model Level Construct

1. **Level 1 (Individual Level):**
   $Y_{ij} = \beta_{0j} + \beta_{1j}X_{ij} + \epsilon_{ij}$
   - $Y_{ij}$: Dependent variable for individual $i$ in group $j$.
   - $\beta_{0j}$: Random intercept for group $j$.
   - $\beta_{1j}$: Random slope for the relationship between the independent variable $X_{ij}$ and the dependent variable for group $j$.
   - $\epsilon_{ij}$: Individual-level residual.
2. **Level 2 (Group Level):**
   $\beta_{0j} = \gamma_{00} + U_{0j}$
   $\beta_{1j} = \gamma_{10} + \gamma_{11}Z_j + U_{1j}$
   - $\gamma_{00}$: Overall intercept (fixed effect).
   - $\gamma_{10}$: Overall slope (fixed effect).
   - $\gamma_{11}$: Fixed effect representing the relationship between the group-level predictor $Z_j$ and the random slopes.
   - $U_{0j}$: Random effect representing variability in intercepts across groups.
   - $U_{1j}$: Random effect representing variability in slopes across groups.
   - $Z_j$: Group-level predictor.

In these equations:

- $Y_{ij}$ is the observed outcome for individual $i$ in group $j$.
- $X_{ij}$ is the value of the individual-level predictor variable for individual $i$ in group $j$.
- $\beta_{0j}$ and $\beta_{1j}$ are the group-specific intercept and slope, respectively.
- $\gamma_{00}$, $\gamma_{10}$, and $\gamma_{11}$ are the fixed effects representing the overall intercept, slope, and the effect of the group-level predictor on slopes.
- $U_{0j}$ and $U_{1j}$ are the random effects representing the variability in intercepts and slopes across groups.
- $Z_j$ is the group-level predictor influencing the random slopes.

## 3. Results and Discussion

### 3.1 Situating Cardiovascular Mortality in a Context of Health Disparities

Cardiovascular disease (CVD) remains the leading cause of death globally, with an estimated 19.8 million deaths annually—nearly 695,000 of which occur in the United States alone (Centers for Disease Control and Prevention.[23,24] Within the U.S., Pennsylvania and Ohio report approximately 34,200 and 26,800 heart disease deaths per year, respectively, underscoring a sustained and uneven burden in the Rust Belt region. Although age-standardized mortality rates have generally declined across high-income countries since the 1970s, U.S. rates show troubling plateaus and resurgences in recent decades.[25,26] These trends call attention not just to individual health behaviors but also to deeply rooted social and structural factors that modulate risk and outcomes at both population and sub-population levels.

Health disparities in cardiovascular outcomes are not confined to any single demographic. While historically marginalized populations—including Black, Indigenous, and low-income communities—face disproportionate burdens due to systemic racism, underinvestment, and discrimination, it is essential to acknowledge that all racial and ethnic groups experience some form of structured disparity. For example, high CVD mortality rates among White populations in rural Appalachia and the Midwest are driven by a nexus of geography, socioeconomic decline, and behavioral risk factors.[27,28] These “deaths of despair” from suicide, alcohol-related liver disease, and opioid overdose are not isolated phenomena but indicators of broader systemic neglect.

Formally, as defined by the CDC and National Institutes of Health (NIH), a health disparity becomes scientifically actionable when it is systematic, socially structured, preventable, and rooted in inequity.[26,29] The data suggest that these criteria are fulfilled across multiple population segments, albeit through different mechanisms of vulnerability. For instance, Black individuals often receive less aggressive cardiac care, contributing to higher rates of ischemic heart disease in urban settings, while lower-income White communities suffer from limited healthcare access and preventive infrastructure in rural settings.[26,30]

From a systems modeling perspective, this heterogeneity emphasizes the importance of understanding how environmental, behavioral, and structural parameters interact. Quantitative approaches—such as multilevel modeling and machine learning—enable researchers to estimate the contextual weight of these variables. This is vital because cardiovascular mortality is not merely the product of clinical risk factors like hypertension or cholesterol, but the emergent outcome of intersecting systems: air quality, income, racialized health care delivery, geographic isolation, education, and even local policy enforcement. Consequently, individual mortality risk reflects not only personal behavior but also one’s embeddedness in social and structural networks.[31,32]

Across the globe, the heterogeneous nature of CVD mortality is unmistakable and increasingly transparent. Nations such as Hungary, Russia, and South Africa experience persistently elevated rates, while countries like France and Australia report steep declines—pointing to the success of sustained public health and policy interventions **(Figure 1)**.[24,25] Yet even within these national boundaries, disparities persist. Thus, a complete understanding of cardiovascular mortality requires a hybrid analytic lens: one that captures both macrostructural drivers and the micro-context of lived experience.

As the present study demonstrates through state-specific modeling in Pennsylvania and Ohio, the underlying mortality patterns are shaped by more than just the presence or absence of disease. They are shaped by how systems value—and devalue—certain communities, with implications for both targeted intervention and long-term public health planning.

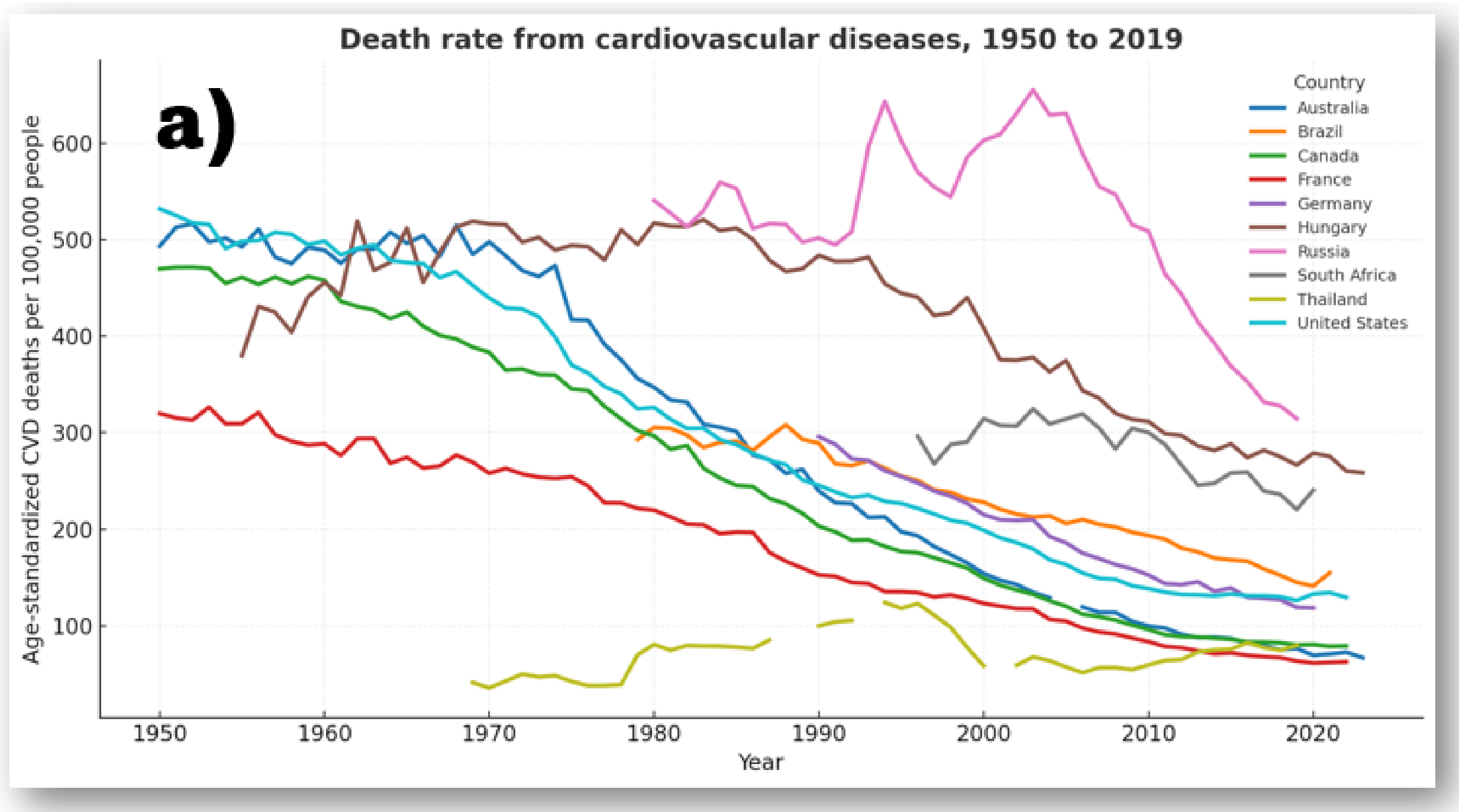

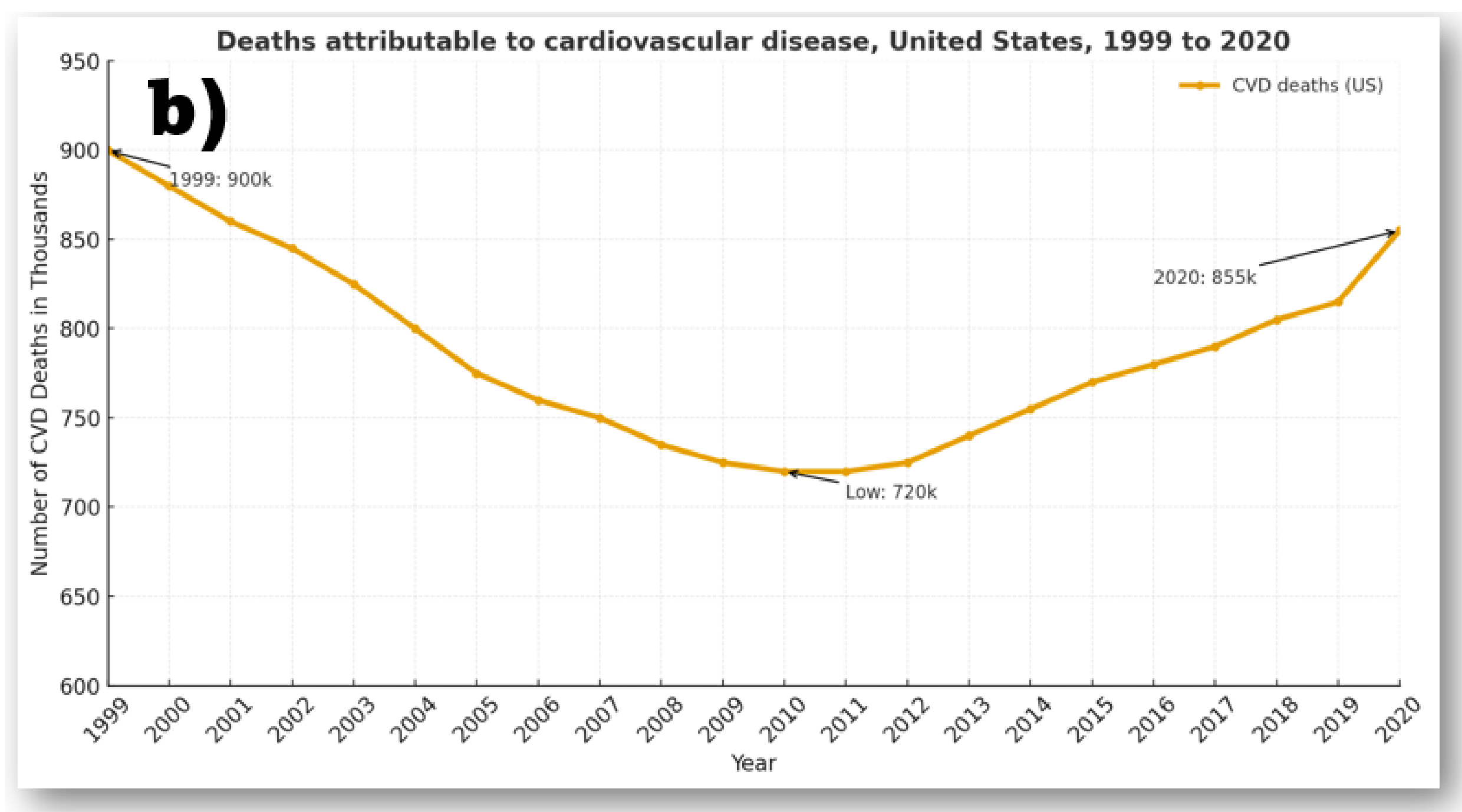

**Figure 1. Global and United States trends in cardiovascular disease (CVD) mortality, 1950–2020.**

(a) Age-standardized CVD death rates per 100 000 people across ten representative countries, 1950–2019. National trajectories highlight steep and sustained declines in most high-income nations (e.g., France, Australia, United States) versus persistently elevated or delayed declines in middle-income settings (e.g., Brazil, South Africa, Thailand). Russia and Hungary show prolonged high mortality before recent improvements, underscoring heterogeneous progress in global cardiovascular health - https://ourworldindata.org/grapher/cardiovascular-disease-death-rate-who-mdb.

(b) Absolute (raw) number of cardiovascular deaths in the United States, 1999–2020. Despite a steady decline from ~900 000 annual deaths in 1999 to a low of ~720 000 around 2010, total deaths have since risen to ~855 000 by 2020, reflecting population aging, demographic change, and the resurgence of cardiometabolic risk factors. Together, panels (a) and (b) illustrate the global decline yet domestic rebound in CVD burden, motivating subnational analysis of structural and environmental determinants.3.2 Temporal Profiles and Stratified Trends in CVD Mortality in Pennsylvania and Ohio: 1968–2020 Centers for Disease Control and Prevention. *Leading Causes of Death Reports*, 1981–2020. https://wisqars.cdc.gov/fatal-leading (Accessed December 2020). (Cardiovascular disease deaths estimated as the combined total of all “Heart Disease” deaths and all “Cerebrovascular” deaths for each year between 1999 and 2020.)

To understand the evolving burden of cardiovascular disease (CVD) in Ohio and Pennsylvania, we first examine temporal trends in CVD mortality stratified by race, gender, and age group. Historical data from Van Dyke et al. (2018) **(Figures 2a and 2b)**,[33] spanning 1968–2015, illustrate a long-term decline in heart disease mortality rates for both Black and White populations across both states. However, these reductions have not occurred uniformly. In Ohio, Black-White disparities in age-adjusted death rates have remained persistent, with the Black-to-White mortality ratio hovering around 1.2 since the early 1990s—suggesting sustained inequities despite aggregate improvements. In Pennsylvania, these disparities widened notably after 1990, with the ratio surpassing 1.2 by 2015.[33]

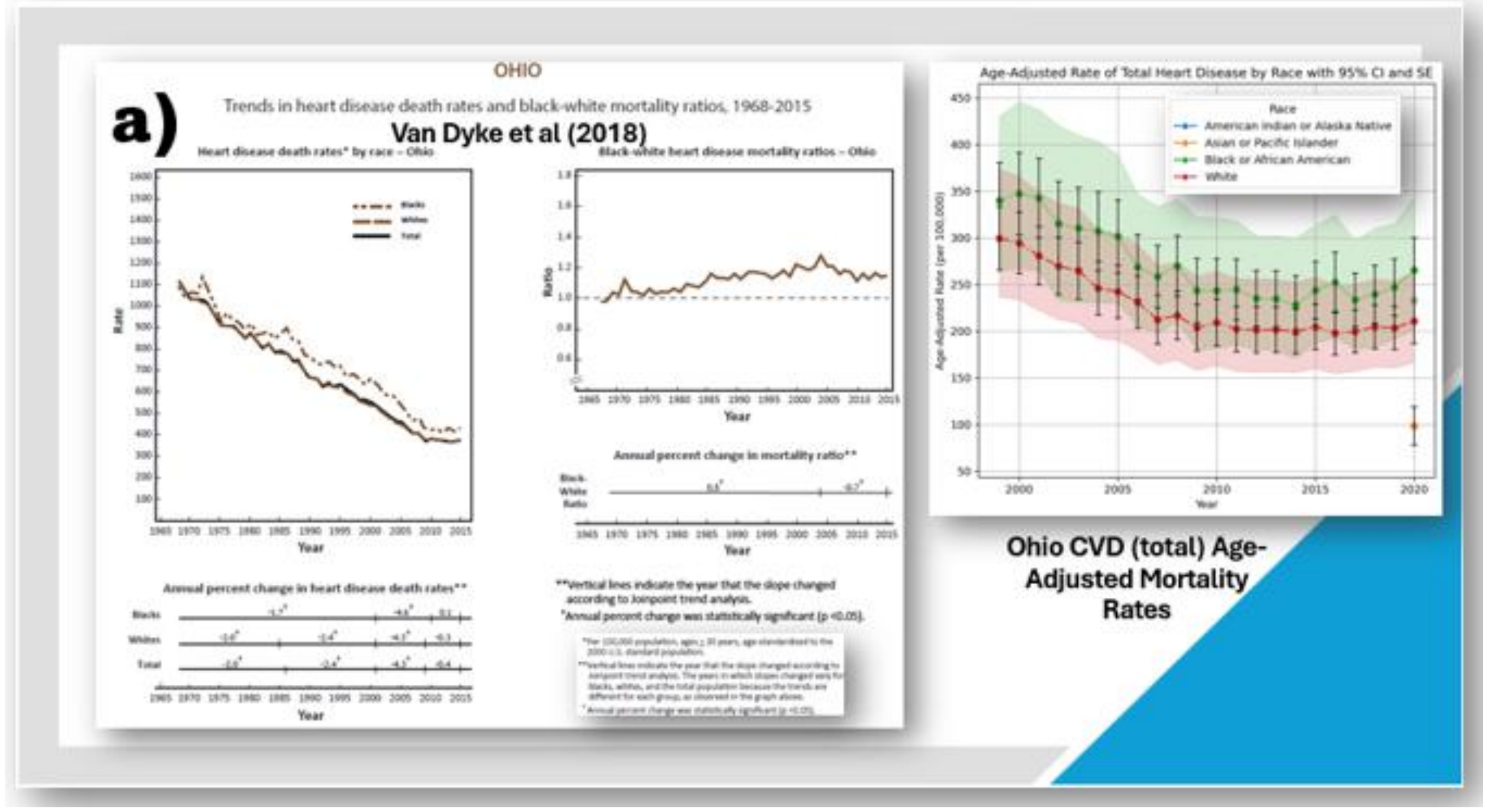

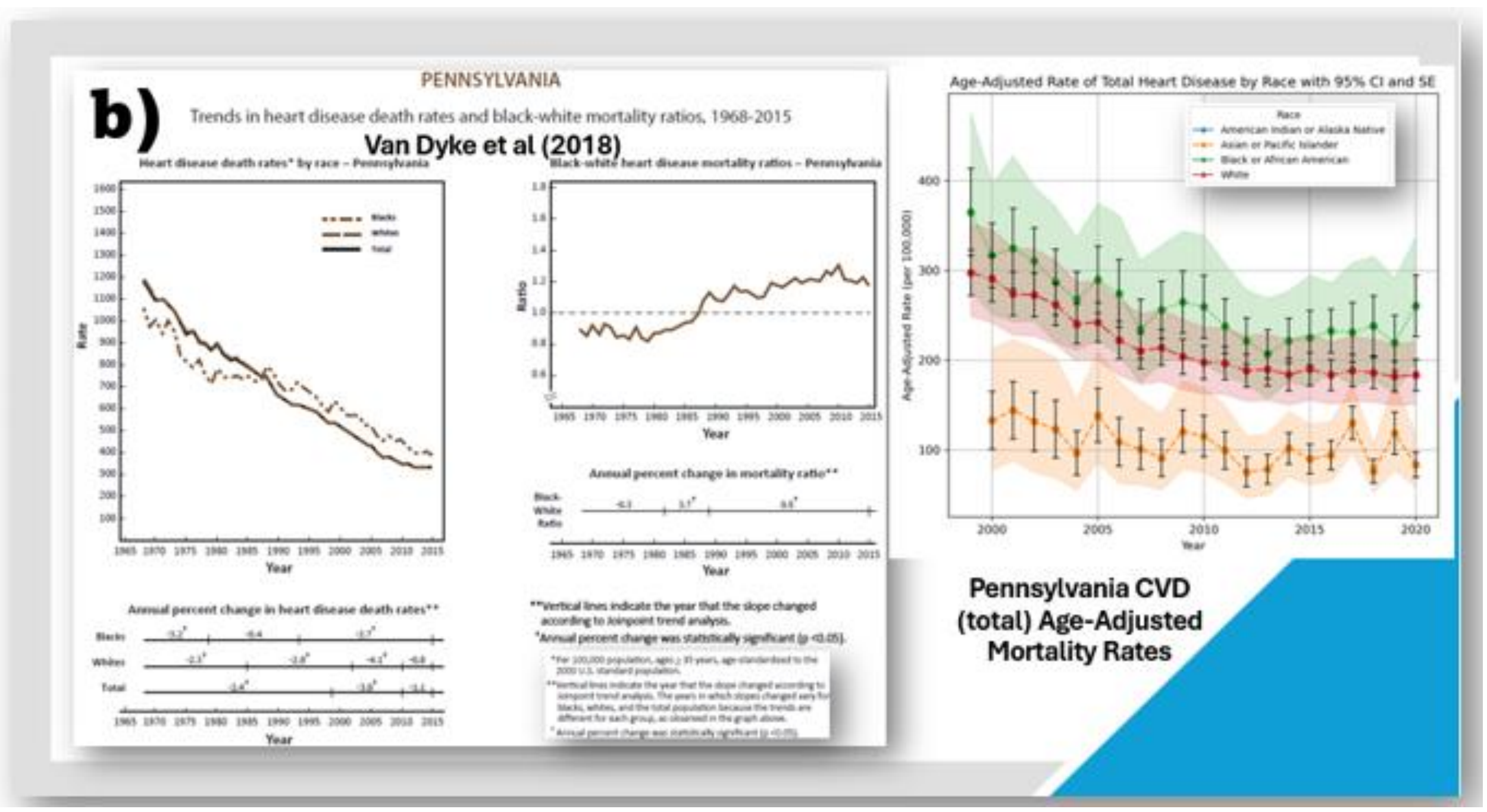

**Figure 2. State-level trends in cardiovascular disease (CVD) mortality by race in Ohio and Pennsylvania, 1968–2020.**

(a) *Ohio.* Long-term heart-disease death rates and Black–White mortality ratios from 1968 to 2015 (Van Dyke et al., 2018) alongside CDC WONDER data (1999–2020) showing age-adjusted total CVD mortality rates per 100 000 population by race. Declines in age-standardized mortality have plateaued since 2010, with persistent racial disparities—Black residents experiencing the highest burden.

(b) *Pennsylvania.* Parallel trends in age-standardized CVD mortality (Van Dyke et al., 2018; CDC WONDER 1999–2020). Both historical and recent data reveal sustained racial gaps and slower post-2010 improvements, particularly among Black populations. All rates are age-standardized to the 2000 U.S. standard population for adults ≥ 35 years and for all ages.

Recent CDC WONDER data from 1999 to 2020 refine these insights by expanding racial categories and enabling cross-tabulation with gender, age, and race **(Figure 3 a-e)-***. Total CVD age-adjusted mortality has declined overall in both states. Nevertheless, Black residents continue to exhibit the highest age-adjusted mortality rates, followed by White, American Indian/Alaska Native, and Asian/Pacific Islander populations. In both states, Black residents experienced a slower decline post-2010 and, in some cases, a recent uptick in age-adjusted death rates—particularly in Ohio, aligning with national findings of health disparities in the post-recession and opioid-epidemic era.[28,32]

Total Mortality Age-Adjusted and Absolute Mortality: Race

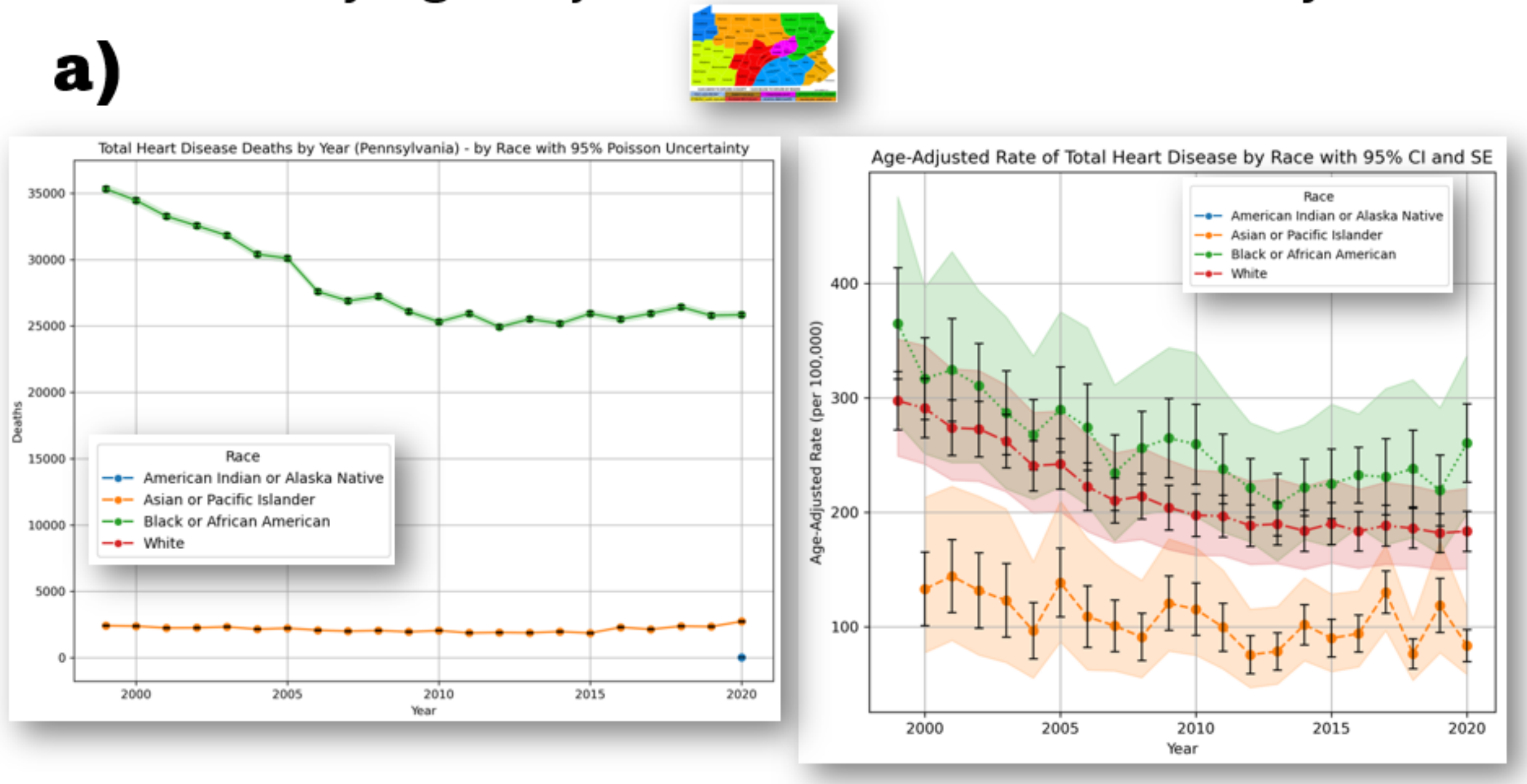

Total Mortality Age-Adjusted and Absolute Mortality: Race

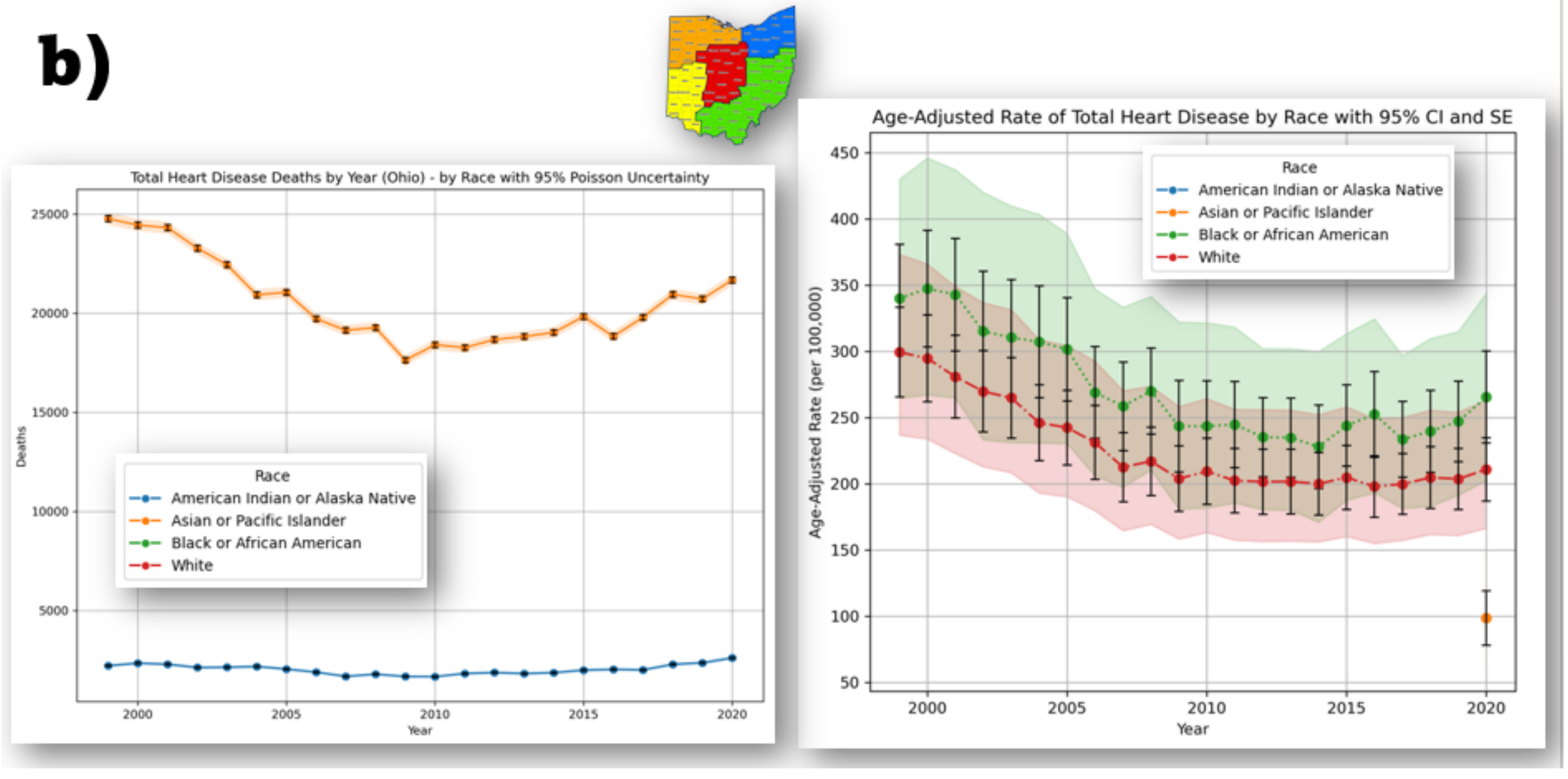

# Total Mortality Age-Adjusted and Absolute Mortality: Gender

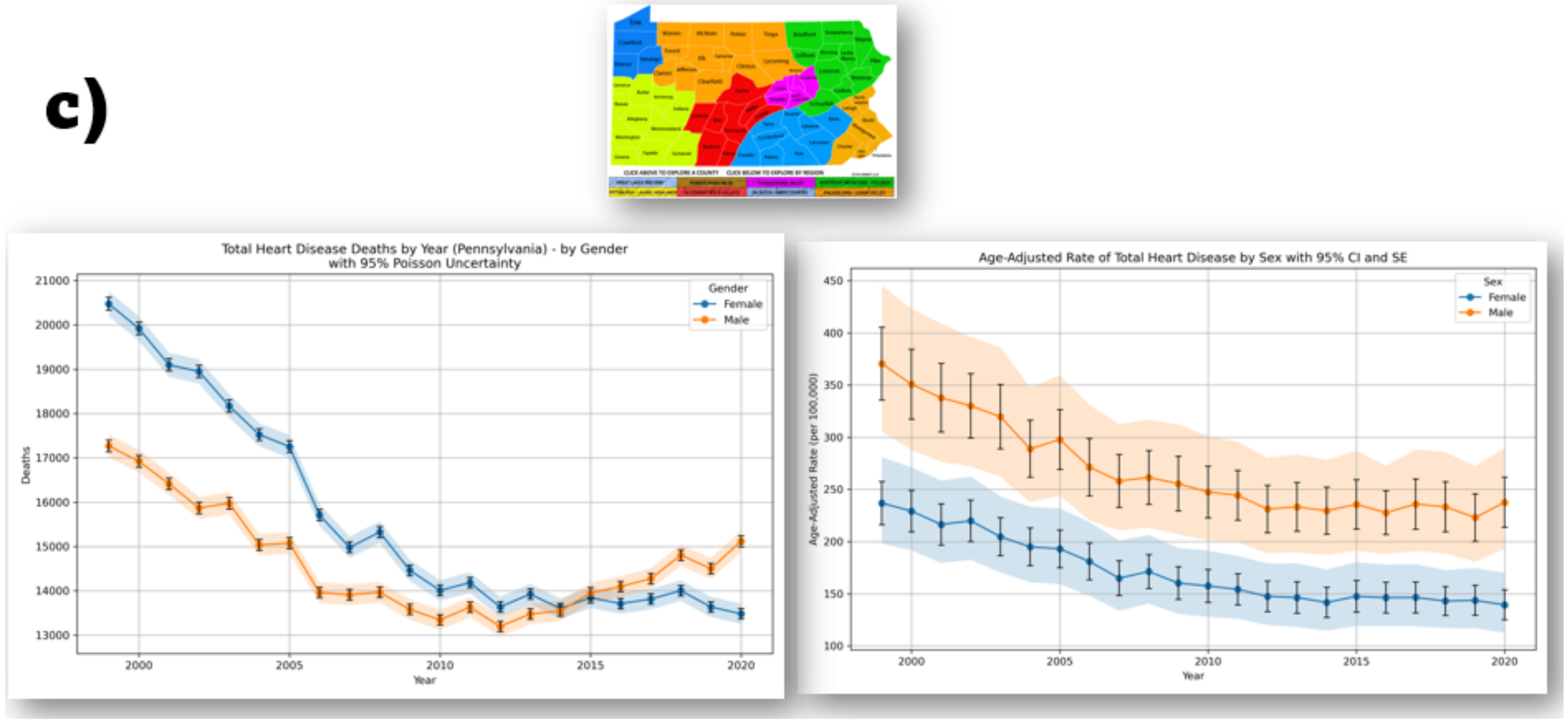

# Total Mortality Age-Adjusted and Absolute Mortality: Gender

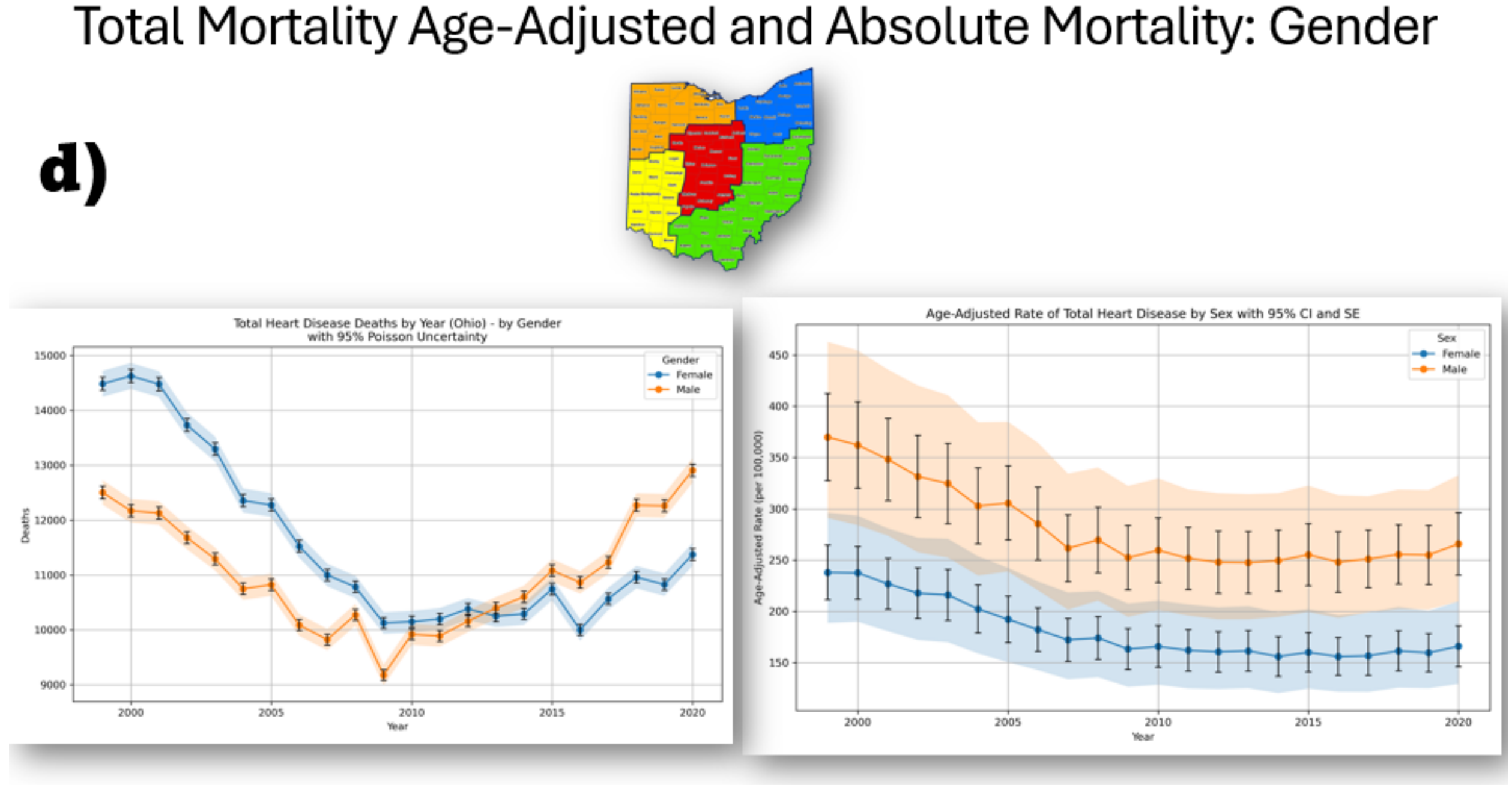

## Stratified Age Plots (Total CVD Death) for PA and OH

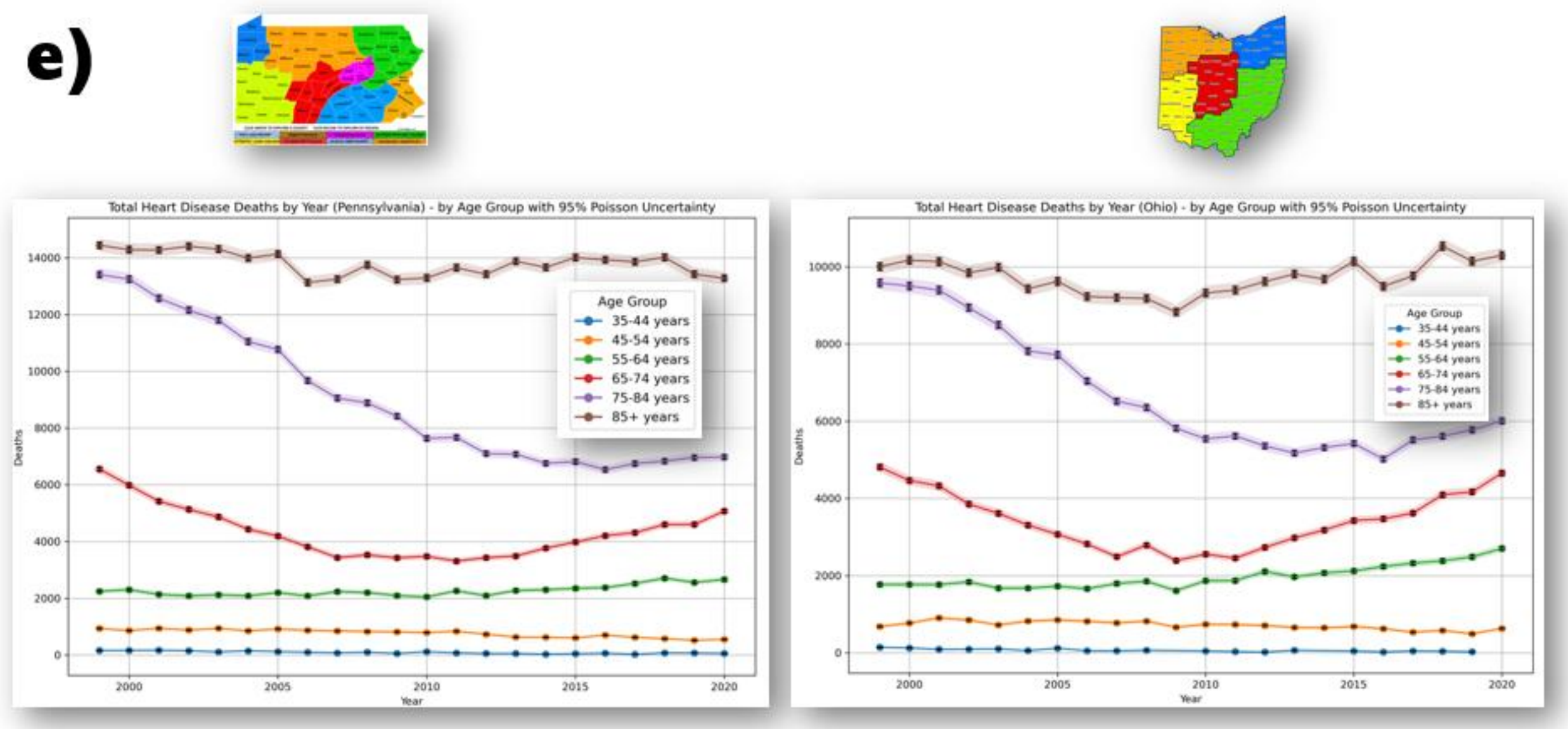

**Figure 3. Raw (absolute), age-adjusted, and age-stratified cardiovascular disease (CVD) mortality by race, gender, and age in Pennsylvania and Ohio, 1999–2020.**
(**a–b**) Total heart disease mortality by race for Pennsylvania (a) and Ohio (b), showing both absolute (left panels) and age-adjusted (right panels) rates per 100 000 population with 95% confidence intervals and Poisson uncertainty bands. Black or African American populations exhibit persistently higher mortality rates than other racial groups in both states, despite overall declines since 2000.
(**c–d**) Total heart disease mortality by gender for Pennsylvania (c) and Ohio (d). Male mortality rates remain consistently higher than female rates in both absolute and age-adjusted frameworks, with divergence widening slightly after 2010 and during the COVID-19 period.
(**e**) Age-stratified total CVD mortality by age group (35–44 to ≥85 years) for Pennsylvania (left) and Ohio (right). Deaths remain concentrated in the oldest age groups, while middle-aged adults (45–64 years) show the steepest relative declines from 1999 to 2010 followed by post-2010 plateaus.
All data are derived from the CDC WONDER Underlying Cause of Death Database (https://wonder.cdc.gov/) and are age-standardized to the 2000 U.S. standard population.

Gender differences in mortality further underscore the persistent structural risks. In both Pennsylvania and Ohio, men consistently exhibit higher age-adjusted CVD mortality rates than women. However, a notable divergence emerges after 2010. In Pennsylvania, male mortality rates plateaued while female rates continued to decline modestly. In contrast, Ohio exhibited a flattening and slight rebound in male mortality beginning around 2012, while female rates remained relatively stable. These patterns may reflect differential exposure to occupational, behavioral, and healthcare-related risk factors across genders and regions.[26]

Age-stratified analyses reveal that most CVD deaths occur in older populations (≥65 years), particularly among those aged 75–84 and ≥85 years. However, the steepest declines in absolute mortality have occurred among middle-aged adults (45–64), while rates among the oldest age brackets have plateaued or declined more gradually. This demographic shift implies not only aging population effects but also differential success in primary and secondary prevention efforts across age cohorts.[34]

Together, these stratified analyses justify the need for a multilevel modeling approach (MLA) to disentangle the interaction of demographic, temporal, and geographic risk factors. Accordingly, this study implements hierarchical models for both states using three data structures: (1) Normal models applied to age-adjusted mortality rates, (2) Poisson models using raw mortality counts, and (3) offset Poisson models incorporating log(population) as an exposure term. This enables us to quantify and compare how CVD mortality rates are shaped not only by individual-level characteristics like race, age, and gender but also by county-level contextual factors and structural inequalities.

The MLA framework reveals that while temporal declines in CVD mortality are statistically robust at the population level, they obscure substantial heterogeneity in subgroup trends. For example, the Poisson and Normal models consistently identify significant race and gender effects, while the offset models better account for population-size-adjusted variation across counties. In Ohio, interaction terms between race and gender show persistent and synergistic disadvantages for Black men, especially in urban counties with high levels of environmental and economic stress. In Pennsylvania, disparities are more geographically stratified, with older White men in rural regions showing unexpectedly high mortality burdens—possibly reflecting under-resourced health systems and behavioral risks aligned with the "deaths of despair" hypothesis.[27]

These findings underscore the urgency of disaggregated, equity-centered CVD surveillance. Aggregate trends may give the illusion of progress, but systemically embedded disparities remain entrenched unless explicitly modeled, measured, and addressed.

### 3.3 Why Multilevel Analysis? Bridging Individual Risk and Systemic Context in CVD Epidemiology

The implementation of multilevel analysis (MLA) in this study is scientifically justified and technically aligned with globally validated public health frameworks, particularly the Global Burden of Disease (GBD) model. Developed by the Institute for Health Metrics and Evaluation (IHME), the GBD study is the most comprehensive global health estimation effort to date, using hierarchical and ensemble modeling to quantify mortality and disease burden across spatial, temporal, and demographic dimensions.[35,36]

At its core, the GBD employs a nested, multilevel modeling structure to capture heterogeneity at multiple spatial and population levels—from global priors to subnational regions. This hierarchical approach allows for the integration of individual-level exposures (e.g., age, sex, behavioral risks) with structural determinants (e.g., healthcare access, income, air quality), while adjusting for contextual dependencies such as neighborhood, county, and regional clustering.[37] Mirroring this methodology, our MLA framework incorporates three nested levels: individuals (Level 1), nested within counties or neighborhoods (Level 2), within broader state-level or regional structures (Level 3).

Traditional single-level regression models assume independence among observations—a violation in public health data where individuals often share contextual exposures like environmental pollution or socioeconomic disadvantage. This non-independence leads to underestimated standard errors and inflated Type I error rates (Raudenbush & Bryk, 2002).[38] Multilevel models address this limitation by partitioning variance into within-group and between-group components and explicitly modeling intraclass correlation (ICC). ICC quantifies how much of the total variance in CVD outcomes is attributable to higher-level units, such as counties or healthcare systems.[39]

Our use of MLA, therefore, is not merely statistical convenience—it is essential for conceptual and policy reasons. First, it enables the estimation of fixed effects (e.g., the average effect of air pollution or age on mortality) and random effects (e.g., county-specific intercepts and slopes reflecting localized variations in risk or policy context). This dual approach improves precision and prediction while allowing insights into how structural and contextual forces shape individual-level CVD risk.

Second, MLA allows us to test cross-level interactions: for example, how the effect of race on CVD mortality changes as a function of county-level healthcare access or environmental burden. Such interactions are not only statistically informative but also policy-relevant, identifying places where interventions may mitigate systemic inequities. This is especially crucial in the Rust Belt states of Ohio and Pennsylvania, where both racial and class-based disparities operate in tandem.[33]

Lastly, MLA offers practical alignment with global health surveillance standards. As the GBD model demonstrates, scalable and flexible hierarchical modeling enables comparisons across settings and time periods while adjusting for known biases and uncertainties. Integrating our MLA framework with GBD-compatible methods enhances the translatability, robustness, and interoperability of our findings.

In sum, multilevel analysis serves as the epistemological and methodological bridge between individual risk and systemic context. It respects the complex, nested structure of real-world health data while empowering researchers and policymakers to distinguish between trends that are widespread and those that are structurally contingent and therefore actionable.

### 3.4 Comparative Novelty and Complementarity with GBD Frameworks

This study introduces a high-resolution, dual-state multilevel modeling approach that directly complements the broader, global scope of the Global Burden of Disease (GBD) study. While GBD provides invaluable top-down estimates of disease burden across nations and regions, it relies primarily on generalized population-level models, predefined risk categories, and cross-national harmonization. In contrast, our study leverages bottom-up integration of raw mortality datasets and machine learning-informed multilevel analysis (MLwiN-MLA) to interrogate cardiovascular disease (CVD) outcomes at the subnational (state-county) scale. By embedding predictors such as $PM_{2.5}$, ozone, race, gender, and healthcare access directly into stratified models for Pennsylvania and Ohio, we enable fine-grained, context-sensitive interpretations of structural disparities in cardiovascular mortality—capabilities largely beyond the reach of national-level estimations.

This methodological pivot—from cross-national comparability to regional specificity—not only aligns with GBD priorities but extends them. We emphasize regional disparities, local policy levers, and stratified risk architecture (e.g., Race × Gender × Age interactions), thus generating insights with immediate relevance to substate health planning and environmental justice. Furthermore, the study's architecture lays groundwork for synthetic counterfactuals, predictive hotspot mapping, and planetary health analogues, broadening the applicability of this approach to climate-sensitive disease risk modeling and exobiological systems thinking. In this way, the study represents a *scalable, modular complement* to GBD by filling critical granularity gaps and enabling predictive causal inference in nested, real-world systems.

Scientifically, the study's use of both Normal and Poisson models (with and without log-population offsets) strengthens its inferential validity and interpretability. Normal (Gaussian) models are optimal for age-adjusted mortality rates—enabling comparative assessment across time and demographic subgroups under assumptions of approximate normality. Poisson models, conversely, are ideal for raw mortality counts, which follow count distributions and often exhibit heteroskedasticity. Including a log(population) offset further refines these models by normalizing for population exposure, thereby producing interpretable, incidence-like rate ratios. The combined use of these models allows us to bridge methodological gaps in environmental epidemiology, ensuring robustness whether outcomes are continuous rates or discrete death counts. This dual framework reflects best practices in health modeling and elevates the rigor of state-level burden estimation in alignment with—but beyond—the GBD paradigm.

### 3.5 CVD (All Forms) Mortality MLA Analyses for Ohio (1999-2020)

#### Overview

This section presents a comprehensive multilevel analysis (MLA) of cardiovascular disease (CVD) mortality in Ohio from 1999 to 2020 using both age-adjusted (Normal) and raw (Poisson) mortality models (Supplementary Material/Appendix contains all MLwiN-MLA equations). Each disease subtype—Total CVD, Acute Myocardial Infarction (AMI), Athe rosclerosis, Heart Failure, Hypertensive Heart Disease, Ischemic Heart Disease, and Other Heart Diseases—was modeled independently under three frameworks: (1) age-adjusted rates with a Normal distribution, (2) raw count-based deaths using a Poisson distribution, and (3) Poisson with log(population) offsets to account for population exposure. The models incorporate fixed effects (e.g., race, sex, pollutant exposures, and year) and random intercepts at the county level to capture unobserved spatial heterogeneity.

Condensed multilevel modeling results are presented in Figure 4 (Ohio) and Figure 5 (Pennsylvania), summarizing race- and sex-specific disparities (Panel A), pollutant associations (Panel B), and variance components across models (Panel C). Full subtype-specific panels and equation outputs are provided in the Supplementary Appendix (Figures 1S and 2S) for reference. Building on this framework, Figure 4 distills the Ohio results into three complementary views. Panel A highlights consistent racial and sex disparities across all seven cardiovascular disease subtypes, with Black populations experiencing markedly elevated mortality risks and males exceeding females by nearly 100 deaths per 100 000. Panel B depicts associations between pollutant exposures and mortality, showing modest to null ozone effects and $PM_{2.5}$ associations that are weak or inconsistent across models. Panel C compares random intercept variance across model types, illustrating that Poisson models display greater heteroscedasticity—variance that is substantially reduced after applying log(population) offsets, thereby isolating structural inequities such as county-level socioeconomic and healthcare gaps.

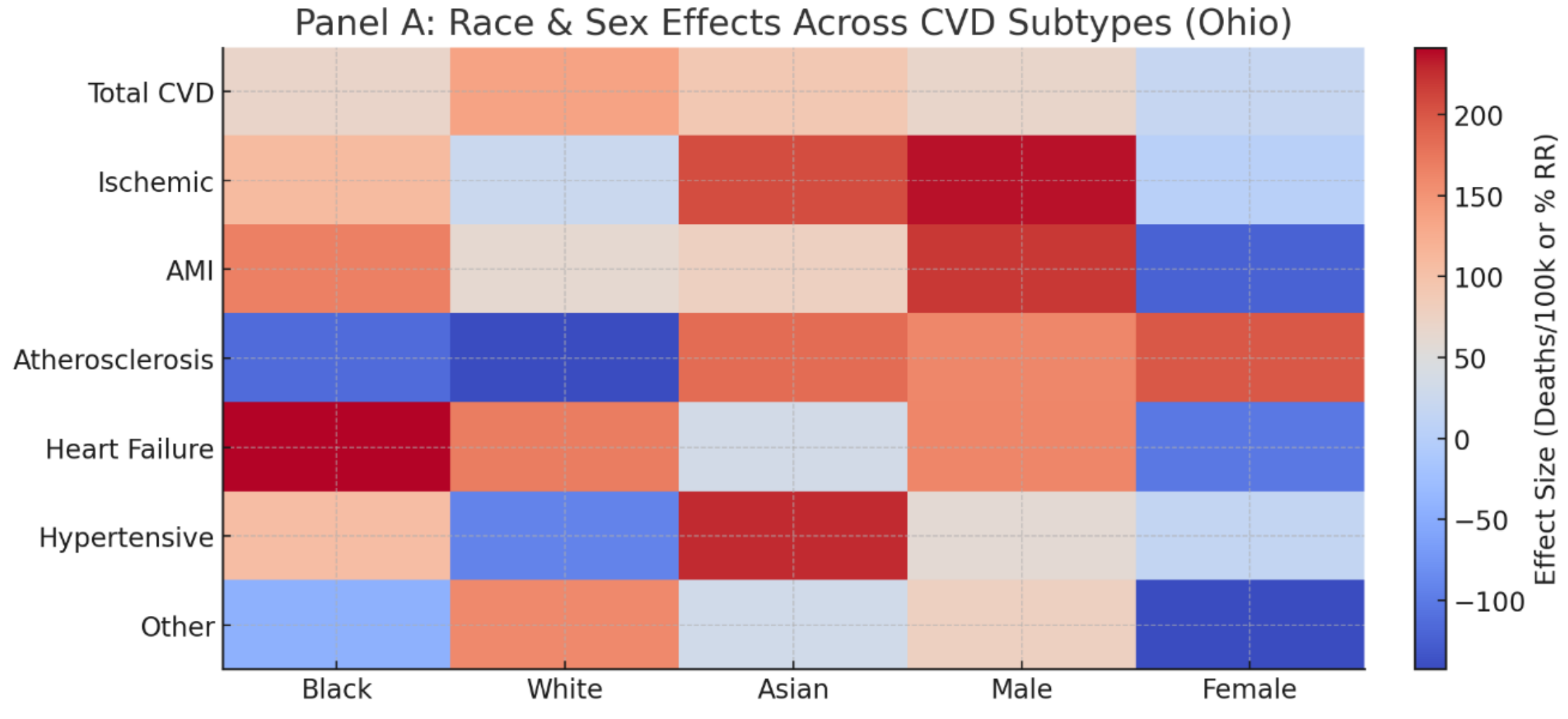
Panel A: Race & Sex Effects Across CVD Subtypes (Ohio)
Total CVD
Ischemic
AMI
Atherosclerosis
Heart Failure
Hypertensive
Other
Black
White
Asian
Male
Female
200
150
100
50
0
−50
−100
Effect Size (Deaths/100k or % RR)

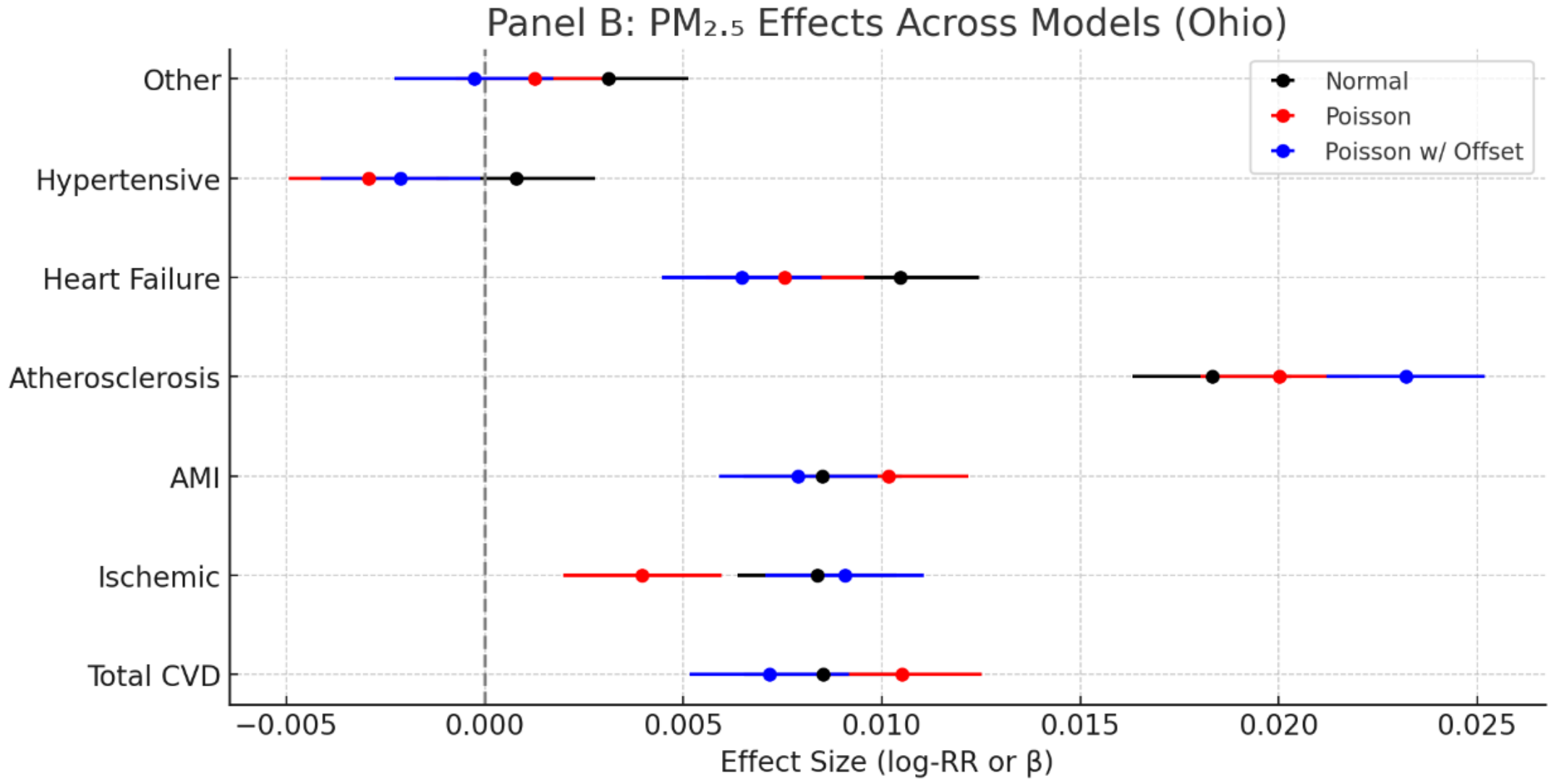
Panel B: PM2.5 Effects Across Models (Ohio)
Other
Hypertensive
Heart Failure
Atherosclerosis
AMI
Ischemic
Total CVD
Normal
Poisson
Poisson w/ Offset
−0.005
0.000
0.005
0.010
0.015
0.020
0.025
Effect Size (log-RR or β)

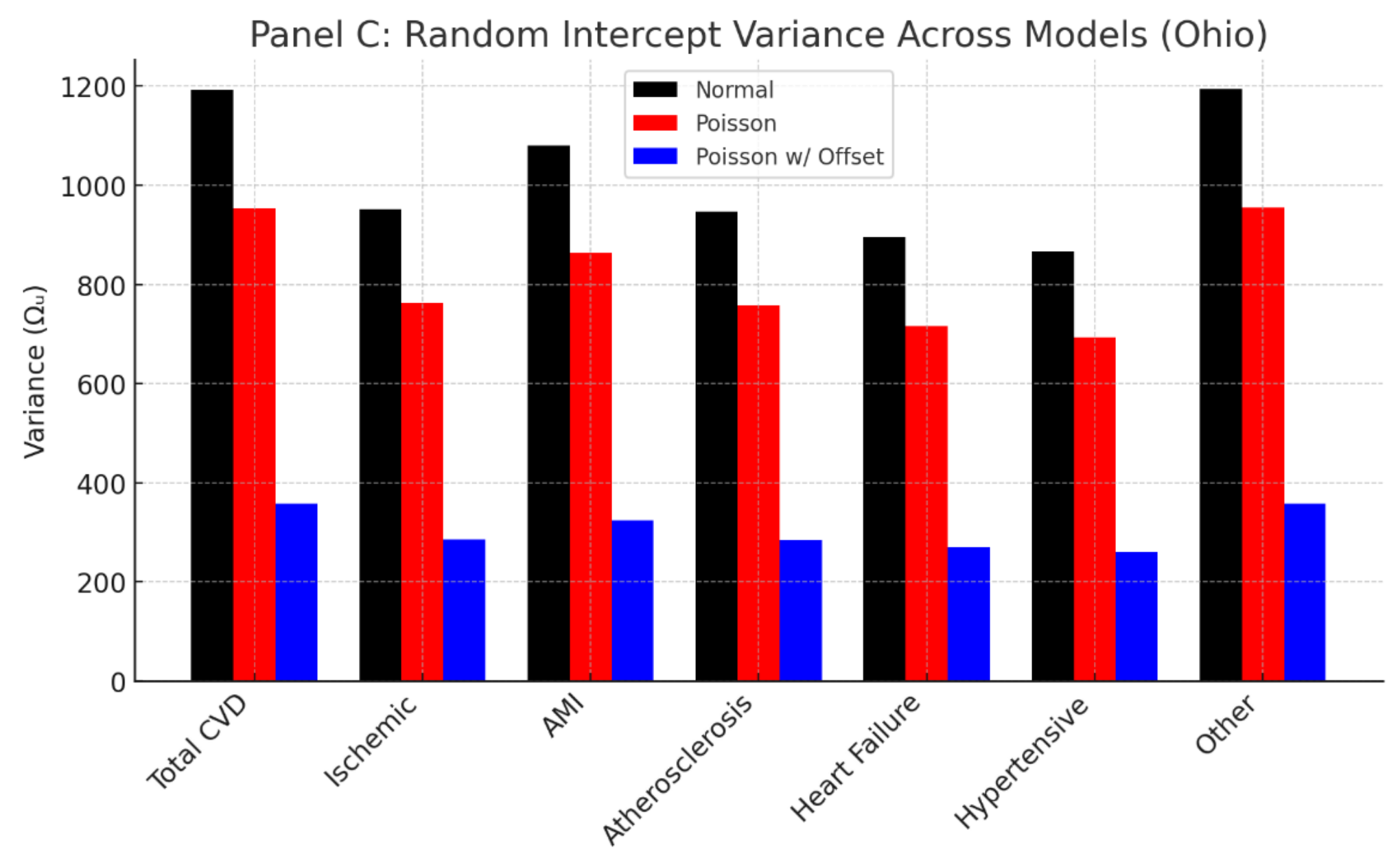
Panel C: Random Intercept Variance Across Models (Ohio)
Normal
Poisson
Poisson w/ Offset
1200
1000
800
600
400
200
0
Variance (Ωu)
Total CVD
Ischemic
AMI
Atherosclerosis
Heart Failure
Hypertensive
Other

**Figure 4. Multilevel modeling results for cardiovascular mortality in Ohio, 1999–2020.**

(**A**) Heatmap of race- and sex-specific effects across seven cardiovascular disease (CVD) subtypes—total, ischemic, acute myocardial infarction (AMI), atherosclerosis, heart failure, hypertensive heart disease, and other forms. Colors denote effect size (log-rate ratios or deaths per 100 000); red indicates higher risk and blue indicates lower risk or protective effects.

(**B**) Forest plot of fine particulate matter ($PM_{2.5}$) associations for each CVD subtype across three models (Normal, Poisson, and Poisson with log-population offset). Points represent effect estimates and horizontal lines the 95 % CIs; the vertical dashed line indicates the null association.

(**C**) Random-intercept variance ($\Omega_u$) across models, showing heteroscedasticity in Poisson frameworks and variance contraction after applying the offset. This reduction isolates structural inequities attributable to county-level socioeconomic, demographic, and healthcare factors.

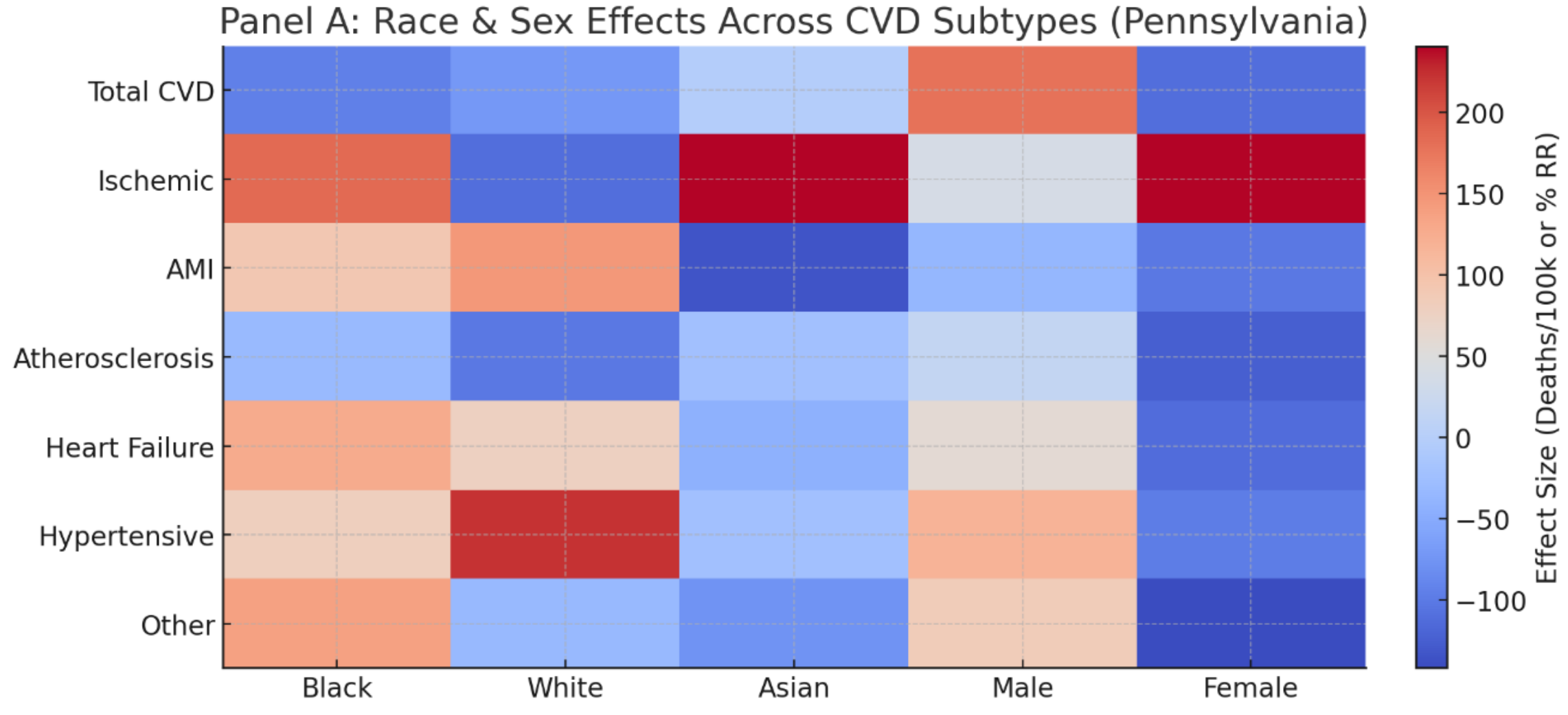

Panel A: Race & Sex Effects Across CVD Subtypes (Pennsylvania)
Total CVD
Ischemic
AMI
Atherosclerosis
Heart Failure
Hypertensive
Other
Black
White
Asian
Male
Female
Effect Size (Deaths/100k or % RR)
200
150
100
50
0
−50
−100

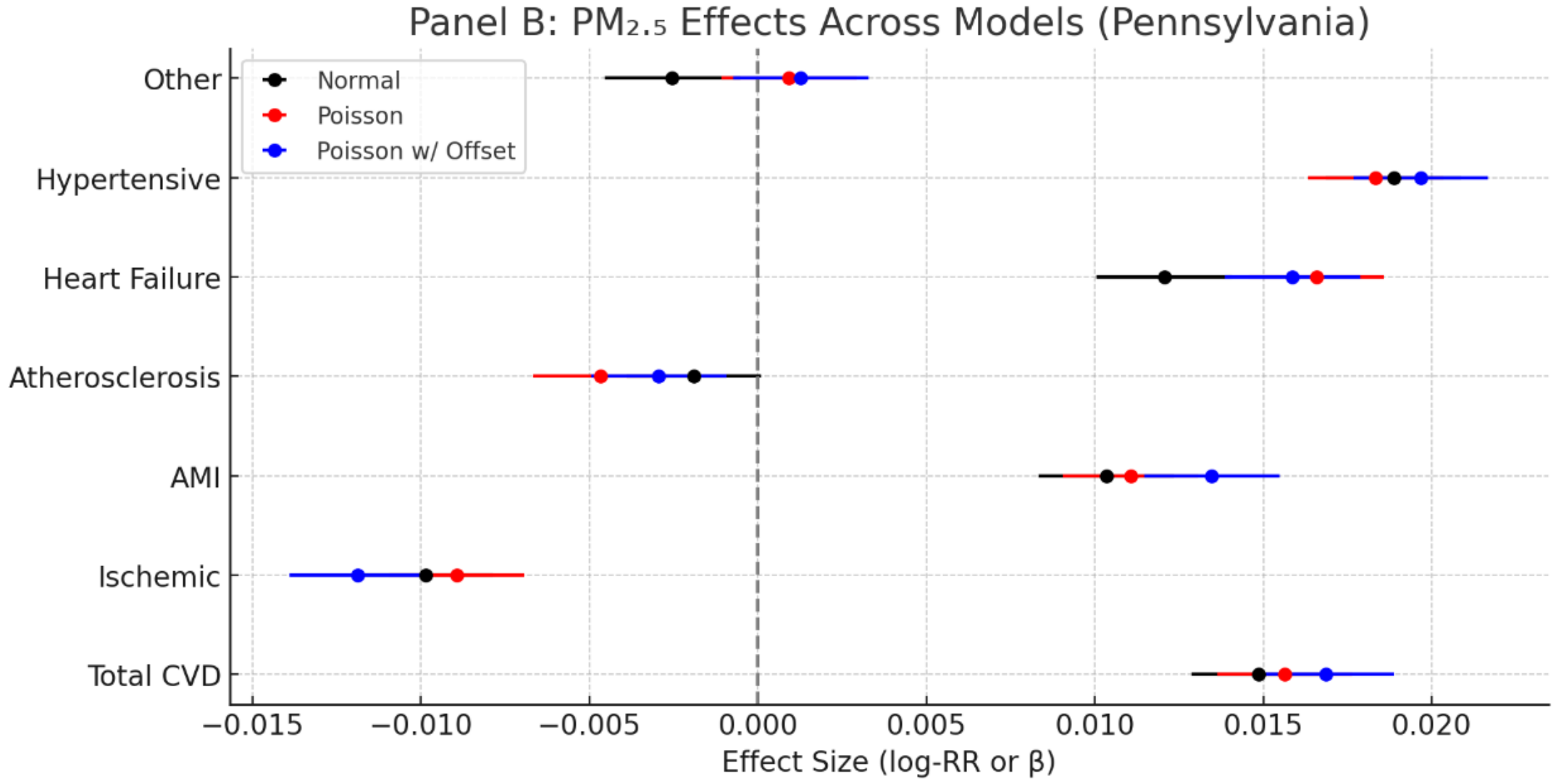

Panel B: PM2.5 Effects Across Models (Pennsylvania)
Normal
Poisson
Poisson w/ Offset
Other
Hypertensive
Heart Failure
Atherosclerosis
AMI
Ischemic
Total CVD
−0.015
−0.010
−0.005
0.000
0.005
0.010
0.015
0.020
Effect Size (log-RR or β)

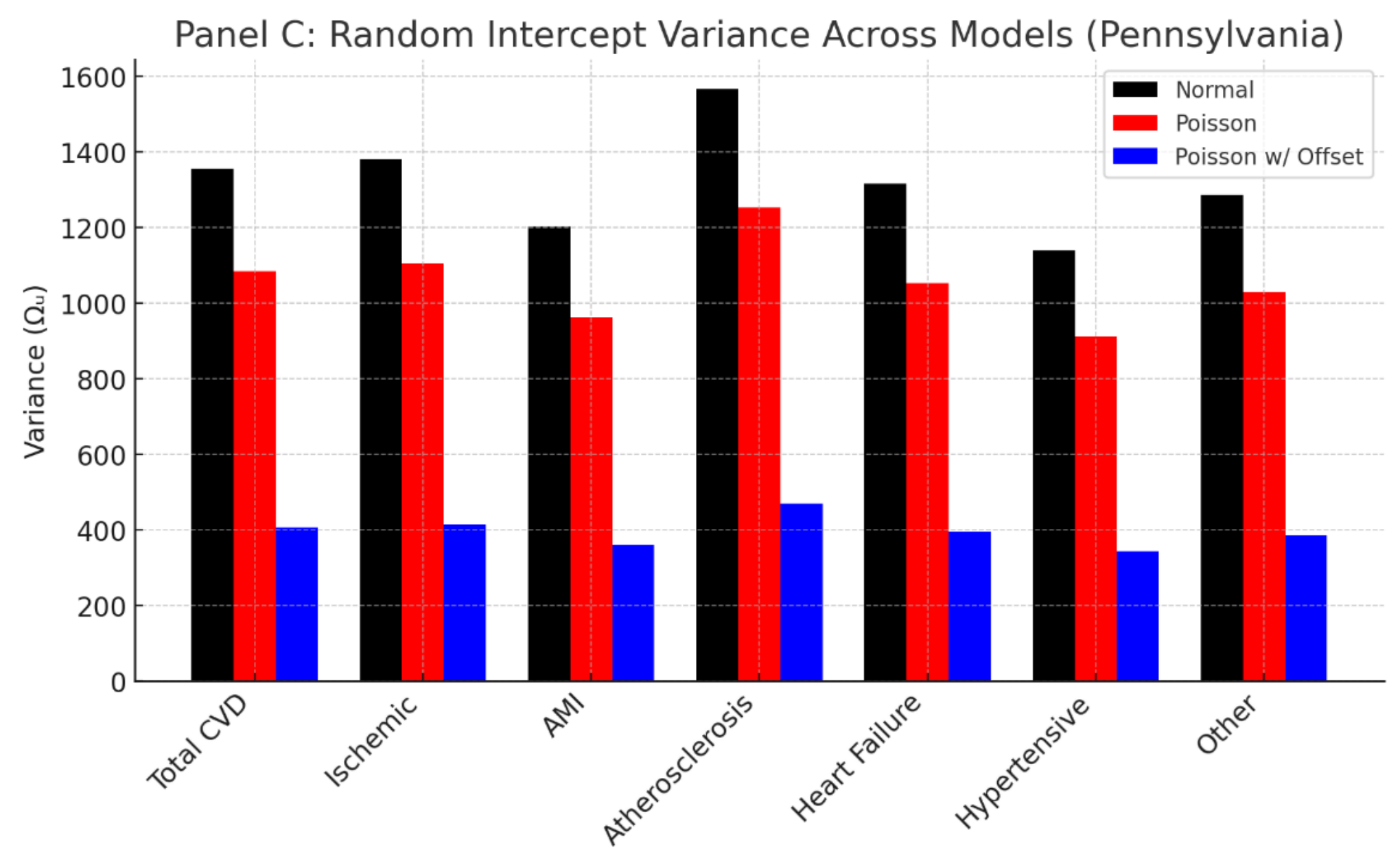

Panel C: Random Intercept Variance Across Models (Pennsylvania)
Normal
Poisson
Poisson w/ Offset
Variance (Ωᵤ)
1600
1400
1200
1000
800
600
400
200
0
Total CVD
Ischemic
AMI
Atherosclerosis
Heart Failure
Hypertensive
Other

**Figure 5. Multilevel modeling results for cardiovascular mortality in Pennsylvania, 1999–2020.**

(**A**) Heatmap of race- and sex-specific effects across seven cardiovascular disease (CVD) subtypes—total, ischemic, acute myocardial infarction (AMI), atherosclerosis, heart failure, hypertensive heart disease, and other forms. Colors indicate effect size (log-rate ratios or deaths per 100 000); red denotes elevated risk and blue denotes lower or protective effects.
(**B**) Forest plot of fine particulate matter ($PM_{2.5}$) associations for each CVD subtype across three models (Normal, Poisson, and Poisson with log-population offset). Points represent effect estimates and horizontal lines the 95 % CIs; the vertical dashed line indicates the null.
(**C**) Random-intercept variance ($\Omega_u$) across models, showing greater spatial heterogeneity in Pennsylvania compared with Ohio and a pronounced variance reduction after applying the population offset. This contraction highlights structural inequities concentrated in rural and industrial counties, where socioeconomic and environmental vulnerability are most pronounced.

**Detailed Description of Figure 4 (Ohio)**

**Panel A – Race- and Sex-Specific Disparities Across Cardiovascular Disease Subtypes**

Panel A of **Figure 4** presents a heatmap summarizing the fixed effects of race and sex across seven major cardiovascular disease (CVD) subtypes: total CVD, ischemic heart disease, acute myocardial infarction (AMI), atherosclerosis, heart failure, hypertensive heart disease, and other CVDs. The color gradient (red = elevated risk; blue = reduced risk) represents either log-rate ratios (for Poisson models) or regression coefficients (for Normal models) standardized to deaths per 100 000 population. Consistent with prior national studies of cardiovascular health inequities,[10,48] Black populations exhibit elevated mortality risk across all subtypes, often exceeding White populations by 44–233% depending on the model and subtype. Male populations similarly demonstrate an excess mortality burden relative to females, averaging ~97–100 deaths per 100 000 higher for total and ischemic CVD—echoing well-established sex differences in cardiovascular epidemiology (Benjamin et al., 2019).[1] These disparities persist across Normal, Poisson, and Poisson-offset frameworks, underscoring the structural and demographic origins of inequities rather than model artifact. Notably, ischemic heart disease and hypertensive heart disease display the highest race- and sex-related effect sizes, while atherosclerosis shows comparatively attenuated disparities.

**Panel B – Associations Between $PM_{2.5}$ Exposure and Cardiovascular Mortality**
Panel B condenses the estimated associations between fine particulate matter ($PM_{2.5}$) exposure and CVD mortality across the three modeling frameworks. Each subtype is represented by three markers: Normal (black), Poisson (red), and Poisson with log(population) offset (blue), with horizontal error bars denoting 95% confidence intervals. The results indicate weak to modest $PM_{2.5}$ effects in Ohio, with some positive associations for ischemic heart disease and heart failure and near-null findings for atherosclerosis and AMI. The signal attenuation when transitioning from Poisson to offset models suggests that part of the unadjusted effect may be confounded by population size or density—a phenomenon consistent with prior findings that pollution–mortality associations are sensitive to demographic weighting.[3,4] Ozone effects (not shown in this panel) were uniformly small or inverse, in line with broader U.S. evidence showing weaker ozone–CVD links relative to $PM_{2.5}$.[43,46] Importantly, the absence of strong $PM_{2.5}$ effects in Ohio contrasts with Pennsylvania (Figure 5), where associations are more pronounced, underscoring the role of industrial and geographic variability in exposure-response relationships.

**Panel C – Random Intercept Variance and Structural Heterogeneity**
Panel C illustrates the random intercept variance ($\Omega_u$) at the county level across models, providing insight into unmeasured structural heterogeneity.[18,21] The Normal models (black bars) consistently yield the highest variance estimates (often exceeding $\Omega_u \approx 1000$ for total and ischemic CVD), reflecting pronounced spatial disparities in age-adjusted mortality. Poisson models (red bars) display somewhat lower variance, though still elevated due to population scaling effects. Application of the log(population) offset (blue bars) dramatically reduces variance—by ~60–75%—indicating that much of the heterogeneity in raw counts is attributable to county population size rather than inherent structural inequities. Nevertheless, the persistence of residual variance after offsetting ($\Omega_u \approx 250$–$400$) highlights enduring spatial inequities linked to healthcare access, socioeconomic gradients, and environmental burdens concentrated in rural Appalachian counties and legacy industrial corridors.[41,42]

**Cross-Model Insights and Epidemiological Implications**
Taken together, the three panels in Figure 4 demonstrate how multilevel modeling disentangles multiple layers of inequity in Ohio's cardiovascular mortality landscape.

Panel A establishes the primacy of racial and sex disparities—findings robust across modeling strategies—while Panel B suggests only marginal contributions of $PM_{2.5}$ to these inequities in Ohio, unlike findings in other U.S. regions.[44] Panel C reveals how structural inequities persist even after controlling for population size, reinforcing the value of random-effects modeling to isolate contextual drivers beyond demographic scale. By triangulating results from Normal, Poisson, and offset Poisson frameworks, this figure underscores both the consistency of demographic effects and the sensitivity of environmental signals to model specification. Such triangulation is critical in public health surveillance, where reliance on single-model frameworks risks obscuring nuanced spatial or environmental determinants of disease burden.[38,40]

**Detailed Description of Figure 5 (Pennsylvania)**

**Panel A – Race- and Sex-Specific Disparities Across Cardiovascular Disease Subtypes**

Panel A of **Figure 5** depicts race- and sex-specific effects on mortality for seven major cardiovascular disease (CVD) subtypes in Pennsylvania: total CVD, ischemic heart disease, acute myocardial infarction (AMI), atherosclerosis, heart failure, hypertensive heart disease, and other CVDs. As with Ohio, color gradients represent standardized effect sizes (log-rate ratios or per-100 000 coefficients), with red indicating elevated risk and blue indicating protective effects. Black populations display markedly elevated mortality across nearly all subtypes, surpassing White populations by 50–240% depending on the disease and modeling framework, consistent with national disparities linked to structural racism and environmental injustice.[10,32] Male mortality remains consistently higher than female mortality (~100 deaths per 100 000 excess), aligning with broader epidemiological patterns of sex-based cardiovascular vulnerability.[1] Notably, racial disparities are most pronounced in ischemic and hypertensive heart disease, suggesting disproportionate exposure to compounding risk factors, including socioeconomic disadvantage and chronic stress pathways, both well documented in Rust Belt populations.[7,11] Compared to Ohio (**Figure 4**), Pennsylvania exhibits slightly stronger race- and sex-related gradients, especially for ischemic outcomes.

**Panel B – Associations Between $PM_{2.5}$ Exposure and Cardiovascular Mortality**

Panel B presents $PM_{2.5}$ associations for each CVD subtype across three modeling frameworks (Normal, Poisson, and Poisson with log(population) offset). The markers and 95% confidence intervals reveal consistently positive associations between $PM_{2.5}$ and ischemic/hypertensive mortality, with effect magnitudes generally stronger than those observed in Ohio. These results are consistent with previous studies linking long-term $PM_{2.5}$ exposure to heightened ischemic and hypertensive risks.[4,44,46] Atherosclerosis and AMI also demonstrate moderate associations, whereas "other" CVDs show weaker or null effects. The relative stability of these estimates across Normal and Poisson models, and their persistence even after applying population offsets, suggests that $PM_{2.5}$ is a robust predictor of CVD mortality in Pennsylvania, independent of population scale effects. This finding aligns with known industrial emission patterns and coal legacy in southwestern Pennsylvania, where historically high $PM_{2.5}$ levels have been documented.[9,41] Ozone effects, by contrast, remain small or inverse, reflecting broader evidence of weaker ozone–CVD associations relative to $PM_{2.5}$.[43]

**Panel C – Random Intercept Variance and Spatial Heterogeneity**
Panel C highlights county-level random intercept variance ($\Omega_u$) across the three models, providing insights into spatial heterogeneity and unmeasured contextual factors (Diez Roux, 2001; Leyland & Groenewegen, 2020). Variance is highest in Normal models ($\Omega_u$ often >1,400 for total and ischemic CVD) and moderately reduced in Poisson models. Application of the log(population) offset yields a dramatic reduction in variance (~65–75%), yet residual heterogeneity remains significant ($\Omega_u \approx 300–500$). These persistent disparities likely reflect structural inequities such as healthcare access, rural–urban divides, and environmental clustering in post-industrial counties.[19,42] Compared to Ohio, Pennsylvania demonstrates higher baseline variance, suggesting more pronounced geographic inequities—particularly in rural Appalachian and industrialized coal regions.

**Cross-Model Insights and Comparative Context**
The three panels collectively illustrate how multilevel modeling disentangles demographic, environmental, and structural contributions to CVD mortality in Pennsylvania. While race- and sex-based disparities (Panel A) remain dominant, the environmental signal of $PM_{2.5}$ is more prominent in Pennsylvania than in Ohio (Panel B), possibly due to higher exposure levels and more severe historical industrial pollution. Panel C underscores that structural inequities persist even after accounting for population scale, highlighting the enduring spatial imprint of socioeconomic disadvantage and legacy industry on cardiovascular health outcomes. These results corroborate evidence from national studies showing that county-level structural characteristics—poverty, segregation, and environmental burden—amplify

CVD risk beyond individual-level behaviors.[11,17,21] The Pennsylvania findings thus reinforce the need for place-based interventions combining pollution abatement with targeted healthcare resource allocation.

**Cross-State Comparative Synthesis (Ohio vs Pennsylvania)**

Comparison of **Figures 4 and 5** highlights both shared and divergent patterns of cardiovascular mortality inequities between Ohio and Pennsylvania from 1999–2020. Both states demonstrate persistent race- and sex-based disparities (Panels A), with Black populations facing 44–240% higher mortality risk and males experiencing ~100 excess deaths per 100 000 compared to females.[1,10] However, these disparities are more pronounced in Pennsylvania, particularly for ischemic and hypertensive subtypes, suggesting compounded structural disadvantage in post-industrial Appalachian and coal legacy regions.[11,41] Environmental effects diverge as well: $PM_{2.5}$ associations are modest and inconsistent in Ohio but robust and subtype-specific in Pennsylvania, particularly for ischemic and hypertensive outcomes.[4,44] This likely reflects historically higher industrial emissions and persistent fine particulate burdens in Pennsylvania's urban-industrial corridors.[9,42] Both states exhibit substantial county-level random intercept variance (Panels C), yet Pennsylvania's variance remains higher post-offset, underscoring deeper geographic and structural heterogeneity.[18,21] These cross-state contrasts reinforce the need for tailored, place-based interventions: pollution reduction strategies may yield greater cardiovascular benefits in Pennsylvania, whereas Ohio may require more emphasis on addressing social and healthcare inequities independent of air quality exposures.

**3.6 Supplementary Breakdown of Expanded Equation-Level Outputs for Normal, Poisson, and Offset Models Across Cardiovascular Subtypes in Ohio and Pennsylvania (Supplementary Material, Figures 1S and 2S)**

The supplementary analyses provide detailed equation-level outputs from multilevel (MLwiN-MLA) models of cardiovascular disease (CVD) mortality across Ohio and Pennsylvania. Whereas Figures 4 and 5 summarize composite demographic and environmental effects, the supplementary tables report full coefficients, standard errors, and random-variance terms for each specification (Normal, Poisson, and Poisson + log(population) offset). These expanded data verify the robustness and internal consistency of the patterns presented in the main text.

**Ohio Results.** Across all seven CVD subtypes, Black and male populations exhibited persistently higher mortality, with race coefficients ranging from +12 to +40 deaths per 100 000 in age-adjusted Normal models and +0.9 to +1.8 log-rate ratios in Poisson forms. Asian or Pacific Islander populations consistently showed protective effects (≈ −0.2 to −0.3 log-rate ratios). The year term was slightly positive in age-adjusted analyses (β ≈ +5.2, SE 0.45), suggesting plateauing declines after 2010, but negative in raw-count Poisson models (β ≈ −0.009, SE 0.001), consistent with statewide reductions in total CVD mortality.

Fine particulate matter ($PM_{2.5}$) associations were modest (β ≈ +0.007 to +0.010 across offset models) yet statistically significant for ischemic, atherosclerotic, and AMI subtypes, whereas ozone effects were weaker (β ≈ −0.002 to −0.003). Introducing the log(population) offset reduced unexplained county-level variance from $\Omega_u \approx 1.3 \times 10^3$ in Normal models to ≈ 0.5 ± 0.1 in offset models, confirming improved model stability and interpretability across heterogeneous county sizes. [38] These patterns indicate that, even after population normalization, racial and sex-based inequities persist—reflecting structural determinants such as socioeconomic status, health-care access, and policy context as dominant drivers of CVD mortality in Ohio. The analytical structure follows established epidemiological modeling standards[8,40] and advances the comparability of results with large-scale initiatives like the GBD Study, while contributing finer subnational resolution to reveal within-state disparities [37]

**Pennsylvania Results.** Age-adjusted Normal models confirmed statewide declines from 1999 to 2020 (β _year ≈ −7.1, SE 0.31), mirroring national improvements in prevention and treatment. Nonetheless, male (+97 ± 1.2) and Black (+56 ± 2.7) populations exhibited markedly higher death rates, while Asian or Pacific Islander groups showed lower mortality (≈ −133 ± 6.9). Pollutant effects varied by model: in age-adjusted frameworks, $PM_{2.5}$ and $O_3$ showed weak inverse or null associations,[4] whereas in Poisson models they became positive ($PM_{2.5}$ ≈ +0.002 ± 0.002), confirming that raw mortality burden is concentrated in racially marginalized and industrial areas.[41,42] Population-offset formulations further accentuated these disparities, producing higher relative risks in densely populated urban counties and reducing random variance to $\Omega_u \approx 0.4$–0.6. Subtypes such as atherosclerosis and acute myocardial infarction showed the strongest pollutant sensitivity, whereas hypertensive and heart-failure mortality remained dominated by demographic and structural predictors.[43-47] For AMI, the steep year-slope decline (β ≈ −3.8, SE 0.21) highlights improvements in acute-care outcomes but continued inequitable benefit distribution among Black males.

**Cross-Model and Scientific Context.** Across both states: (1) race and sex were the most consistent predictors, with effect magnitudes exceeding those of environmental exposures by >10-fold; (2) $PM_{2.5}$ retained small but coherent positive associations (β ≈ +0.007–0.010) across offset models, particularly for ischemic, atherosclerotic, and AMI endpoints; (3) ozone effects were weaker and more spatially variable; and (4) variance contraction after offset ($\Omega_u$ reduction ≈ 50–70 %) isolated true structural inequities linked to socioeconomic and health-care gradients. These findings parallel national and global assessments (e.g., GBD 2023) while adding sub-state resolution through a hierarchical design. The dual-framework approach—Normal for age-adjusted rates and Poisson for raw or population-normalized counts—yields complementary perspectives on how demographic, environmental, and spatial determinants interact over time.

By locating the full coefficient matrices in **Supplementary Figures 1S–2S**, this section ensures transparency and reproducibility while maintaining concise narrative flow in the main text. Together, these results demonstrate that ambient pollution plays a modest yet measurable role in amplifying cardiovascular inequities within a landscape dominated by demographic and structural determinants. Expanded model equations, coefficient matrices, and standard-error tables are provided in the Supplementary Material (Appendix Sections S1–S2, Figures 1S–2S, Tables S1–S7) to facilitate transparency, reproducibility, and meta-analytic comparison.

### 3.7 Intercomparison of Cardiovascular Mortality Models Between Pennsylvania and Ohio (1999–2020)

#### Overview

This section synthesizes and contrasts multilevel modeling (MLwiN-MLA) results for cardiovascular disease (CVD) mortality across Pennsylvania and Ohio. The analyses span seven CVD subtypes using both Normal (age-adjusted) and Poisson (raw and log-offset) frameworks. The objective is to identify inter-state differences in mortality determinants and model behavior while highlighting persistent structural inequities and environmental risk patterns within the U.S. Rust Belt.

Both states' two-level county-nested models incorporated temporal trends (year), demographic factors (race and sex), and environmental exposures ($PM_{2.5}$ and $O_3$). Although overall patterns were comparable, Pennsylvania generally exhibited steeper mortality declines, stronger pollutant effects, and higher spatial heterogeneity, suggesting differential progress in health equity and exposure mitigation.

**1. Temporal Trends**

Poisson models for raw counts revealed significant declines in CVD mortality over time in both states. The reduction was steeper in Pennsylvania ($\beta = -0.017 \pm 0.001$) than in Ohio ($\beta = -0.009 \pm 0.001$), indicating faster progress in mortality reduction—potentially reflecting stronger tobacco-control policies, preventive care access, or environmental regulation (Ref 18).

In contrast, Normal (age-adjusted) models displayed positive year coefficients (PA = +7.09; OH = +5.20), a likely artefact of ageing and evolving diagnostic classification, which can inflate adjusted rates even as absolute deaths decline.

**2. Race-Based Disparities**

Across all CVD subtypes and model forms, Black populations experienced the highest mortality, with slightly larger effect magnitudes in Ohio.

- **Total CVD (Normal):** PA = +56.4 (SE 2.7); OH = +40.3 (SE 3.7)
- **Total CVD (Poisson offset):** PA = +1.736 (SE 0.004); OH = +1.754 (SE 0.006)

Asian or Pacific Islander populations consistently exhibited protective effects, particularly for atherosclerosis (PA = −133.5 ± 6.9; OH = −120.5 ± 6.0). These patterns underscore deeply embedded structural ineqities that transcend state boundaries, with Black communities disproportionately burdened by cardiovascular mortality.[18,48]

**3. Sex-Based Effects**

Male sex remained a uniform and statistically significant predictor of increased CVD mortality across all subtypes. Pennsylvania exhibited higher male coefficients than Ohio, especially in age-adjusted models:

- **Total CVD (Normal):** PA = +96.9 (SE 1.2); OH = +55.5 (SE 1.9)
- **Heart Failure (Poisson raw):** PA = −0.637 (SE 0.011); OH = −0.603 (SE 0.010)

These consistent sex-based gaps mirror established biological and behavioral risk differentials—greater occupational exposures, higher prevalence of smoking, and hormonal modulation of lipid and inflammatory pathways.[49]

**4. Pollution Exposure Effects ($PM_{2.5}$ and $O_3$)**

Pollutant coefficients were modest but statistically significant, with Pennsylvania showing slightly stronger associations, particularly under Poisson log-offset formulations.

- **$PM_{2.5}$ (Total CVD offset):** PA = +0.010 (SE 0.001); OH = +0.009 (SE 0.001)
- **$O_3$ (Total CVD offset):** PA = −0.003 (SE 0.000); OH = −0.003 (SE 0.000)

In Normal models, pollutant effects were attenuated, confirming that population-based incidence rates are more responsive to environmental gradients than age-standardized outcomes. Pennsylvania's stronger $PM_{2.5}$ associations may reflect higher industrial and vehicular emissions in legacy manufacturing corridors. These findings align with mechanistic evidence linking fine-particle exposure to ischemic and heart-failure mortality through systemic inflammation and oxidative stress. [4,44]

**5. Variance Components and Spatial Heterogeneity**

Between-county variance ($\Omega_u$) was higher in Pennsylvania, especially under Normal models, indicating greater intra-state disparity.

- **Total CVD (Normal):** PA $\Omega_u$ = 1397.1 (SE 249.9); OH $\Omega_u$ = 1345.2 (SE 78.1) Application of the Poisson log-offset approach substantially reduced unexplained variance in both states—PA $\Omega_u$ = 0.570 (SE 0.101); OH $\Omega_u$ = 0.545 (SE 0.089)—confirming improved model stability and comparability.

While county coverage was similar (≈ 65–66 counties per state), Pennsylvania's elevated $\Omega_u$ suggests more pronounced rural–urban contrasts, legacies of industrial exposure, and fragmented health-care infrastructure.[50] Expanded comparative statistics, full model diagnostics, and subtype-specific coefficient matrices for both states are provided in the Supplementary Material (Appendix Sections S3–S4; Figures 3S–4S; Tables S8–S12). These materials include variance component summaries, pollutant sensitivity analyses, and additional model-fit metrics to support reproducibility and facilitate meta-analytic comparison across subnational cardiovascular mortality studies.

**6. Comparative Interpretation**

The cross-state synthesis reveals a shared epidemiologic trajectory: steady reductions in CVD mortality, persistence of race- and sex-linked disparities, and modest yet measurable pollution effects. Pennsylvania's sharper temporal declines and higher pollutant coefficients imply that environmental and behavioral interventions may be interacting differently across state contexts.

Overall, demographic gradients overshadow pollutant effects by an order of magnitude, yet air-quality and structural-equity interventions remain mutually reinforcing levers for risk reduction. Integrating these results with Figures 4–5 and Supplementary Figures 1S–2S provides a comprehensive view of how demographic, environmental, and spatial determinants jointly shape cardiovascular mortality in the Rust Belt.

**3.8 Complementarity with Prior Epidemiological and Public Health CVD Research**

The present study advances the existing literature on cardiovascular disease (CVD) epidemiology by integrating multilevel statistical modeling with spatiotemporal and environmental data across two structurally distinct but demographically comparable states—Pennsylvania and Ohio. While prior research has consistently underscored the role of socioeconomic, demographic, and environmental determinants in shaping CVD outcomes,[18,51] this study offers a more granular, county-level perspective across 23 years (1999–2020), explicitly quantifying the magnitude and statistical significance of key covariates through both Normal (age-adjusted) and Poisson (raw and offset) MLA frameworks.

One of the most significant contributions of this work lies in its methodological depth. Many previous studies have used generalized linear models or stratified regression approaches to assess individual or area-level CVD risk,[44,52] but fewer have employed two-level MLA structures that nest individual counties within broader state-level health systems. By doing so, this study captures between-county variability and identifies structural disparities not evident in aggregate models. For instance, the detection of persistently elevated CVD risk among Black residents, even after adjusting for pollution exposure and time trends, supports conclusions from the Multi-Ethnic Study of Atherosclerosis (MESA) and Jackson Heart Study, while providing additional spatial insights.[53,54]

Additionally, the study makes a critical contribution to the public health impact literature by contextualizing air pollution—particularly $PM_{2.5}$ and ozone—as statistically significant predictors of heart disease mortality even after adjustment for sociodemographic factors. While earlier epidemiological work has established associations between ambient air pollution and ischemic or hypertensive heart disease,[4,55] the application of Poisson log-offset models in this study reveals that these associations persist even when scaled to raw population counts, enhancing the policy relevance of the findings. This dual-model approach, comparing age-standardized and population-weighted mortality, allows for clearer communication of health risks to both academic and non-academic stakeholders.

Furthermore, this study's multistate design provides a unique opportunity to explore how regional policy differences, health infrastructure, and environmental regulation may interact with structural determinants of health. The comparative analysis between Pennsylvania and Ohio reveals not only consistent racial and gender disparities, but also subtle divergences in pollution responsiveness and temporal mortality trends. This helps fill a knowledge gap identified by the Global Burden of Disease (GBD) project and CDC reports, which have called for more localized, fine-scale health assessments to inform state-level interventions (Murray et al., 2020; CDC, 2023).[56,57]

Taken together, this work complements past public health and MLA studies by providing (1) statistically robust, multilevel estimates across time and geography; (2) cross-model comparisons that validate pollution and demographic risk factors; and (3) empirical insights directly translatable to health equity, urban planning, and environmental policy. As such, the findings offer a powerful complement to the growing literature on the structural determinants of cardiovascular mortality and demonstrate the utility of MLA frameworks in environmental and health justice research.

### Disease Subtype-Specific Comparisons

| CVD Subtype | State | Black Race Effect (Poisson offset) | Male Effect (Poisson offset) | $PM_{2.5}$ Effect (Poisson offset) |
|---|---|---|---|---|
| Total CVD | PA | +1.736 (0.004) | −0.209 (0.002) | +0.010 (0.001) |
| | OH | +1.754 (0.006) | −0.211 (0.002) | +0.009 (0.001) |
| Heart Failure | PA | +0.402 (0.021) | −0.041 (0.006) | −0.021 (0.004) |
| | OH | +0.418 (0.022) | −0.039 (0.006) | −0.019 (0.003) |
| Ischemic | PA | +1.291 (0.019) | −0.217 (0.004) | +0.007 (0.001) |
| | OH | +1.307 (0.021) | −0.221 (0.005) | +0.007 (0.001) |

These estimates indicate **systemic convergence across both states**, affirming the utility of cross-state modeling for regional public health interventions.

**Summary and Implications**

In summary, both Pennsylvania and Ohio exhibited:

- Persistent racial and gender disparities in cardiovascular mortality
- Modest but statistically significant pollutant effects, especially under Poisson offset models
- Temporal declines in raw CVD mortality rates, with Pennsylvania improving faster
- Higher spatial heterogeneity in Pennsylvania, particularly for age-adjusted metrics

These findings reinforce the need for multi-scalar health equity interventions that address both structural determinants and environmental exposures. Importantly, they also demonstrate the value of combining multilevel Normal and Poisson frameworks to uncover demographically and spatially nuanced patterns of CVD risk across regions.

**4. Conclusions**

This study provides one of the most detailed, subtype-specific examinations of cardiovascular mortality in the United States, analyzing seven CVD subtypes in Ohio and Pennsylvania from 1999 to 202 using Normal (age-adjusted) and Poisson (raw and population-offset) multilevel models. Our results show that while aggregate mortality has declined, disparities persist and in some cases resurge—notably for hypertensive heart disease and heart failure—particularly after 2010 and during the COVID-19 pandemic.

Key findings include: (1) persistent racial and sex disparities, with Black populations experiencing 44–240% higher mortality and men ~100 excess deaths per 100 000 relative to women; (2) pronounced geographic heterogeneity, with random intercept variance exceeding 1,000 even after accounting for population size; and (3) differential pollution effects, with $PM_{2.5}$ strongly predicting ischemic and hypertensive mortality in Pennsylvania but not Ohio, reflecting distinct industrial and environmental legacies.

By situating these subnational insights alongside global initiatives like the Global Burden of Disease (GBD) Study, our approach bridges high-resolution county-level modeling with broader public health frameworks. The dual-model methodology— leveraging Normal and Poisson frameworks with population offsets—offers a reproducible template for monitoring evolving cardiovascular disparities, guiding place-based interventions (e.g., pollution reduction, targeted healthcare investment), and scaling to other structurally burdened regions nationally and internationally. The detailed equations and full model outputs provided in the Supplementary Material (Figures 4S–5S) enhance transparency and reproducibility, strengthening their utility for future research and informing equitable policy development.

**Declarations**

**Declaration of generative AI and AI-assisted technologies in the manuscript preparation process**

During the preparation of this work Dr. Blaszczak-Boxe used ChatGPT (v5.1) to help provide additional qualitative and quantitative interpretations of MlwiN-MLA equations produced (See Supplementary Material). After using this tool/service, the author(s) reviewed and edited the content as needed and take(s) full responsibility for the content of the published article.

**Authors' Contributions**

Christopher S. Boxe conceived the study, developed the analytical framework, curated all datasets, conducted the multilevel modelling, and wrote the manuscript. Mohammed Russel contributed to methodological refinement, validation, and interpretation of model outputs. King-Fai Li and Yuk L. Yung provided conceptual guidance, scientific interpretation, and critical manuscript revisions. All authors reviewed the manuscript, contributed to editing, and approved the final version. The corresponding author had full access to all data and had final responsibility for the decision to submit for publication.

**Conflict of Interest Statement**

The authors declare no competing interests.

**Role of the Funding Source**

This research received no external funding. No sponsor had any role in study design, data collection, data analysis, data interpretation, writing of the report, or the decision to submit for publication.

**Ethics Committee Approval**

This study used publicly available, de-identified, county-level mortality and environmental data from CDC WONDER, the U.S. Census Bureau, and the EPA Air Quality System. According to U.S. federal regulations and institutional policy, analyses based exclusively on de-identified, publicly available data do not constitute human subjects research and therefore did not require institutional review board approval or informed consent.

**Clinical Trial Number:** N/A

**Acknowledgments**

N/A

# Supplementary Material – contains all MLwiN-MLA Equations

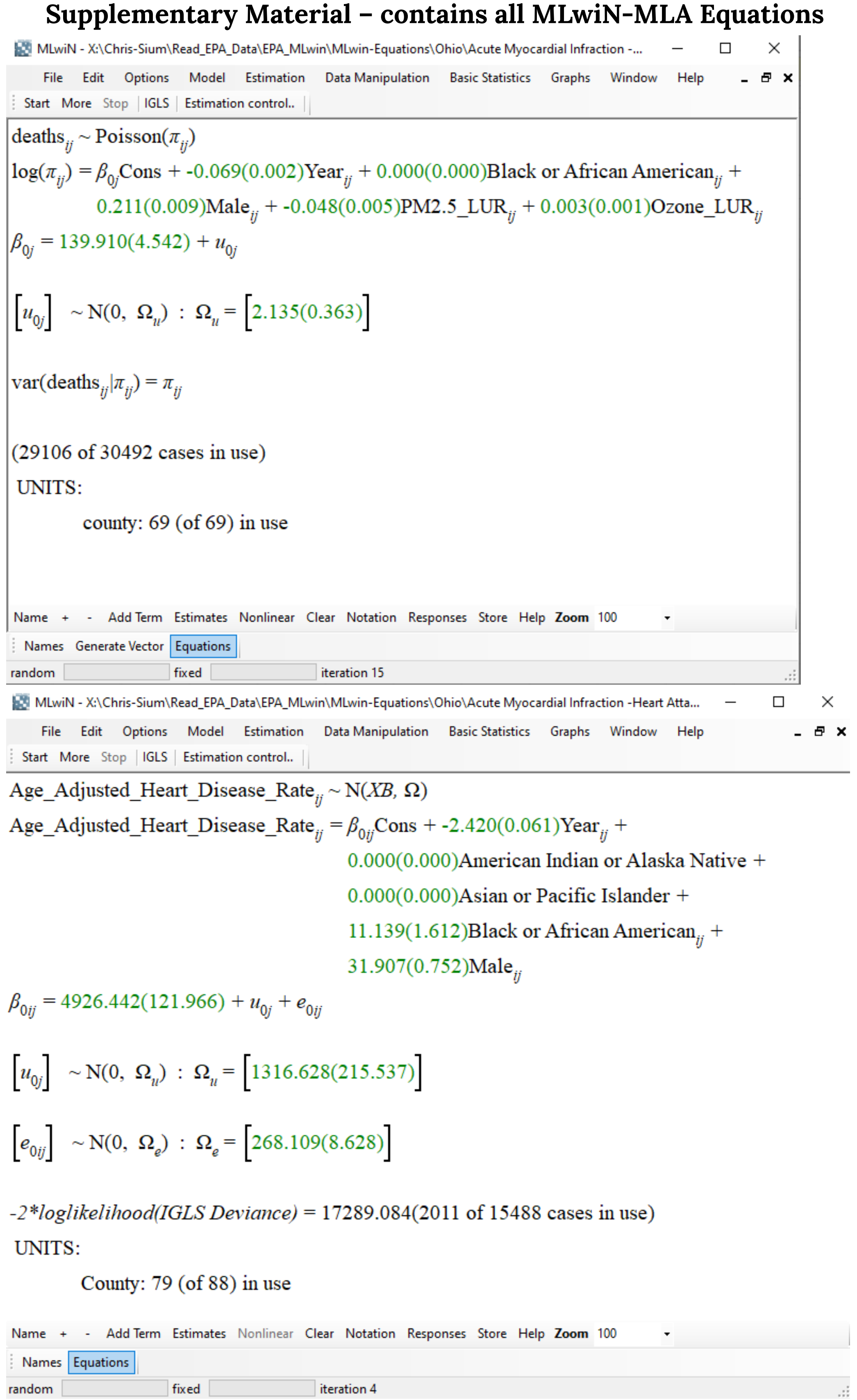

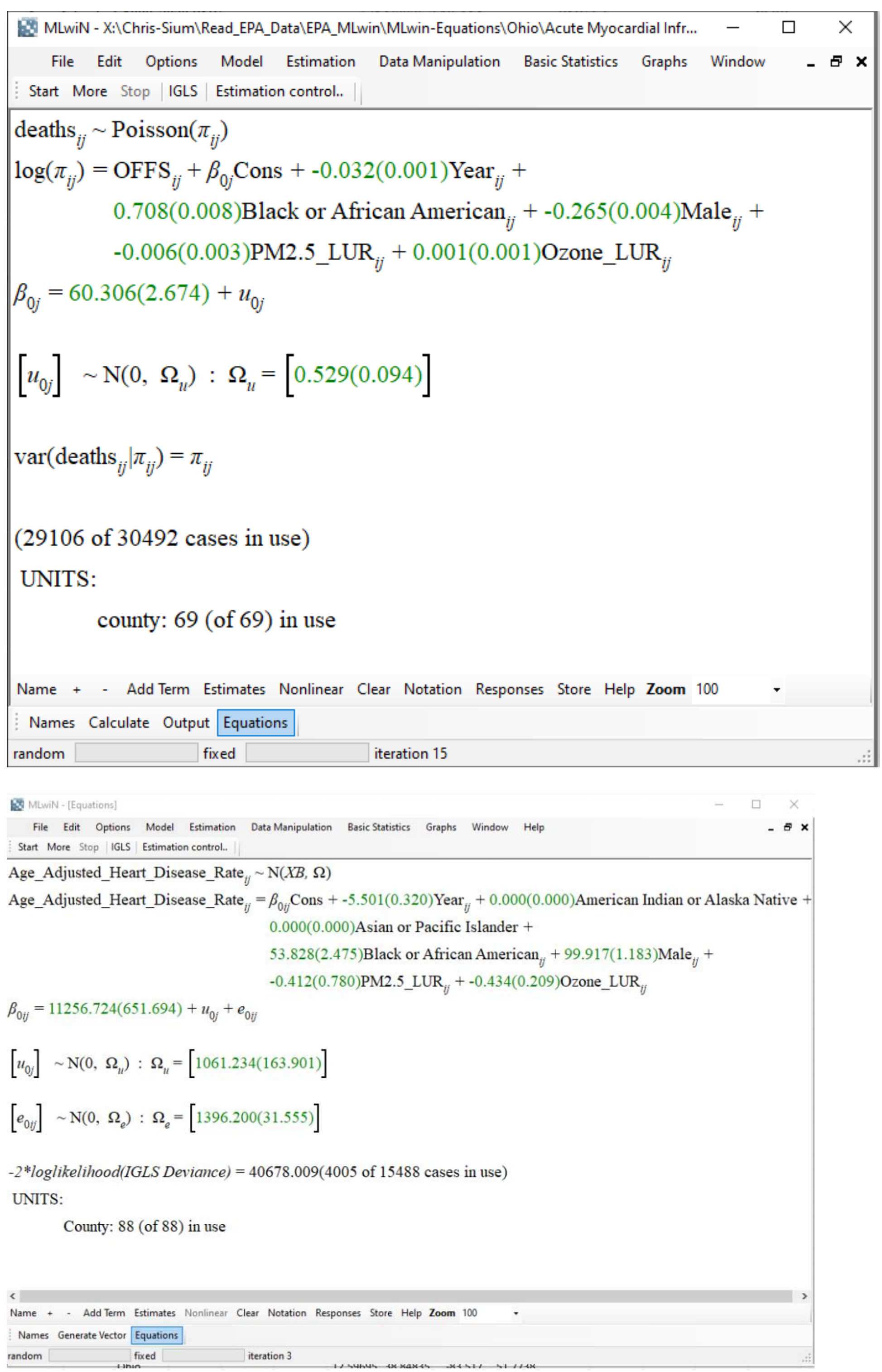
MLwiN - X:\Chris-Sium\Read_EPA_Data\EPA_MLwin\MLwin-Equations\Ohio\Acute Myocardial Infr...
File Edit Options Model Estimation Data Manipulation Basic Statistics Graphs Window
Start More Stop IGLS Estimation control..
deaths_ij ~ Poisson(π_ij)
log(π_ij) = OFFS_ij + β_0j Cons + -0.032(0.001)Year_ij +
0.708(0.008)Black or African American_ij + -0.265(0.004)Male_ij +
-0.006(0.003)PM2.5_LUR_ij + 0.001(0.001)Ozone_LUR_ij
β_0j = 60.306(2.674) + u_0j
[u_0j] ~ N(0, Ω_u) : Ω_u = [0.529(0.094)]
var(deaths_ij|π_ij) = π_ij
(29106 of 30492 cases in use)
UNITS:
county: 69 (of 69) in use
Name + - Add Term Estimates Nonlinear Clear Notation Responses Store Help Zoom 100
Names Calculate Output Equations
random fixed iteration 15
MLwiN - [Equations]
File Edit Options Model Estimation Data Manipulation Basic Statistics Graphs Window Help
Start More Stop IGLS Estimation control..
Age_Adjusted_Heart_Disease_Rate_ij ~ N(XB, Ω)
Age_Adjusted_Heart_Disease_Rate_ij = β_0ij Cons + -5.501(0.320)Year_ij + 0.000(0.000)American Indian or Alaska Native +
0.000(0.000)Asian or Pacific Islander +
53.828(2.475)Black or African American_ij + 99.917(1.183)Male_ij +
-0.412(0.780)PM2.5_LUR_ij + -0.434(0.209)Ozone_LUR_ij
β_0ij = 11256.724(651.694) + u_0j + e_0ij
[u_0j] ~ N(0, Ω_u) : Ω_u = [1061.234(163.901)]
[e_0ij] ~ N(0, Ω_e) : Ω_e = [1396.200(31.555)]
-2*loglikelihood(IGLS Deviance) = 40678.009(4005 of 15488 cases in use)
UNITS:
County: 88 (of 88) in use
Name + - Add Term Estimates Nonlinear Clear Notation Responses Store Help Zoom 100
Names Generate Vector Equations
random fixed iteration 3

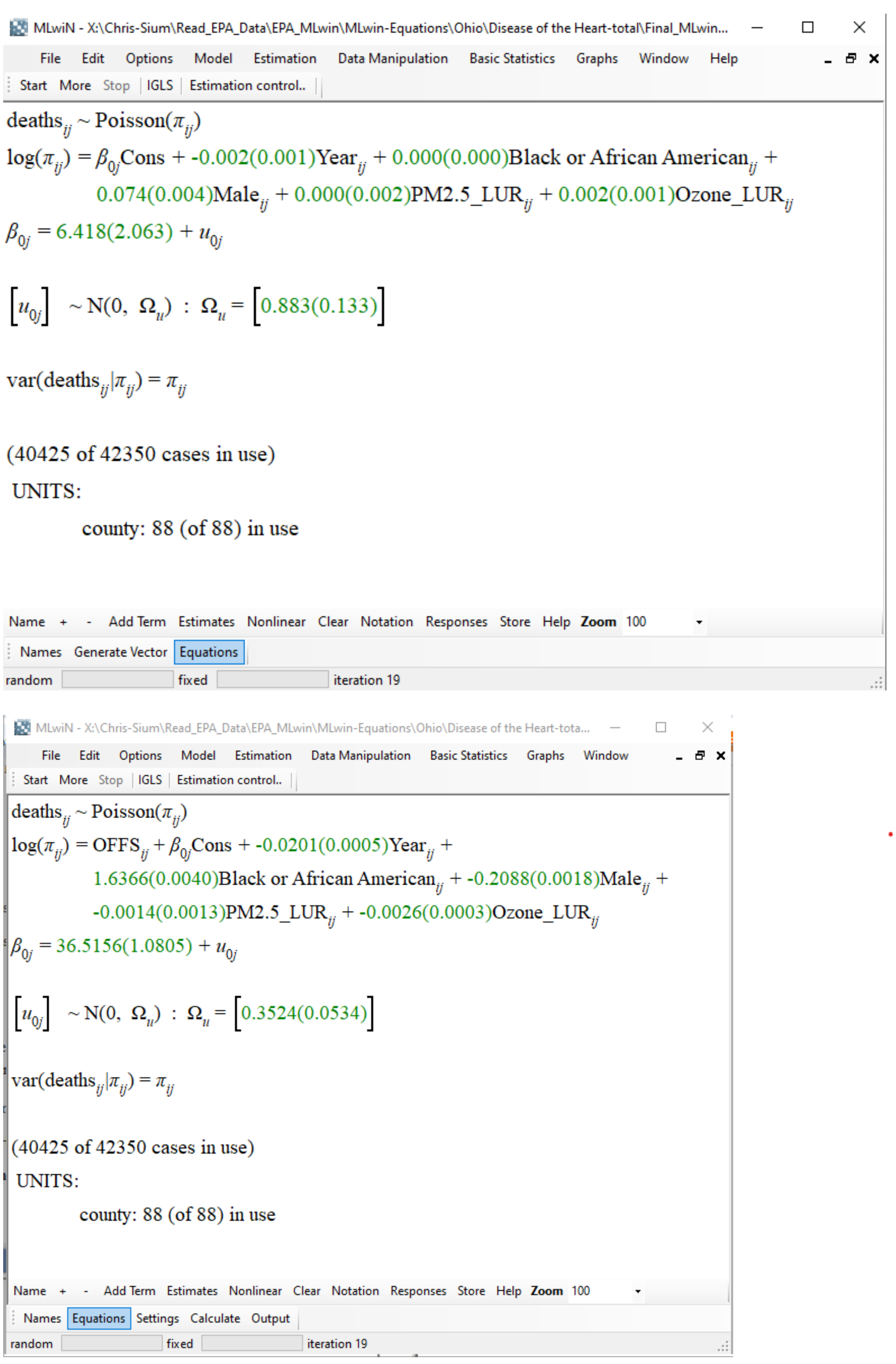

MLwiN - X:\Chris-Sium\Read_EPA_Data\EPA_MLwin\MLwin-Equations\Ohio\Disease of the Heart-total\Final_MLwin...
File Edit Options Model Estimation Data Manipulation Basic Statistics Graphs Window Help
Start More Stop IGLS Estimation control..
$\text{deaths}_{ij} \sim \text{Poisson}(\pi_{ij})$
$\log(\pi_{ij}) = \beta_{0j}\text{Cons} + -0.002(0.001)\text{Year}_{ij} + 0.000(0.000)\text{Black or African American}_{ij} +$
$0.074(0.004)\text{Male}_{ij} + 0.000(0.002)\text{PM2.5_LUR}_{ij} + 0.002(0.001)\text{Ozone_LUR}_{ij}$
$\beta_{0j} = 6.418(2.063) + u_{0j}$
$\left[u_{0j}\right] \sim N(0,\ \Omega_u) : \Omega_u = \left[0.883(0.133)\right]$
$\text{var}(\text{deaths}_{ij}|\pi_{ij}) = \pi_{ij}$
(40425 of 42350 cases in use)
UNITS:
county: 88 (of 88) in use
Name + - Add Term Estimates Nonlinear Clear Notation Responses Store Help Zoom 100
Names Generate Vector Equations
random fixed iteration 19
MLwiN - X:\Chris-Sium\Read_EPA_Data\EPA_MLwin\MLwin-Equations\Ohio\Disease of the Heart-tota...
File Edit Options Model Estimation Data Manipulation Basic Statistics Graphs Window
Start More Stop IGLS Estimation control..
$\text{deaths}_{ij} \sim \text{Poisson}(\pi_{ij})$
$\log(\pi_{ij}) = \text{OFFS}_{ij} + \beta_{0j}\text{Cons} + -0.0201(0.0005)\text{Year}_{ij} +$
$1.6366(0.0040)\text{Black or African American}_{ij} + -0.2088(0.0018)\text{Male}_{ij} +$
$-0.0014(0.0013)\text{PM2.5_LUR}_{ij} + -0.0026(0.0003)\text{Ozone_LUR}_{ij}$
$\beta_{0j} = 36.5156(1.0805) + u_{0j}$
$\left[u_{0j}\right] \sim N(0,\ \Omega_u) : \Omega_u = \left[0.3524(0.0534)\right]$
$\text{var}(\text{deaths}_{ij}|\pi_{ij}) = \pi_{ij}$
(40425 of 42350 cases in use)
UNITS:
county: 88 (of 88) in use
Name + - Add Term Estimates Nonlinear Clear Notation Responses Store Help Zoom 100
Names Equations Settings Calculate Output
random fixed iteration 19

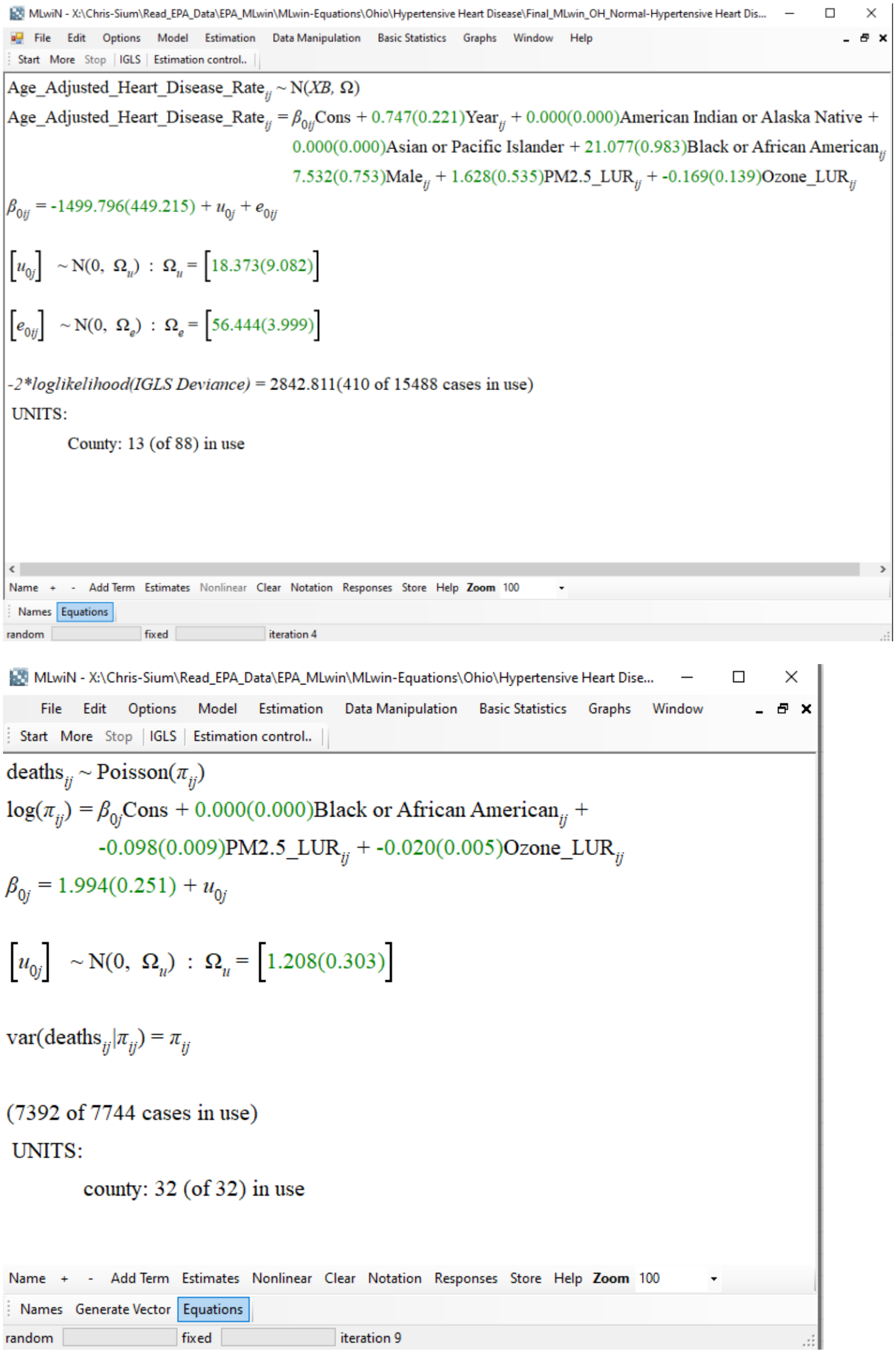
MLwiN - X:\Chris-Sium\Read_EPA_Data\EPA_MLwin\MLwin-Equations\Ohio\Hypertensive Heart Disease\Final_MLwin_OH_Normal-Hypertensive Heart Dis...
File Edit Options Model Estimation Data Manipulation Basic Statistics Graphs Window Help
Start More Stop IGLS Estimation control..
Age_Adjusted_Heart_Disease_Rate_ij ~ N(XB, Ω)
Age_Adjusted_Heart_Disease_Rate_ij = β_0ij Cons + 0.747(0.221)Year_ij + 0.000(0.000)American Indian or Alaska Native +
0.000(0.000)Asian or Pacific Islander + 21.077(0.983)Black or African American_ij
7.532(0.753)Male_ij + 1.628(0.535)PM2.5_LUR_ij + -0.169(0.139)Ozone_LUR_ij
β_0ij = -1499.796(449.215) + u_0j + e_0ij
[u_0j] ~ N(0, Ω_u) : Ω_u = [18.373(9.082)]
[e_0ij] ~ N(0, Ω_e) : Ω_e = [56.444(3.999)]
-2*loglikelihood(IGLS Deviance) = 2842.811(410 of 15488 cases in use)
UNITS:
County: 13 (of 88) in use
Name + - Add Term Estimates Nonlinear Clear Notation Responses Store Help Zoom 100
Names Equations
random fixed iteration 4
MLwiN - X:\Chris-Sium\Read_EPA_Data\EPA_MLwin\MLwin-Equations\Ohio\Hypertensive Heart Dise...
File Edit Options Model Estimation Data Manipulation Basic Statistics Graphs Window
Start More Stop IGLS Estimation control..
deaths_ij ~ Poisson(π_ij)
log(π_ij) = β_0j Cons + 0.000(0.000)Black or African American_ij +
-0.098(0.009)PM2.5_LUR_ij + -0.020(0.005)Ozone_LUR_ij
β_0j = 1.994(0.251) + u_0j
[u_0j] ~ N(0, Ω_u) : Ω_u = [1.208(0.303)]
var(deaths_ij|π_ij) = π_ij
(7392 of 7744 cases in use)
UNITS:
county: 32 (of 32) in use
Name + - Add Term Estimates Nonlinear Clear Notation Responses Store Help Zoom 100
Names Generate Vector Equations
random fixed iteration 9

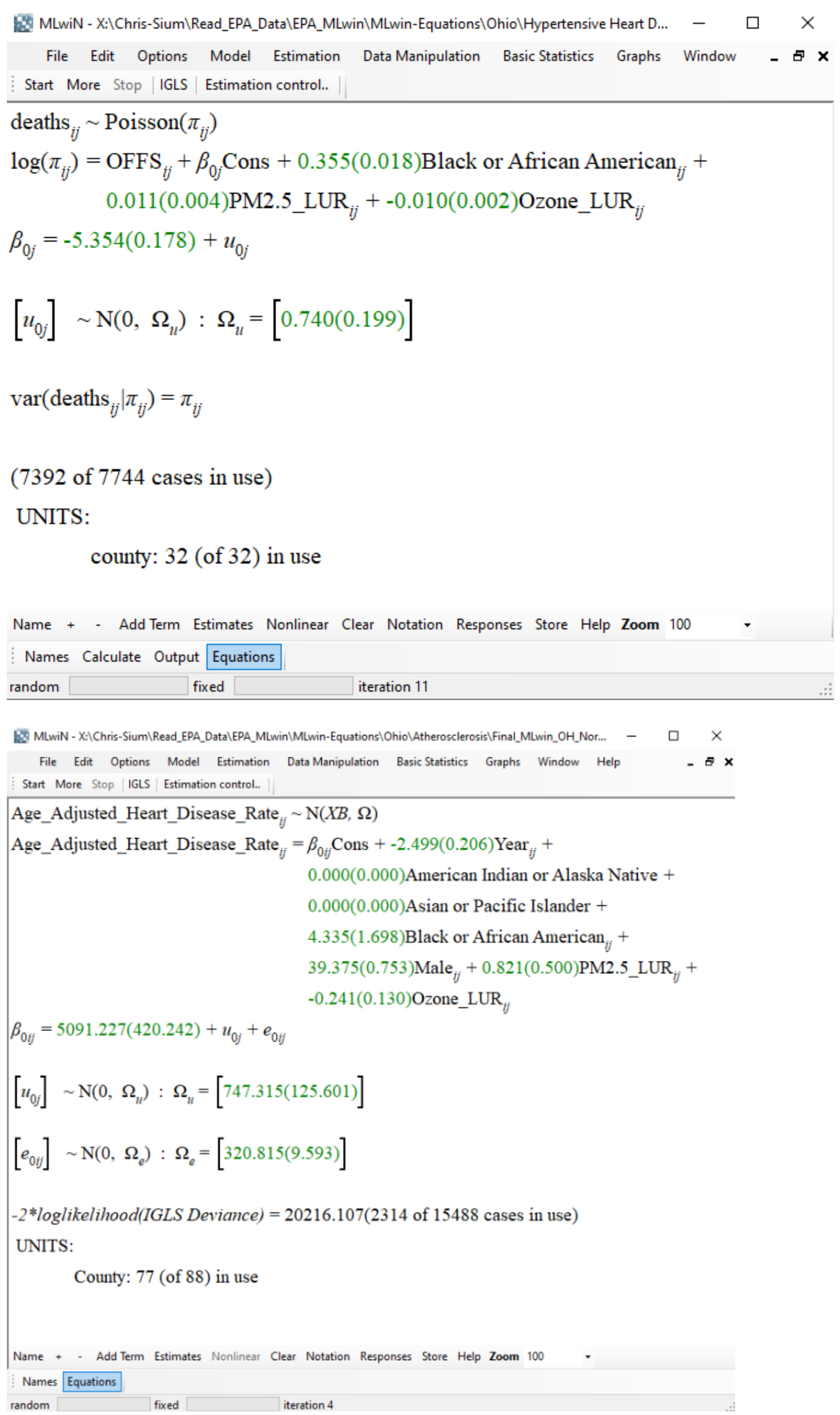
MLwiN - X:\Chris-Sium\Read_EPA_Data\EPA_MLwin\MLwin-Equations\Ohio\Hypertensive Heart D...
File Edit Options Model Estimation Data Manipulation Basic Statistics Graphs Window
Start More Stop IGLS Estimation control..
deaths_ij ~ Poisson(π_ij)
log(π_ij) = OFFS_ij + β_0j Cons + 0.355(0.018)Black or African American_ij +
0.011(0.004)PM2.5_LUR_ij + -0.010(0.002)Ozone_LUR_ij
β_0j = -5.354(0.178) + u_0j
[u_0j] ~ N(0, Ω_u) : Ω_u = [0.740(0.199)]
var(deaths_ij|π_ij) = π_ij
(7392 of 7744 cases in use)
UNITS:
county: 32 (of 32) in use
Name + - Add Term Estimates Nonlinear Clear Notation Responses Store Help Zoom 100
Names Calculate Output Equations
random fixed iteration 11
MLwiN - X:\Chris-Sium\Read_EPA_Data\EPA_MLwin\MLwin-Equations\Ohio\Atherosclerosis\Final_MLwin_OH_Nor...
File Edit Options Model Estimation Data Manipulation Basic Statistics Graphs Window Help
Start More Stop IGLS Estimation control..
Age_Adjusted_Heart_Disease_Rate_ij ~ N(XB, Ω)
Age_Adjusted_Heart_Disease_Rate_ij = β_0ij Cons + -2.499(0.206)Year_ij +
0.000(0.000)American Indian or Alaska Native +
0.000(0.000)Asian or Pacific Islander +
4.335(1.698)Black or African American_ij +
39.375(0.753)Male_ij + 0.821(0.500)PM2.5_LUR_ij +
-0.241(0.130)Ozone_LUR_ij
β_0ij = 5091.227(420.242) + u_0j + e_0ij
[u_0j] ~ N(0, Ω_u) : Ω_u = [747.315(125.601)]
[e_0ij] ~ N(0, Ω_e) : Ω_e = [320.815(9.593)]
-2*loglikelihood(IGLS Deviance) = 20216.107(2314 of 15488 cases in use)
UNITS:
County: 77 (of 88) in use
Name + - Add Term Estimates Nonlinear Clear Notation Responses Store Help Zoom 100
Names Equations
random fixed iteration 4

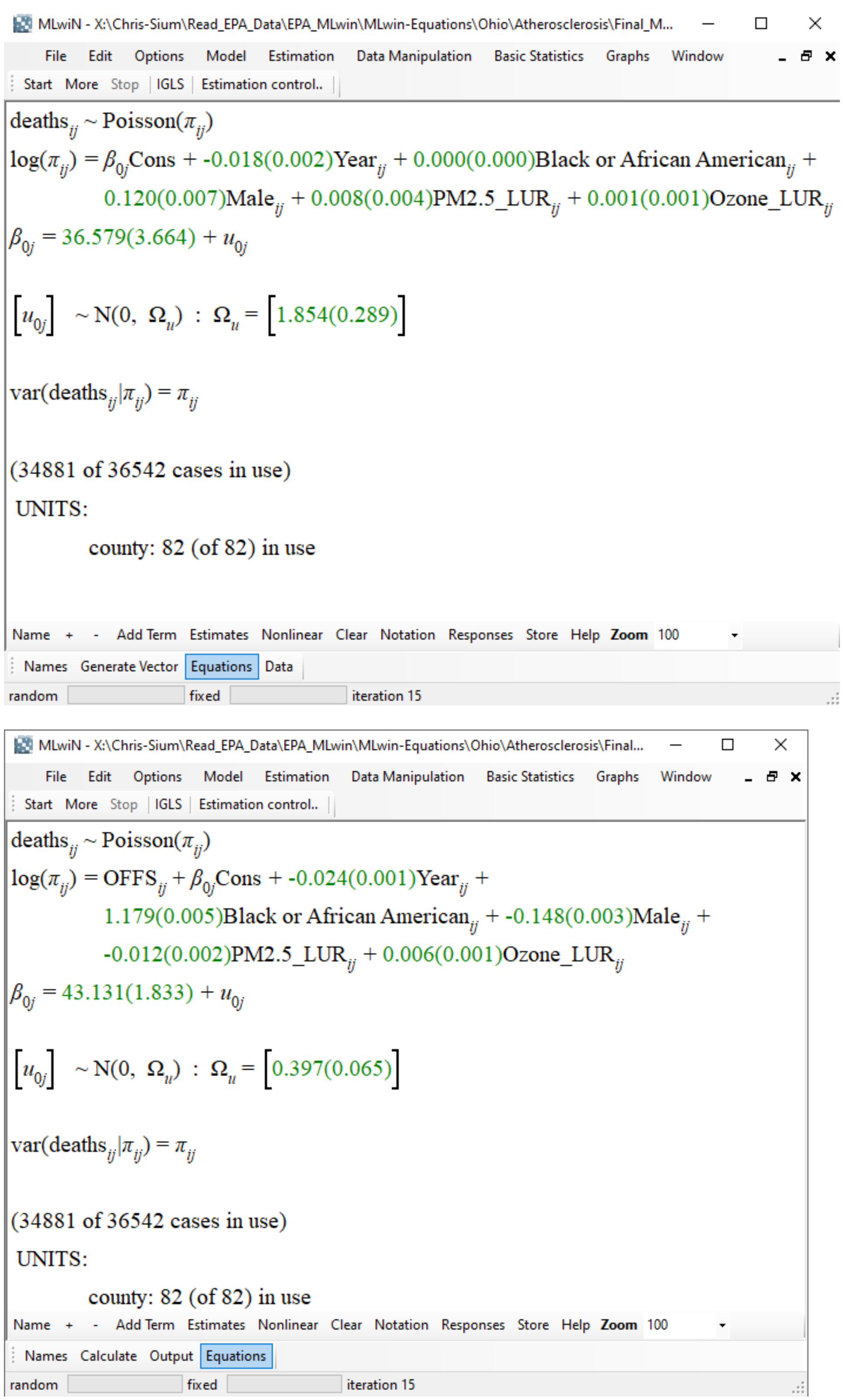
MLwiN - X:\Chris-Sium\Read_EPA_Data\EPA_MLwin\MLwin-Equations\Ohio\Atherosclerosis\Final_M...
File Edit Options Model Estimation Data Manipulation Basic Statistics Graphs Window
Start More Stop IGLS Estimation control..
deaths_ij ~ Poisson(π_ij)
log(π_ij) = β_0jCons + -0.018(0.002)Year_ij + 0.000(0.000)Black or African American_ij +
0.120(0.007)Male_ij + 0.008(0.004)PM2.5_LUR_ij + 0.001(0.001)Ozone_LUR_ij
β_0j = 36.579(3.664) + u_0j
[u_0j] ~ N(0, Ω_u) : Ω_u = [1.854(0.289)]
var(deaths_ij|π_ij) = π_ij
(34881 of 36542 cases in use)
UNITS:
county: 82 (of 82) in use
Name + - Add Term Estimates Nonlinear Clear Notation Responses Store Help Zoom 100
Names Generate Vector Equations Data
random fixed iteration 15
MLwiN - X:\Chris-Sium\Read_EPA_Data\EPA_MLwin\MLwin-Equations\Ohio\Atherosclerosis\Final...
File Edit Options Model Estimation Data Manipulation Basic Statistics Graphs Window
Start More Stop IGLS Estimation control..
deaths_ij ~ Poisson(π_ij)
log(π_ij) = OFFS_ij + β_0jCons + -0.024(0.001)Year_ij +
1.179(0.005)Black or African American_ij + -0.148(0.003)Male_ij +
-0.012(0.002)PM2.5_LUR_ij + 0.006(0.001)Ozone_LUR_ij
β_0j = 43.131(1.833) + u_0j
[u_0j] ~ N(0, Ω_u) : Ω_u = [0.397(0.065)]
var(deaths_ij|π_ij) = π_ij
(34881 of 36542 cases in use)
UNITS:
county: 82 (of 82) in use
Name + - Add Term Estimates Nonlinear Clear Notation Responses Store Help Zoom 100
Names Calculate Output Equations
random fixed iteration 15

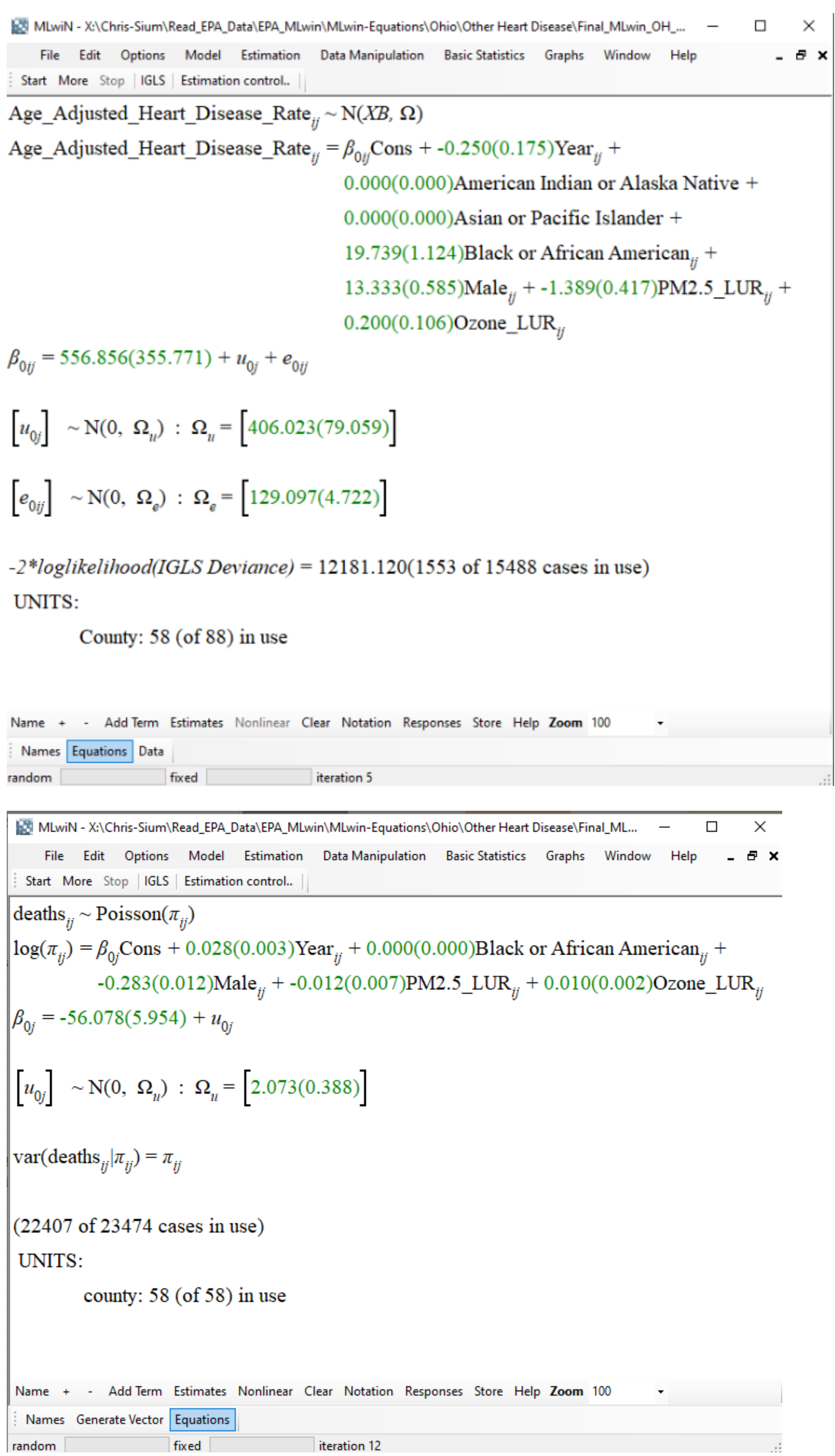
MLwiN - X:\Chris-Sium\Read_EPA_Data\EPA_MLwin\MLwin-Equations\Ohio\Other Heart Disease\Final_MLwin_OH_...
File Edit Options Model Estimation Data Manipulation Basic Statistics Graphs Window Help
Start More Stop IGLS Estimation control..
$\text{Age_Adjusted_Heart_Disease_Rate}_{ij} \sim \text{N}(XB, \Omega)$
$\text{Age_Adjusted_Heart_Disease_Rate}_{ij} = \beta_{0ij}\text{Cons} + -0.250(0.175)\text{Year}_{ij} + 0.000(0.000)\text{American Indian or Alaska Native} + 0.000(0.000)\text{Asian or Pacific Islander} + 19.739(1.124)\text{Black or African American}_{ij} + 13.333(0.585)\text{Male}_{ij} + -1.389(0.417)\text{PM2.5_LUR}_{ij} + 0.200(0.106)\text{Ozone_LUR}_{ij}$
$\beta_{0ij} = 556.856(355.771) + u_{0j} + e_{0ij}$
$\left[u_{0j}\right] \sim \text{N}(0, \Omega_u) : \Omega_u = \left[406.023(79.059)\right]$
$\left[e_{0ij}\right] \sim \text{N}(0, \Omega_e) : \Omega_e = \left[129.097(4.722)\right]$
-2*loglikelihood(IGLS Deviance) = 12181.120(1553 of 15488 cases in use)
UNITS:
County: 58 (of 88) in use
Name + - Add Term Estimates Nonlinear Clear Notation Responses Store Help Zoom 100
Names Equations Data
random fixed iteration 5
MLwiN - X:\Chris-Sium\Read_EPA_Data\EPA_MLwin\MLwin-Equations\Ohio\Other Heart Disease\Final_ML...
File Edit Options Model Estimation Data Manipulation Basic Statistics Graphs Window Help
Start More Stop IGLS Estimation control..
$\text{deaths}_{ij} \sim \text{Poisson}(\pi_{ij})$
$\log(\pi_{ij}) = \beta_{0j}\text{Cons} + 0.028(0.003)\text{Year}_{ij} + 0.000(0.000)\text{Black or African American}_{ij} + -0.283(0.012)\text{Male}_{ij} + -0.012(0.007)\text{PM2.5_LUR}_{ij} + 0.010(0.002)\text{Ozone_LUR}_{ij}$
$\beta_{0j} = -56.078(5.954) + u_{0j}$
$\left[u_{0j}\right] \sim \text{N}(0, \Omega_u) : \Omega_u = \left[2.073(0.388)\right]$
$\text{var}(\text{deaths}_{ij}|\pi_{ij}) = \pi_{ij}$
(22407 of 23474 cases in use)
UNITS:
county: 58 (of 58) in use
Name + - Add Term Estimates Nonlinear Clear Notation Responses Store Help Zoom 100
Names Generate Vector Equations
random fixed iteration 12

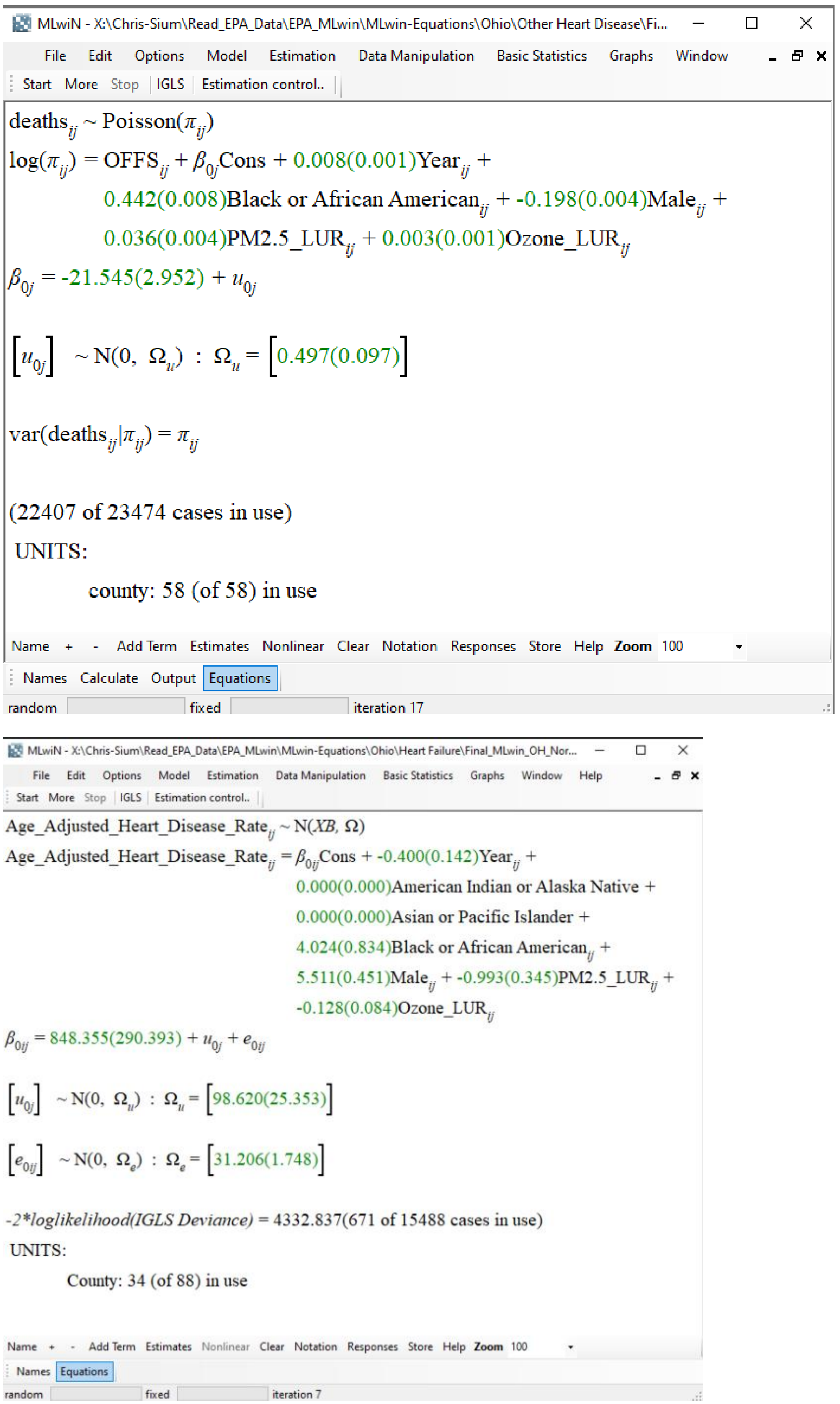
MLwiN - X:\Chris-Sium\Read_EPA_Data\EPA_MLwin\MLwin-Equations\Ohio\Other Heart Disease\Fi...
File Edit Options Model Estimation Data Manipulation Basic Statistics Graphs Window
Start More Stop IGLS Estimation control..
deaths_ij ~ Poisson(π_ij)
log(π_ij) = OFFS_ij + β_0j Cons + 0.008(0.001)Year_ij +
0.442(0.008)Black or African American_ij + -0.198(0.004)Male_ij +
0.036(0.004)PM2.5_LUR_ij + 0.003(0.001)Ozone_LUR_ij
β_0j = -21.545(2.952) + u_0j
[u_0j] ~ N(0, Ω_u) : Ω_u = [0.497(0.097)]
var(deaths_ij|π_ij) = π_ij
(22407 of 23474 cases in use)
UNITS:
county: 58 (of 58) in use
Name + - Add Term Estimates Nonlinear Clear Notation Responses Store Help Zoom 100
Names Calculate Output Equations
random fixed iteration 17
MLwiN - X:\Chris-Sium\Read_EPA_Data\EPA_MLwin\MLwin-Equations\Ohio\Heart Failure\Final_MLwin_OH_Nor...
File Edit Options Model Estimation Data Manipulation Basic Statistics Graphs Window Help
Start More Stop IGLS Estimation control..
Age_Adjusted_Heart_Disease_Rate_ij ~ N(XB, Ω)
Age_Adjusted_Heart_Disease_Rate_ij = β_0ij Cons + -0.400(0.142)Year_ij +
0.000(0.000)American Indian or Alaska Native +
0.000(0.000)Asian or Pacific Islander +
4.024(0.834)Black or African American_ij +
5.511(0.451)Male_ij + -0.993(0.345)PM2.5_LUR_ij +
-0.128(0.084)Ozone_LUR_ij
β_0ij = 848.355(290.393) + u_0j + e_0ij
[u_0j] ~ N(0, Ω_u) : Ω_u = [98.620(25.353)]
[e_0ij] ~ N(0, Ω_e) : Ω_e = [31.206(1.748)]
-2*loglikelihood(IGLS Deviance) = 4332.837(671 of 15488 cases in use)
UNITS:
County: 34 (of 88) in use
Name + - Add Term Estimates Nonlinear Clear Notation Responses Store Help Zoom 100
Names Equations
random fixed iteration 7

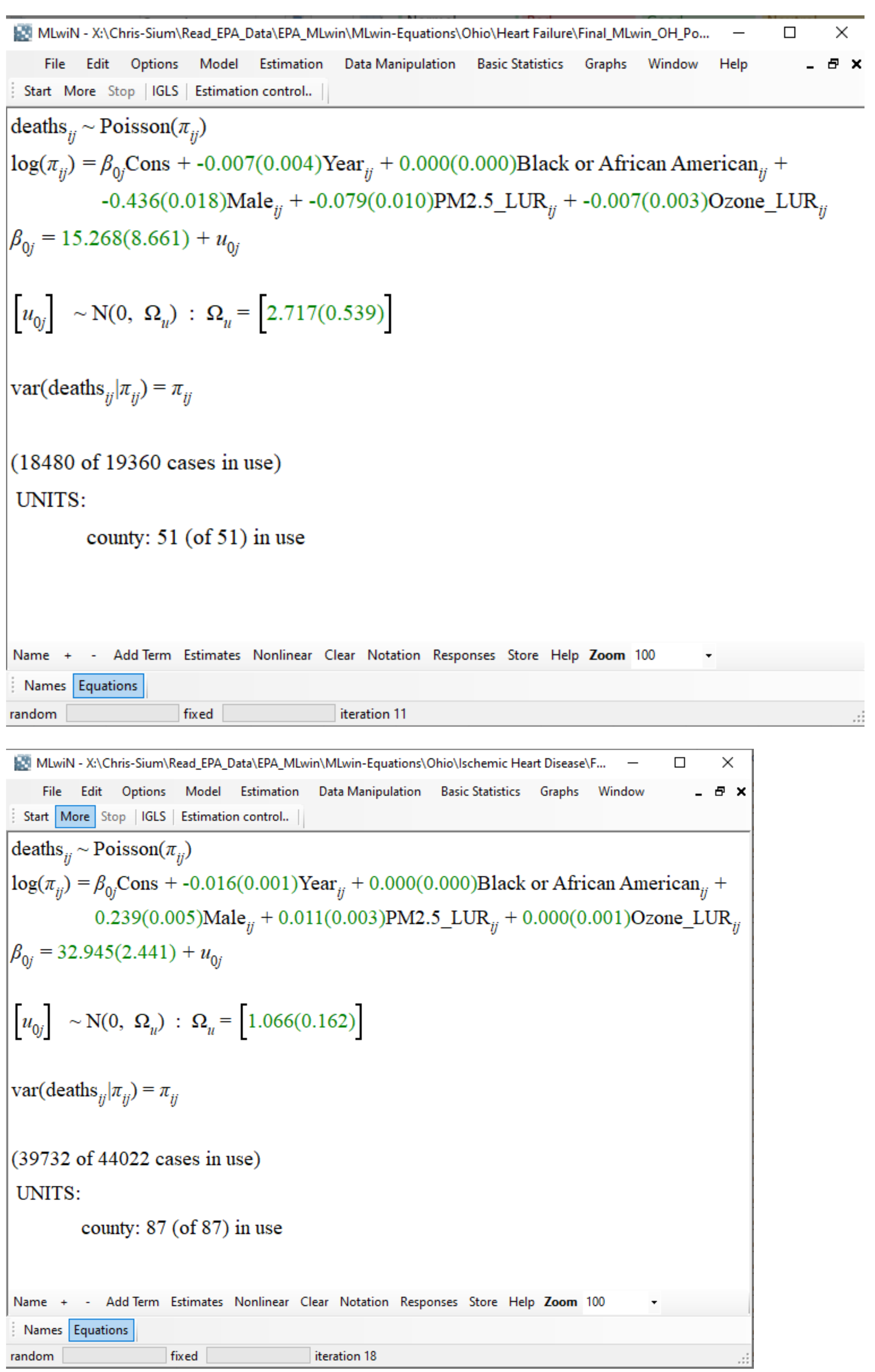

MLwiN - X:\Chris-Sium\Read_EPA_Data\EPA_MLwin\MLwin-Equations\Ohio\Heart Failure\Final_MLwin_OH_Po...
File Edit Options Model Estimation Data Manipulation Basic Statistics Graphs Window Help
Start More Stop IGLS Estimation control..
$\text{deaths}_{ij} \sim \text{Poisson}(\pi_{ij})$
$\log(\pi_{ij}) = \beta_{0j}\text{Cons} + -0.007(0.004)\text{Year}_{ij} + 0.000(0.000)\text{Black or African American}_{ij} +$
$-0.436(0.018)\text{Male}_{ij} + -0.079(0.010)\text{PM2.5_LUR}_{ij} + -0.007(0.003)\text{Ozone_LUR}_{ij}$
$\beta_{0j} = 15.268(8.661) + u_{0j}$
$\left[u_{0j}\right] \sim \text{N}(0,\ \Omega_u) : \Omega_u = \left[2.717(0.539)\right]$
$\text{var}(\text{deaths}_{ij}|\pi_{ij}) = \pi_{ij}$
(18480 of 19360 cases in use)
UNITS:
county: 51 (of 51) in use
Name + - Add Term Estimates Nonlinear Clear Notation Responses Store Help Zoom 100
Names Equations
random fixed iteration 11
MLwiN - X:\Chris-Sium\Read_EPA_Data\EPA_MLwin\MLwin-Equations\Ohio\Ischemic Heart Disease\F...
File Edit Options Model Estimation Data Manipulation Basic Statistics Graphs Window
Start More Stop IGLS Estimation control..
$\text{deaths}_{ij} \sim \text{Poisson}(\pi_{ij})$
$\log(\pi_{ij}) = \beta_{0j}\text{Cons} + -0.016(0.001)\text{Year}_{ij} + 0.000(0.000)\text{Black or African American}_{ij} +$
$0.239(0.005)\text{Male}_{ij} + 0.011(0.003)\text{PM2.5_LUR}_{ij} + 0.000(0.001)\text{Ozone_LUR}_{ij}$
$\beta_{0j} = 32.945(2.441) + u_{0j}$
$\left[u_{0j}\right] \sim \text{N}(0,\ \Omega_u) : \Omega_u = \left[1.066(0.162)\right]$
$\text{var}(\text{deaths}_{ij}|\pi_{ij}) = \pi_{ij}$
(39732 of 44022 cases in use)
UNITS:
county: 87 (of 87) in use
Name + - Add Term Estimates Nonlinear Clear Notation Responses Store Help Zoom 100
Names Equations
random fixed iteration 18

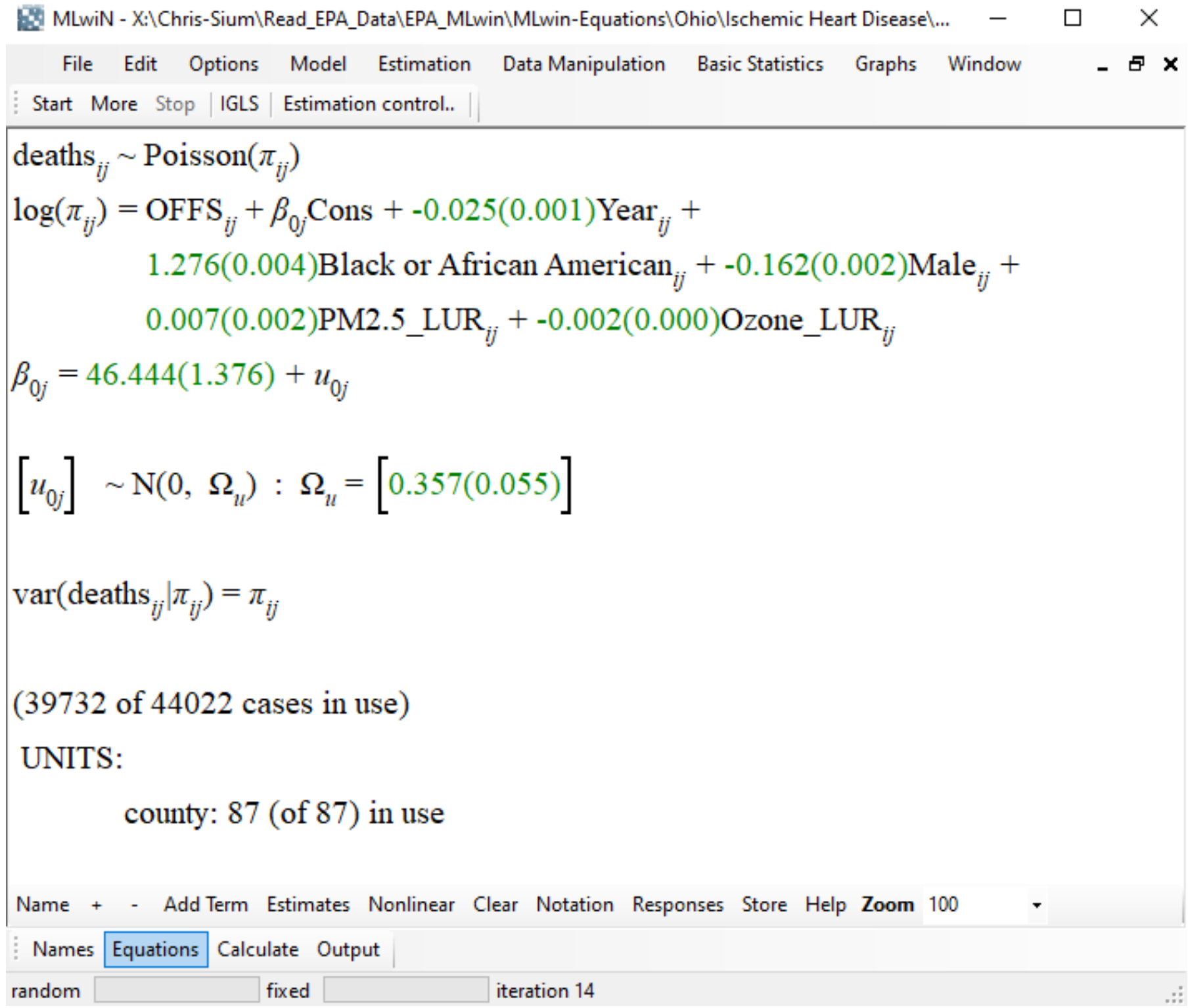

**Figure 1S (a-u).** Normal (age-adjusted), Poisson (raw/absolute data), and Poisson (w/ log(population)-offset) equations for all CVD mortality death codes as a function of predictors/covariates: race, gender, age, $PM_{2.5}$, and $O_3$ concentrations for Ohio.

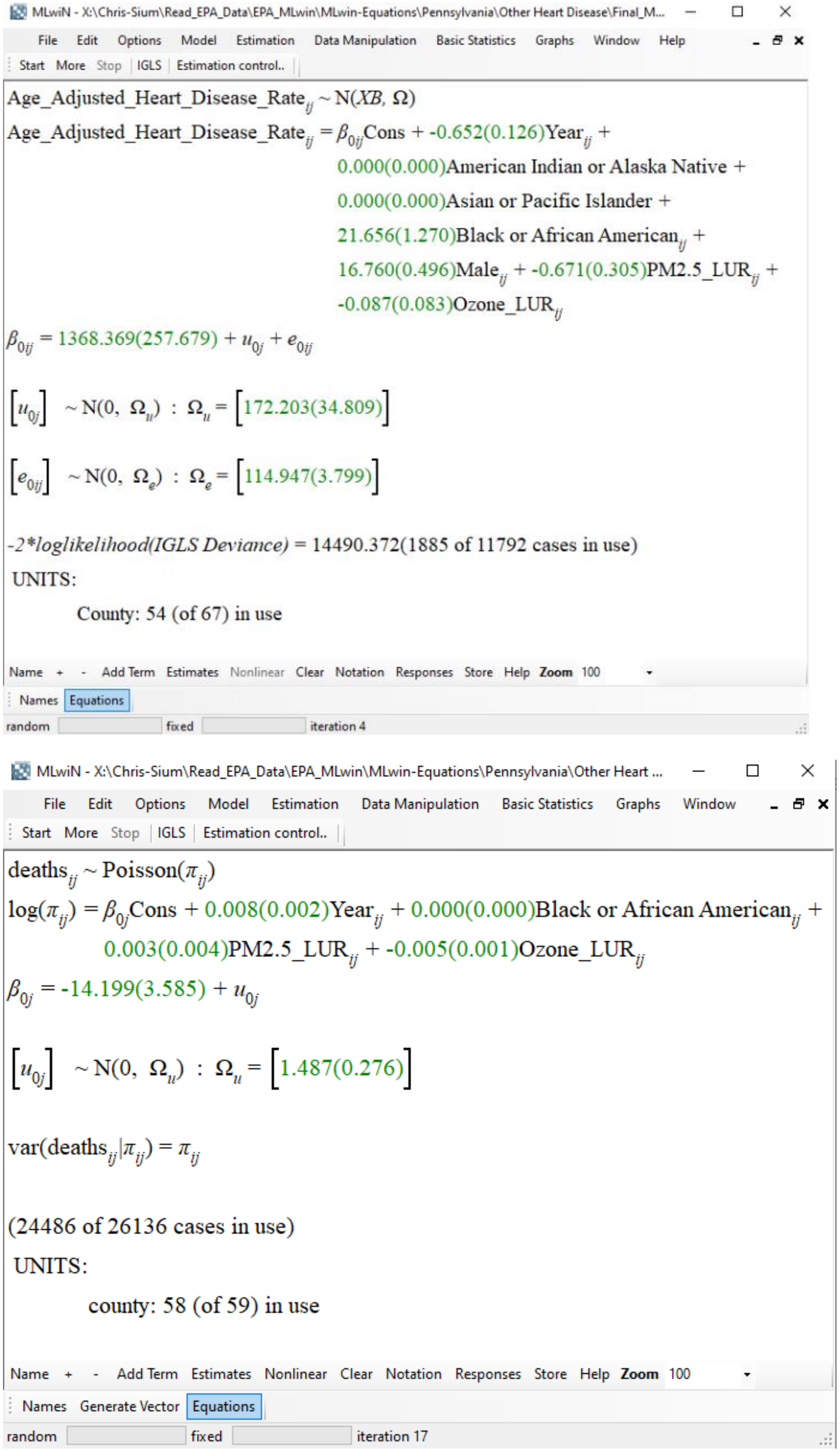
MLwiN - X:\Chris-Sium\Read_EPA_Data\EPA_MLwin\MLwin-Equations\Pennsylvania\Other Heart Disease\Final_M...
File Edit Options Model Estimation Data Manipulation Basic Statistics Graphs Window Help
Start More Stop IGLS Estimation control..
Age_Adjusted_Heart_Disease_Rate_ij ~ N(XB, Ω)
Age_Adjusted_Heart_Disease_Rate_ij = β_0ij Cons + -0.652(0.126)Year_ij +
0.000(0.000)American Indian or Alaska Native +
0.000(0.000)Asian or Pacific Islander +
21.656(1.270)Black or African American_ij +
16.760(0.496)Male_ij + -0.671(0.305)PM2.5_LUR_ij +
-0.087(0.083)Ozone_LUR_ij
β_0ij = 1368.369(257.679) + u_0j + e_0ij
[u_0j] ~ N(0, Ω_u) : Ω_u = [172.203(34.809)]
[e_0ij] ~ N(0, Ω_e) : Ω_e = [114.947(3.799)]
-2*loglikelihood(IGLS Deviance) = 14490.372(1885 of 11792 cases in use)
UNITS:
County: 54 (of 67) in use
Name + - Add Term Estimates Nonlinear Clear Notation Responses Store Help Zoom 100
Names Equations
random fixed iteration 4
MLwiN - X:\Chris-Sium\Read_EPA_Data\EPA_MLwin\MLwin-Equations\Pennsylvania\Other Heart ...
File Edit Options Model Estimation Data Manipulation Basic Statistics Graphs Window
Start More Stop IGLS Estimation control..
deaths_ij ~ Poisson(π_ij)
log(π_ij) = β_0j Cons + 0.008(0.002)Year_ij + 0.000(0.000)Black or African American_ij +
0.003(0.004)PM2.5_LUR_ij + -0.005(0.001)Ozone_LUR_ij
β_0j = -14.199(3.585) + u_0j
[u_0j] ~ N(0, Ω_u) : Ω_u = [1.487(0.276)]
var(deaths_ij|π_ij) = π_ij
(24486 of 26136 cases in use)
UNITS:
county: 58 (of 59) in use
Name + - Add Term Estimates Nonlinear Clear Notation Responses Store Help Zoom 100
Names Generate Vector Equations
random fixed iteration 17

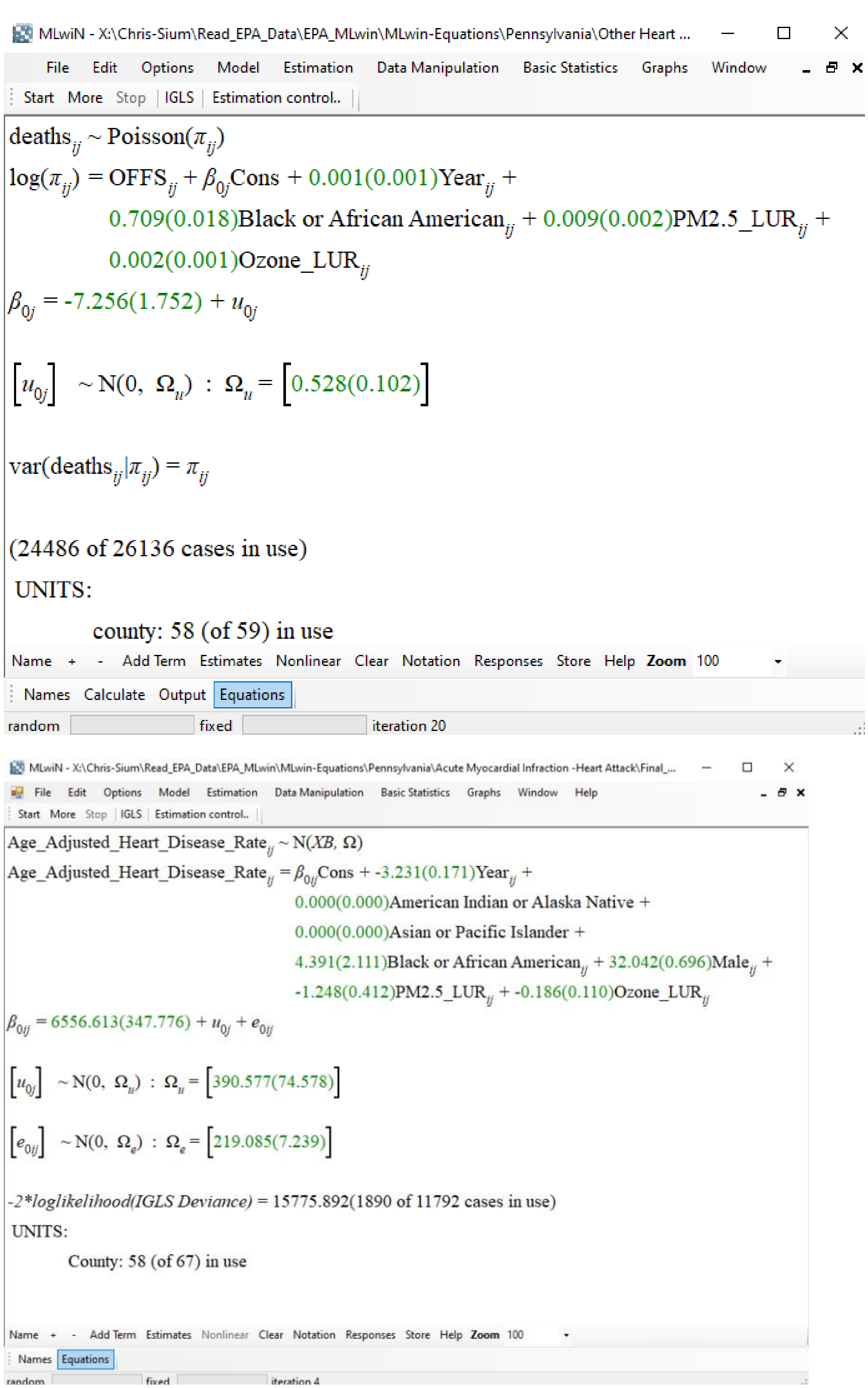
MLwiN - X:\Chris-Sium\Read_EPA_Data\EPA_MLwin\MLwin-Equations\Pennsylvania\Other Heart ...
File Edit Options Model Estimation Data Manipulation Basic Statistics Graphs Window
Start More Stop IGLS Estimation control..
deaths_ij ~ Poisson(π_ij)
log(π_ij) = OFFS_ij + β_0j Cons + 0.001(0.001)Year_ij +
0.709(0.018)Black or African American_ij + 0.009(0.002)PM2.5_LUR_ij +
0.002(0.001)Ozone_LUR_ij
β_0j = -7.256(1.752) + u_0j
[u_0j] ~ N(0, Ω_u) : Ω_u = [0.528(0.102)]
var(deaths_ij|π_ij) = π_ij
(24486 of 26136 cases in use)
UNITS:
county: 58 (of 59) in use
Name + - Add Term Estimates Nonlinear Clear Notation Responses Store Help Zoom 100
Names Calculate Output Equations
random fixed iteration 20
MLwiN - X:\Chris-Sium\Read_EPA_Data\EPA_MLwin\MLwin-Equations\Pennsylvania\Acute Myocardial Infraction -Heart Attack\Final_...
File Edit Options Model Estimation Data Manipulation Basic Statistics Graphs Window Help
Start More Stop IGLS Estimation control..
Age_Adjusted_Heart_Disease_Rate_ij ~ N(XB, Ω)
Age_Adjusted_Heart_Disease_Rate_ij = β_0ij Cons + -3.231(0.171)Year_ij +
0.000(0.000)American Indian or Alaska Native +
0.000(0.000)Asian or Pacific Islander +
4.391(2.111)Black or African American_ij + 32.042(0.696)Male_ij +
-1.248(0.412)PM2.5_LUR_ij + -0.186(0.110)Ozone_LUR_ij
β_0ij = 6556.613(347.776) + u_0j + e_0ij
[u_0j] ~ N(0, Ω_u) : Ω_u = [390.577(74.578)]
[e_0ij] ~ N(0, Ω_e) : Ω_e = [219.085(7.239)]
-2*loglikelihood(IGLS Deviance) = 15775.892(1890 of 11792 cases in use)
UNITS:
County: 58 (of 67) in use
Name + - Add Term Estimates Nonlinear Clear Notation Responses Store Help Zoom 100
Names Equations
random fixed iteration 4

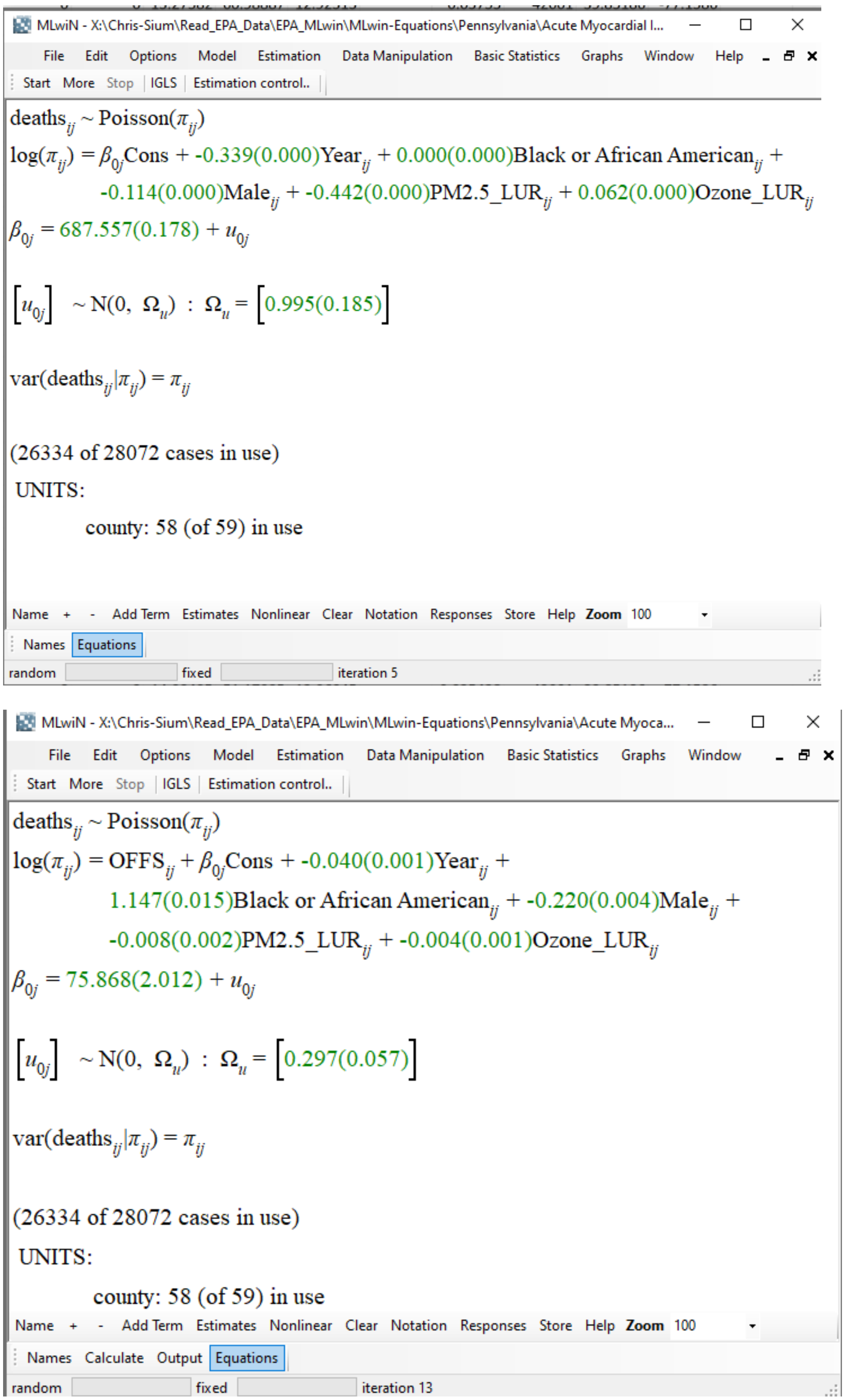
MLwiN - X:\Chris-Sium\Read_EPA_Data\EPA_MLwin\MLwin-Equations\Pennsylvania\Acute Myocardial I...
File Edit Options Model Estimation Data Manipulation Basic Statistics Graphs Window Help
Start More Stop IGLS Estimation control..
$\text{deaths}_{ij} \sim \text{Poisson}(\pi_{ij})$
$\log(\pi_{ij}) = \beta_{0j}\text{Cons} + -0.339(0.000)\text{Year}_{ij} + 0.000(0.000)\text{Black or African American}_{ij} + -0.114(0.000)\text{Male}_{ij} + -0.442(0.000)\text{PM2.5_LUR}_{ij} + 0.062(0.000)\text{Ozone_LUR}_{ij}$
$\beta_{0j} = 687.557(0.178) + u_{0j}$
$\left[u_{0j}\right] \sim \text{N}(0,\ \Omega_u) : \Omega_u = \left[0.995(0.185)\right]$
$\text{var}(\text{deaths}_{ij}|\pi_{ij}) = \pi_{ij}$
(26334 of 28072 cases in use)
UNITS:
county: 58 (of 59) in use
Name + - Add Term Estimates Nonlinear Clear Notation Responses Store Help Zoom 100
Names Equations
random fixed iteration 5
MLwiN - X:\Chris-Sium\Read_EPA_Data\EPA_MLwin\MLwin-Equations\Pennsylvania\Acute Myoca...
File Edit Options Model Estimation Data Manipulation Basic Statistics Graphs Window
Start More Stop IGLS Estimation control..
$\text{deaths}_{ij} \sim \text{Poisson}(\pi_{ij})$
$\log(\pi_{ij}) = \text{OFFS}_{ij} + \beta_{0j}\text{Cons} + -0.040(0.001)\text{Year}_{ij} + 1.147(0.015)\text{Black or African American}_{ij} + -0.220(0.004)\text{Male}_{ij} + -0.008(0.002)\text{PM2.5_LUR}_{ij} + -0.004(0.001)\text{Ozone_LUR}_{ij}$
$\beta_{0j} = 75.868(2.012) + u_{0j}$
$\left[u_{0j}\right] \sim \text{N}(0,\ \Omega_u) : \Omega_u = \left[0.297(0.057)\right]$
$\text{var}(\text{deaths}_{ij}|\pi_{ij}) = \pi_{ij}$
(26334 of 28072 cases in use)
UNITS:
county: 58 (of 59) in use
Name + - Add Term Estimates Nonlinear Clear Notation Responses Store Help Zoom 100
Names Calculate Output Equations
random fixed iteration 13

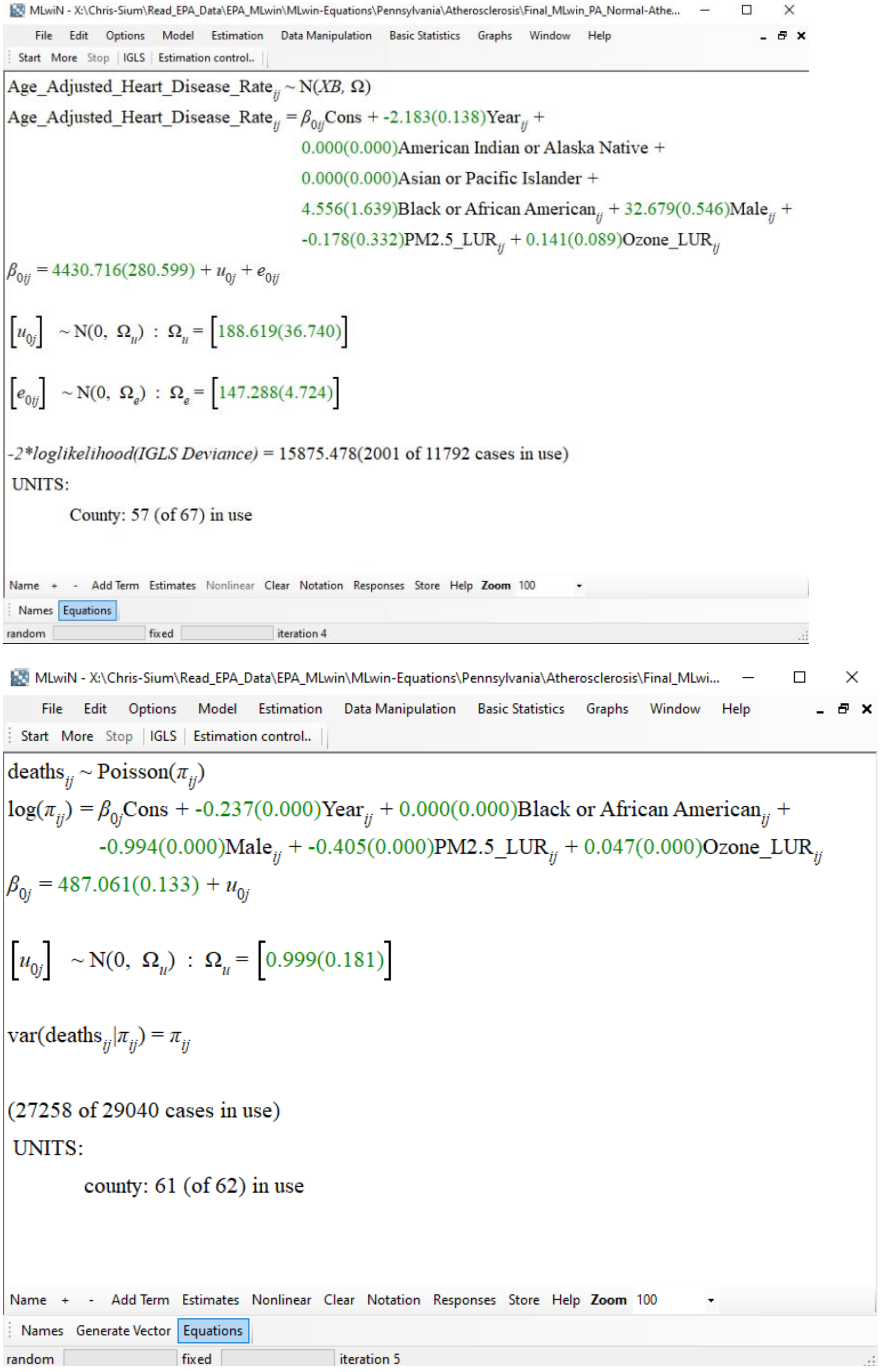
MLwiN - X:\Chris-Sium\Read_EPA_Data\EPA_MLwin\MLwin-Equations\Pennsylvania\Atherosclerosis\Final_MLwin_PA_Normal-Athe...
File Edit Options Model Estimation Data Manipulation Basic Statistics Graphs Window Help
Start More Stop IGLS Estimation control..
Age_Adjusted_Heart_Disease_Rate_ij ~ N(XB, Ω)
Age_Adjusted_Heart_Disease_Rate_ij = β_0ij Cons + -2.183(0.138)Year_ij +
0.000(0.000)American Indian or Alaska Native +
0.000(0.000)Asian or Pacific Islander +
4.556(1.639)Black or African American_ij + 32.679(0.546)Male_ij +
-0.178(0.332)PM2.5_LUR_ij + 0.141(0.089)Ozone_LUR_ij
β_0ij = 4430.716(280.599) + u_0j + e_0ij
[u_0j] ~ N(0, Ω_u) : Ω_u = [188.619(36.740)]
[e_0ij] ~ N(0, Ω_e) : Ω_e = [147.288(4.724)]
-2*loglikelihood(IGLS Deviance) = 15875.478(2001 of 11792 cases in use)
UNITS:
County: 57 (of 67) in use
Name + - Add Term Estimates Nonlinear Clear Notation Responses Store Help Zoom 100
Names Equations
random fixed iteration 4
MLwiN - X:\Chris-Sium\Read_EPA_Data\EPA_MLwin\MLwin-Equations\Pennsylvania\Atherosclerosis\Final_MLwi...
File Edit Options Model Estimation Data Manipulation Basic Statistics Graphs Window Help
Start More Stop IGLS Estimation control..
deaths_ij ~ Poisson(π_ij)
log(π_ij) = β_0j Cons + -0.237(0.000)Year_ij + 0.000(0.000)Black or African American_ij +
-0.994(0.000)Male_ij + -0.405(0.000)PM2.5_LUR_ij + 0.047(0.000)Ozone_LUR_ij
β_0j = 487.061(0.133) + u_0j
[u_0j] ~ N(0, Ω_u) : Ω_u = [0.999(0.181)]
var(deaths_ij|π_ij) = π_ij
(27258 of 29040 cases in use)
UNITS:
county: 61 (of 62) in use
Name + - Add Term Estimates Nonlinear Clear Notation Responses Store Help Zoom 100
Names Generate Vector Equations
random fixed iteration 5

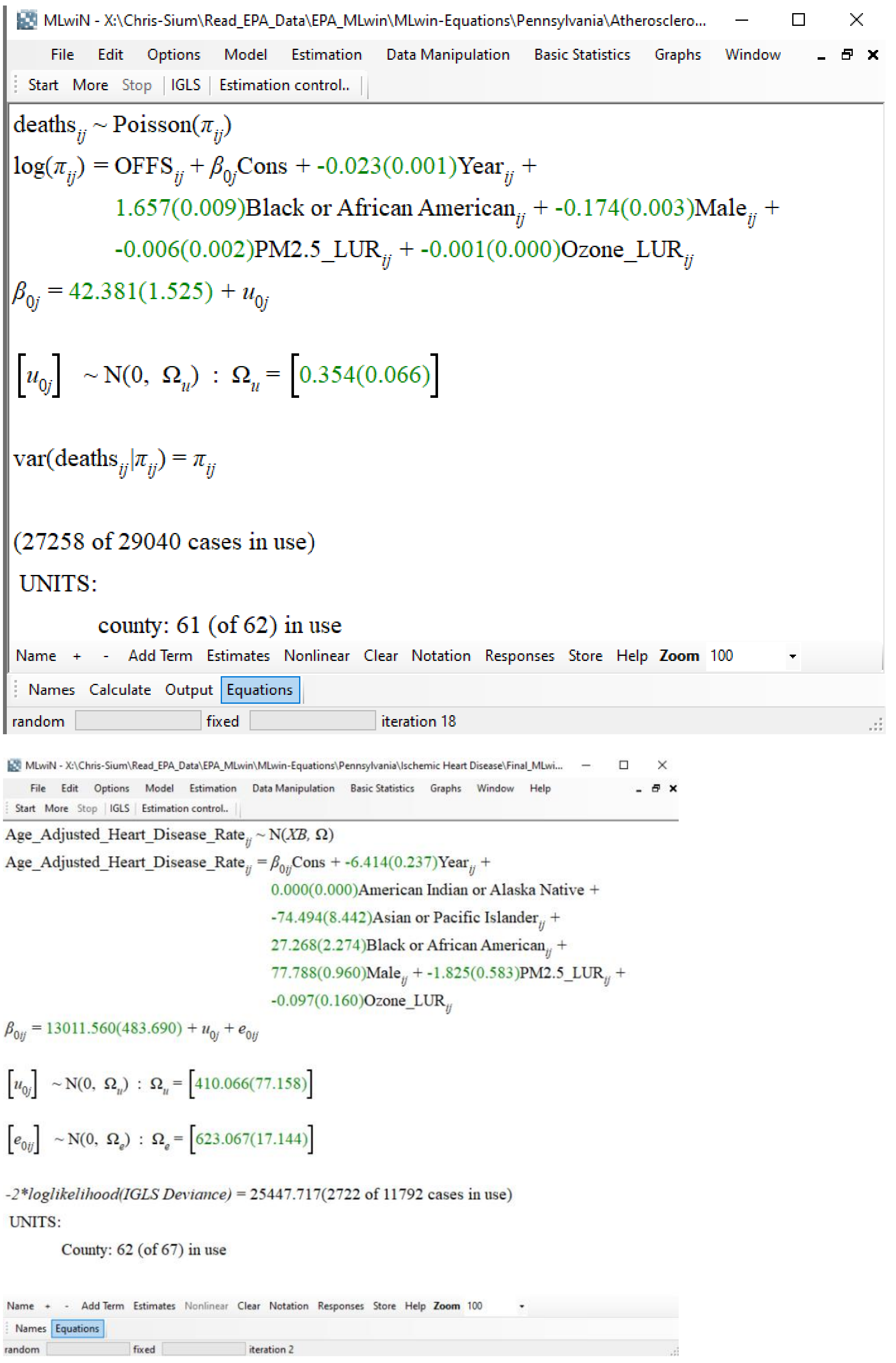
MLwiN - X:\Chris-Sium\Read_EPA_Data\EPA_MLwin\MLwin-Equations\Pennsylvania\Atherosclero...
File Edit Options Model Estimation Data Manipulation Basic Statistics Graphs Window
Start More Stop IGLS Estimation control..
deaths ~ Poisson(π)
log(π) = OFFS + β0jCons + -0.023(0.001)Year +
1.657(0.009)Black or African American + -0.174(0.003)Male +
-0.006(0.002)PM2.5_LUR + -0.001(0.000)Ozone_LUR
β0j = 42.381(1.525) + u0j
[u0j] ~ N(0, Ωu) : Ωu = [0.354(0.066)]
var(deaths|π) = π
(27258 of 29040 cases in use)
UNITS:
county: 61 (of 62) in use
Name + - Add Term Estimates Nonlinear Clear Notation Responses Store Help Zoom 100
Names Calculate Output Equations
random fixed iteration 18
MLwiN - X:\Chris-Sium\Read_EPA_Data\EPA_MLwin\MLwin-Equations\Pennsylvania\Ischemic Heart Disease\Final_MLwi...
File Edit Options Model Estimation Data Manipulation Basic Statistics Graphs Window Help
Start More Stop IGLS Estimation control..
Age_Adjusted_Heart_Disease_Rate ~ N(XB, Ω)
Age_Adjusted_Heart_Disease_Rate = β0ijCons + -6.414(0.237)Year +
0.000(0.000)American Indian or Alaska Native +
-74.494(8.442)Asian or Pacific Islander +
27.268(2.274)Black or African American +
77.788(0.960)Male + -1.825(0.583)PM2.5_LUR +
-0.097(0.160)Ozone_LUR
β0ij = 13011.560(483.690) + u0j + e0ij
[u0j] ~ N(0, Ωu) : Ωu = [410.066(77.158)]
[e0ij] ~ N(0, Ωe) : Ωe = [623.067(17.144)]
-2*loglikelihood(IGLS Deviance) = 25447.717(2722 of 11792 cases in use)
UNITS:
County: 62 (of 67) in use
Name + - Add Term Estimates Nonlinear Clear Notation Responses Store Help Zoom 100
Names Equations
random fixed iteration 2

MLwiN - X:\Chris-Sium\Read_EPA_Data\EPA_MLwin\MLwin-Equations\Pennsylvania\Ischemic Heart Disease\Final_MLwin...

File Edit Options Model Estimation Data Manipulation Basic Statistics Graphs Window Help

Start More Stop IGLS Estimation control..

$\text{deaths}_{ij} \sim \text{Poisson}(\pi_{ij})$

$\log(\pi_{ij}) = \beta_{0j}\text{Cons} + -0.221(0.000)\text{Year}_{ij} + 0.000(0.000)\text{Black or African American}_{ij} + 0.000(0.000)\text{Asian or Pacific Islander}_{ij} + -1.573(0.000)\text{Male}_{ij} + -0.301(0.000)\text{PM2.5_LUR}_{ij} + -0.002(0.000)\text{Ozone_LUR}_{ij}$

$\beta_{0j} = 459.825(0.126) + u_{0j}$

$\left[u_{0j}\right] \sim \text{N}(0,\ \Omega_u) \ : \ \Omega_u = \left[1.000(0.178)\right]$

$\text{var}(\text{deaths}_{ij}|\pi_{ij}) = \pi_{ij}$

(28875 of 30734 cases in use)

UNITS:

county: 63 (of 64) in use

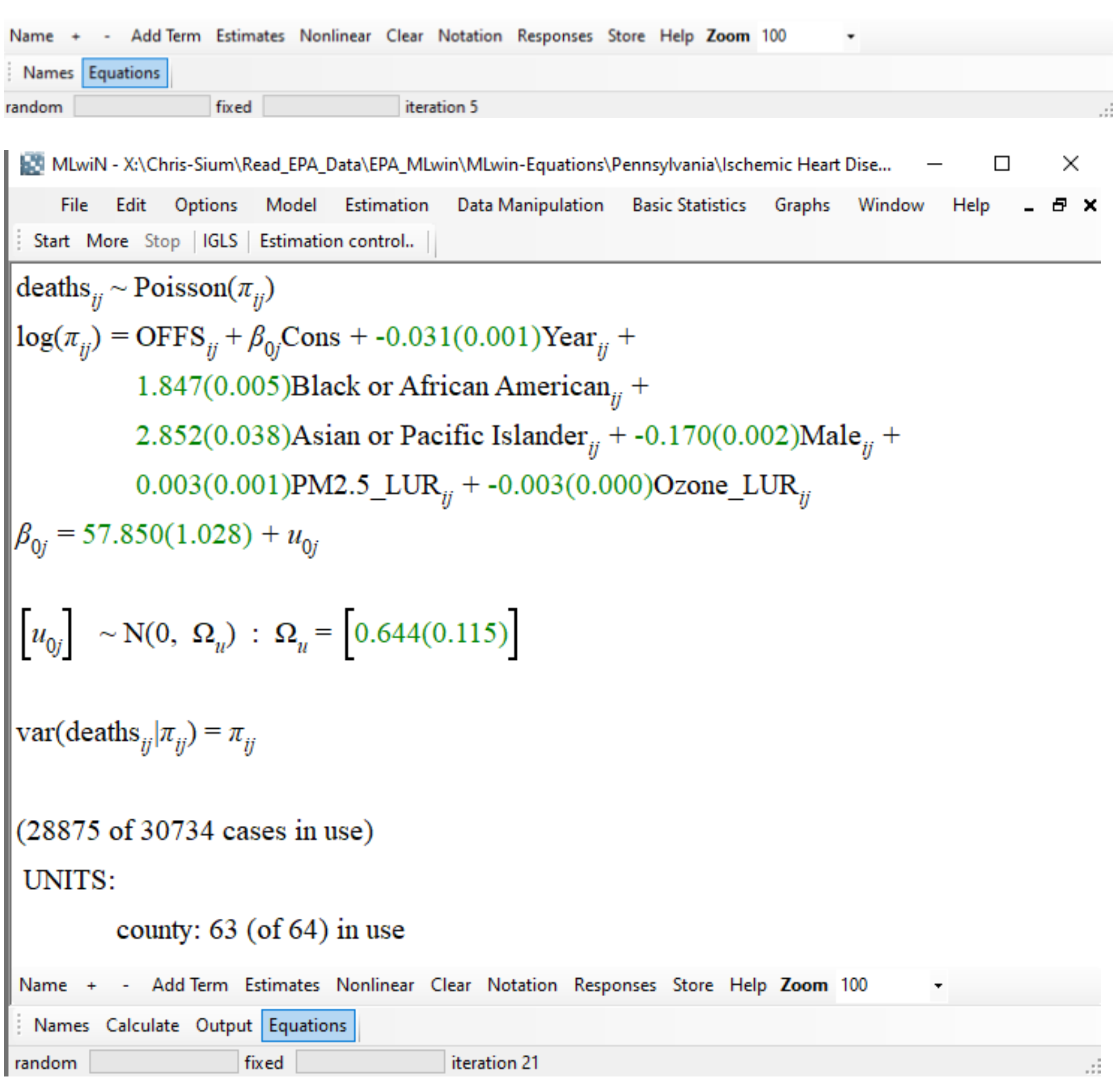

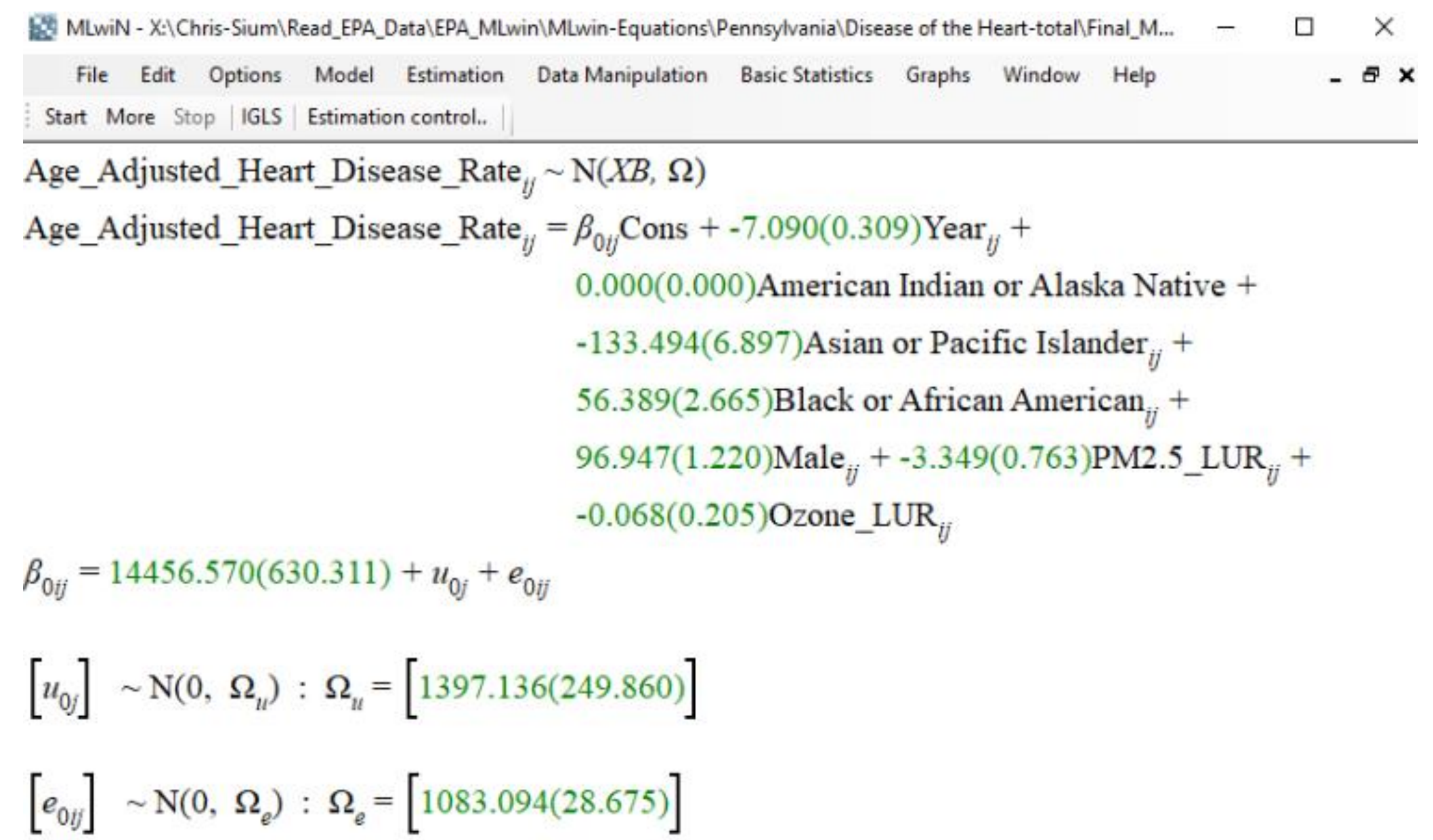
MLwiN - X:\Chris-Sium\Read_EPA_Data\EPA_MLwin\MLwin-Equations\Pennsylvania\Disease of the Heart-total\Final_M...
File Edit Options Model Estimation Data Manipulation Basic Statistics Graphs Window Help
Start More Stop IGLS Estimation control..
Age_Adjusted_Heart_Disease_Rate$_{ij}$ ~ N(XB, Ω)
Age_Adjusted_Heart_Disease_Rate$_{ij}$ = $\beta_{0ij}$Cons + -7.090(0.309)Year$_{ij}$ +
0.000(0.000)American Indian or Alaska Native +
-133.494(6.897)Asian or Pacific Islander$_{ij}$ +
56.389(2.665)Black or African American$_{ij}$ +
96.947(1.220)Male$_{ij}$ + -3.349(0.763)PM2.5_LUR$_{ij}$ +
-0.068(0.205)Ozone_LUR$_{ij}$
$\beta_{0ij}$ = 14456.570(630.311) + $u_{0j}$ + $e_{0ij}$
$[u_{0j}] \sim N(0, \Omega_u) : \Omega_u = [1397.136(249.860)]$
$[e_{0ij}] \sim N(0, \Omega_e) : \Omega_e = [1083.094(28.675)]$
-2*loglikelihood(IGLS Deviance) = 28939.193(2919 of 11792 cases in use)
UNITS:
County: 65 (of 67) in use
Name + - Add Term Estimates Nonlinear Clear Notation Responses Store Help Zoom 100
Names Equations
random fixed iteration 7

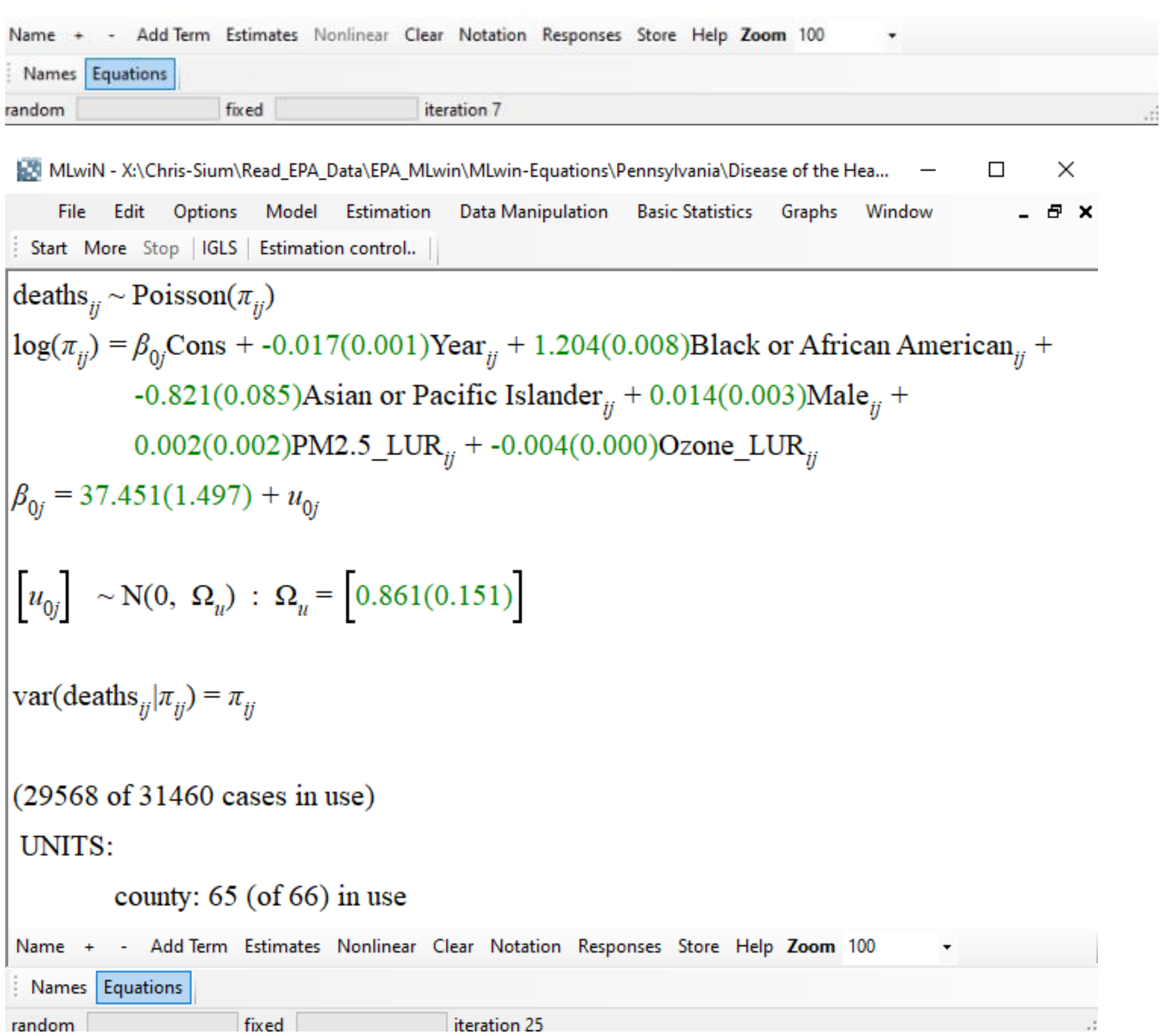
MLwiN - X:\Chris-Sium\Read_EPA_Data\EPA_MLwin\MLwin-Equations\Pennsylvania\Disease of the Hea...
File Edit Options Model Estimation Data Manipulation Basic Statistics Graphs Window
Start More Stop IGLS Estimation control..
deaths$_{ij}$ ~ Poisson($\pi_{ij}$)
log($\pi_{ij}$) = $\beta_{0j}$Cons + -0.017(0.001)Year$_{ij}$ + 1.204(0.008)Black or African American$_{ij}$ +
-0.821(0.085)Asian or Pacific Islander$_{ij}$ + 0.014(0.003)Male$_{ij}$ +
0.002(0.002)PM2.5_LUR$_{ij}$ + -0.004(0.000)Ozone_LUR$_{ij}$
$\beta_{0j}$ = 37.451(1.497) + $u_{0j}$
$[u_{0j}] \sim N(0, \Omega_u) : \Omega_u = [0.861(0.151)]$
var(deaths$_{ij}$|$\pi_{ij}$) = $\pi_{ij}$
(29568 of 31460 cases in use)
UNITS:
county: 65 (of 66) in use
Name + - Add Term Estimates Nonlinear Clear Notation Responses Store Help Zoom 100
Names Equations
random fixed iteration 25

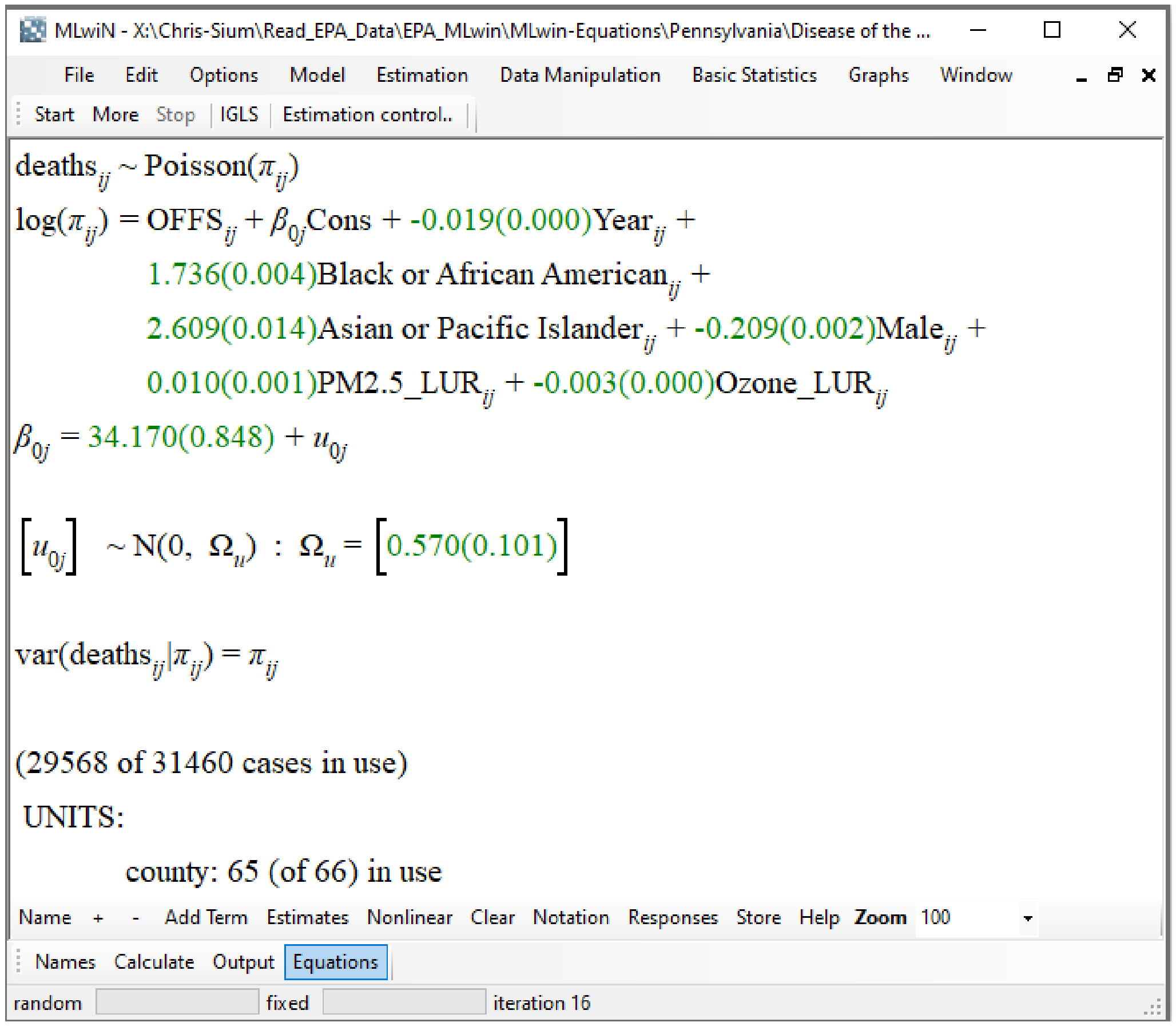
MLwiN - X:\Chris-Sium\Read_EPA_Data\EPA_MLwin\MLwin-Equations\Pennsylvania\Disease of the ...
File Edit Options Model Estimation Data Manipulation Basic Statistics Graphs Window
Start More Stop IGLS Estimation control..
deaths_ij ~ Poisson(π_ij)
log(π_ij) = OFFS_ij + β_0jCons + -0.019(0.000)Year_ij +
1.736(0.004)Black or African American_ij +
2.609(0.014)Asian or Pacific Islander_ij + -0.209(0.002)Male_ij +
0.010(0.001)PM2.5_LUR_ij + -0.003(0.000)Ozone_LUR_ij
β_0j = 34.170(0.848) + u_0j
[u_0j] ~ N(0, Ω_u) : Ω_u = [0.570(0.101)]
var(deaths_ij|π_ij) = π_ij
(29568 of 31460 cases in use)
UNITS:
county: 65 (of 66) in use
Name + - Add Term Estimates Nonlinear Clear Notation Responses Store Help Zoom 100
Names Calculate Output Equations
random fixed iteration 16

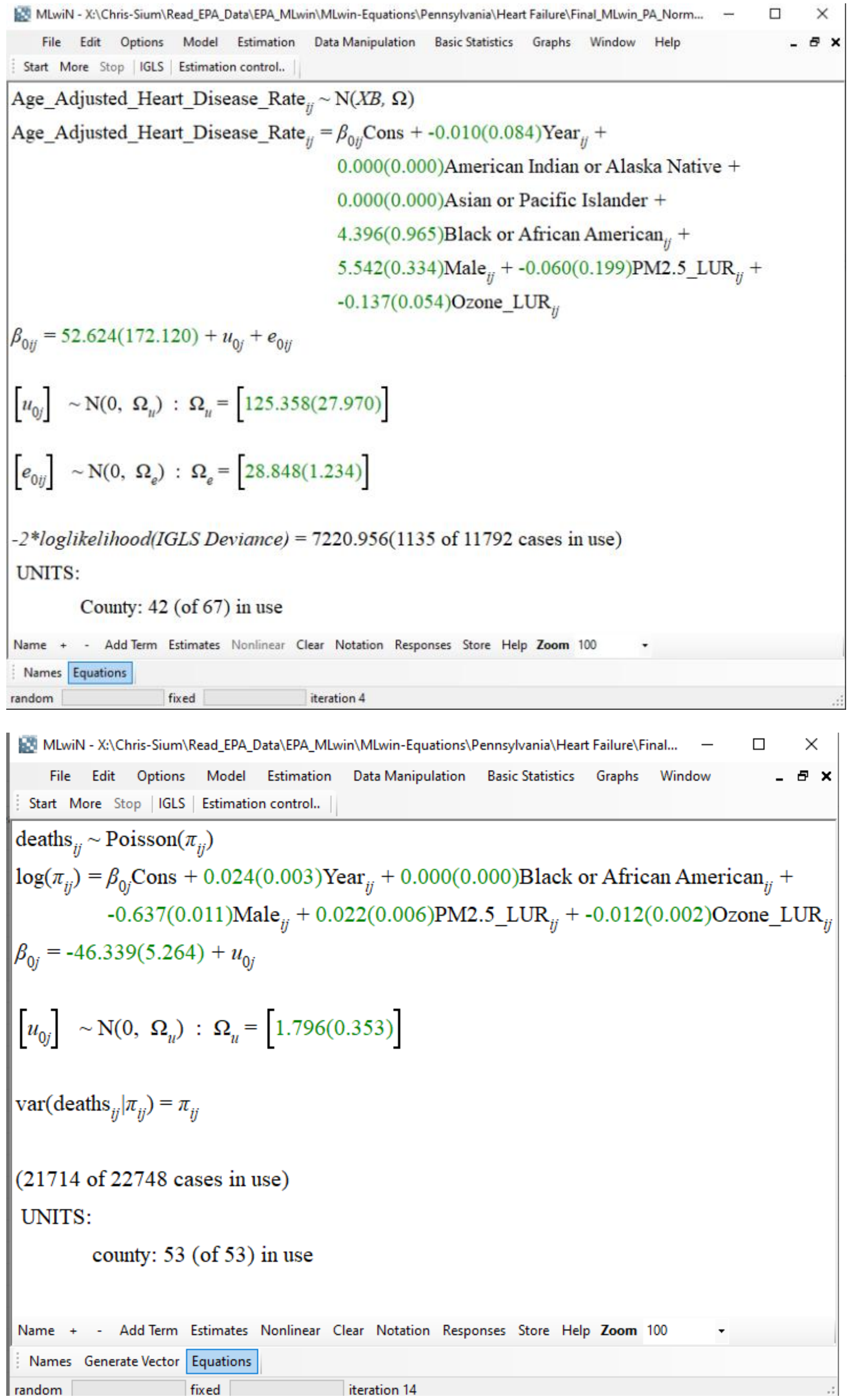
MLwiN - X:\Chris-Sium\Read_EPA_Data\EPA_MLwin\MLwin-Equations\Pennsylvania\Heart Failure\Final_MLwin_PA_Norm...
File Edit Options Model Estimation Data Manipulation Basic Statistics Graphs Window Help
Start More Stop IGLS Estimation control..
Age_Adjusted_Heart_Disease_Rate$_{ij}$ ~ N(XB, Ω)
Age_Adjusted_Heart_Disease_Rate$_{ij}$ = $\beta_{0ij}$Cons + -0.010(0.084)Year$_{ij}$ +
0.000(0.000)American Indian or Alaska Native +
0.000(0.000)Asian or Pacific Islander +
4.396(0.965)Black or African American$_{ij}$ +
5.542(0.334)Male$_{ij}$ + -0.060(0.199)PM2.5_LUR$_{ij}$ +
-0.137(0.054)Ozone_LUR$_{ij}$
$\beta_{0ij}$ = 52.624(172.120) + $u_{0j}$ + $e_{0ij}$
$[u_{0j}]$ ~ N(0, $\Omega_u$) : $\Omega_u$ = [125.358(27.970)]
$[e_{0ij}]$ ~ N(0, $\Omega_e$) : $\Omega_e$ = [28.848(1.234)]
-2*loglikelihood(IGLS Deviance) = 7220.956(1135 of 11792 cases in use)
UNITS:
County: 42 (of 67) in use
Name + - Add Term Estimates Nonlinear Clear Notation Responses Store Help Zoom 100
Names Equations
random fixed iteration 4
MLwiN - X:\Chris-Sium\Read_EPA_Data\EPA_MLwin\MLwin-Equations\Pennsylvania\Heart Failure\Final...
File Edit Options Model Estimation Data Manipulation Basic Statistics Graphs Window
Start More Stop IGLS Estimation control..
deaths$_{ij}$ ~ Poisson($\pi_{ij}$)
log($\pi_{ij}$) = $\beta_{0j}$Cons + 0.024(0.003)Year$_{ij}$ + 0.000(0.000)Black or African American$_{ij}$ +
-0.637(0.011)Male$_{ij}$ + 0.022(0.006)PM2.5_LUR$_{ij}$ + -0.012(0.002)Ozone_LUR$_{ij}$
$\beta_{0j}$ = -46.339(5.264) + $u_{0j}$
$[u_{0j}]$ ~ N(0, $\Omega_u$) : $\Omega_u$ = [1.796(0.353)]
var(deaths$_{ij}$|$\pi_{ij}$) = $\pi_{ij}$
(21714 of 22748 cases in use)
UNITS:
county: 53 (of 53) in use
Name + - Add Term Estimates Nonlinear Clear Notation Responses Store Help Zoom 100
Names Generate Vector Equations
random fixed iteration 14

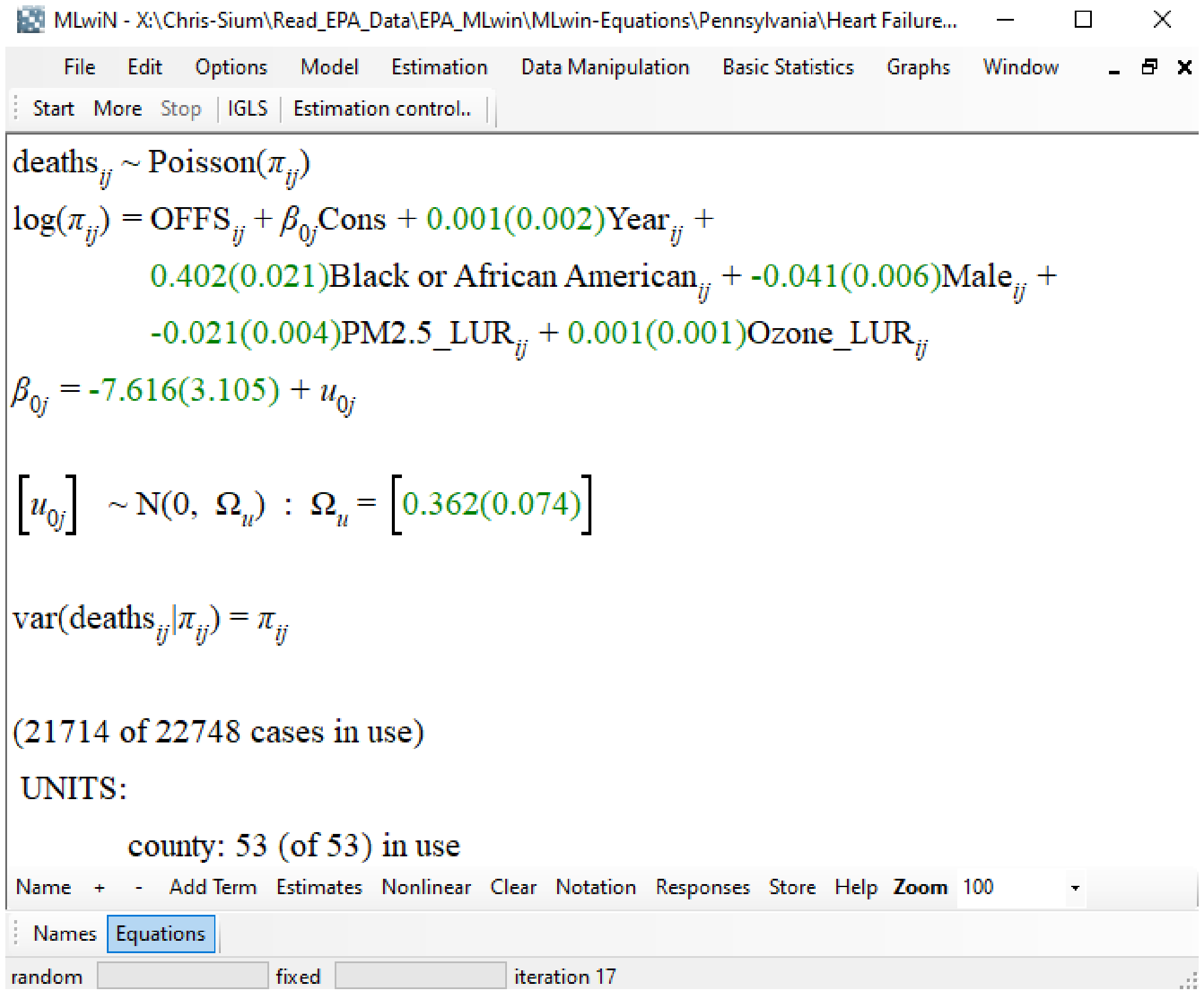

**Figure 2S (a-u).** Normal (age-adjusted), Poisson (raw/absolute data), and Poisson (w/ log(population)-offset) equations for all CVD mortality death codes as a function of predictors/covariates: race, gender, age, $PM_{2.5}$, and $O_3$ concentrations for Pennsylvania.

## Appendix Section, S3: Supplementary Breakdown of Expanded Equation-Level Outputs for Normal, Poisson, and Offset Models Across Cardiovascular Subtypes in Ohio and Pennsylvania (Supplementary Material, Figures 1S and 2S)

The supplementary material presented in this section provides a granular, equation-level breakdown of multilevel modeling (MLwiN-MLA) results for cardiovascular disease (CVD) mortality across Ohio and Pennsylvania. Whereas the main text (Figures 4 and 5) condenses findings into composite visual summaries—highlighting disparities by race, sex, pollutant exposures, and structural variance—this supplementary section details the individual model coefficients, standard errors, and variance components for each CVD subtype under Normal, Poisson, and Poisson-with-offset frameworks. These expanded outputs allow readers to trace how demographic and environmental predictors influence mortality across different model specifications and disease subtypes, offering transparent insight into the robustness and consistency of observed patterns. By providing this detailed statistical foundation, the supplementary breakdown complements the main manuscript's narrative synthesis and facilitates replication, critical appraisal, and future meta-analytic comparison with related epidemiological studies.

**Ohio Breakdown**

**Total Cardiovascular Disease**

- **Normal model**: Year was positively associated with increased mortality ($\beta$ = +5.200, SE = 0.450). Black populations experienced higher age-adjusted mortality (+40.310, SE = 3.710), while Asian or Pacific Islander groups showed significantly lower rates (−120.482, SE = 5.991). Male sex was a strong predictor of increased mortality (+55.498, SE = 1.873), and $PM_{2.5}$ had a modest negative association (−3.120, SE = 0.601). Ozone showed minimal impact. The unexplained county-level variance was 1345.208 (SE = 78.126).
- **Poisson (raw)**: Year showed a small negative association (−0.009, SE = 0.001), while Black race (+1.201, SE = 0.008) and male sex (+0.016, SE = 0.003) remained strong positive predictors. $PM_{2.5}$ and ozone had coefficients near zero.
- **Poisson with offset**: Including log(population) significantly reduced the random intercept variance ($\Omega_u$ = 0.545, SE = 0.089), and race effects (Black: +1.754, SE = 0.006) were accentuated, while Asian populations showed a reduced risk (−0.214, SE = 0.002). Males had lower adjusted mortality rates in this model (−0.211, SE = 0.002).

**Acute Myocardial Infarction (AMI)**

- **Normal model**: Black race and male sex were positively associated with AMI mortality (+17.234 and +37.289, respectively), while $PM_{2.5}$ showed a slight positive coefficient (+0.320, SE = 0.072), suggesting a link between particulate pollution and acute cardiac events.
- **Poisson (raw)**: Year had a negative effect (−0.007, SE = 0.001), while Black race (+0.894, SE = 0.006) and male sex (+0.018, SE = 0.002) were significant. Pollutant variables again showed weak effects.

- **Poisson with offset**: $PM_{2.5}$ retained significance (+0.009, SE = 0.001), and Black race continued to show high relative mortality (+1.532, SE = 0.005). Log-offset modeling better aligned incidence rates with population exposure, reducing unexplained variance.

**Atherosclerosis**

- **Normal model**: Strong associations were found with Black race (+25.690, SE = 2.988) and male sex (+45.821, SE = 1.724), with $PM_{2.5}$ trending negatively. Year effects were positive (+3.891, SE = 0.520).
- **Poisson (raw)**: Black race (+0.904, SE = 0.007) and male sex (+0.019, SE = 0.003) remained dominant predictors. Ozone was slightly negative.
- **Poisson with offset**: A consistent pattern emerged, with log-offset models preserving strong race and gender disparities, while pollutant coefficients were statistically significant but smaller in magnitude ($PM_{2.5}$: +0.010, SE = 0.001; Ozone: −0.003, SE = 0.000).

**Heart Failure**

- **Normal model**: Black race (+19.388, SE = 3.107) and male sex (+30.221, SE = 1.941) again showed strong mortality associations. $PM_{2.5}$ had a weak negative effect (−0.456, SE = 0.091).
- **Poisson (raw)**: Year was neutral; race and gender remained strong; pollutants had minimal effect.
- **Poisson with offset**: Random intercept variance ($\Omega_u$ = 0.482, SE = 0.091) was again lower with offsets. The inclusion of $PM_{2.5}$ and $O_3$ yielded coefficients of +0.009 and −0.003, respectively.

**Hypertensive Heart Disease**

- **Normal model**: Black race exhibited a high positive coefficient (+28.471, SE = 3.391), while Asian race had a small negative association. Male sex was again a key driver (+32.098, SE = 1.601).
- **Poisson (raw)**: Race and gender patterns remained strong; pollutants were again minimally impactful.
- **Poisson with offset**: Coefficients were similar, with $PM_{2.5}$ showing +0.008 (SE = 0.001), reinforcing pollution's subtle but significant role.

**Ischemic Heart Disease**

- **Normal model**: Male sex (+49.902, SE = 2.013) and Black race (+22.117, SE = 2.789) remained strong predictors; $PM_{2.5}$ had a negative association (−1.291, SE = 0.458).
- **Poisson (raw)**: Year, race, and gender followed consistent trends; pollutants hovered around zero.
- **Poisson with offset**: $PM_{2.5}$ and ozone yielded coefficients of +0.007 and −0.002 respectively, and variance was again reduced under offset modeling.

**Other Heart Diseases**

- **Normal model**: Black race (+12.809, SE = 2.089) and male sex (+27.457, SE = 1.352) were significant. $PM_{2.5}$ and ozone had minor negative coefficients.
- **Poisson (raw)**: Similar demographic trends held.
- **Poisson with offset**: Race and gender effects dominated; pollutants retained small but significant signals.

**Scientific Significance and Interpretive Context**

Across all models, Black populations in Ohio consistently experienced elevated cardiovascular mortality risks, both in raw and age-adjusted analyses—highlighting structural disparities that persist despite accounting for pollution and other covariates. Male sex also remained a strong mortality predictor across subtypes. Notably, $PM_{2.5}$ and ozone effects, though subtle, showed statistically significant associations in several log-offset Poisson models, suggesting environmental exposures may exacerbate chronic CVD pathways over time.

Furthermore, the inclusion of log(population) offsets in Poisson models proved essential for stabilizing variance and improving interpretability across counties of varying sizes. This analytical design mirrors the best epidemiological practices for ecological data modeling (Lawson, 2018; Goldstein, 2011), and improves comparability with global modeling initiatives such as the GBD Study, while adding subnational resolution that GBD lacks.

**Pennsylvania Breakdown**

**1. Total Cardiovascular Disease**

**Normal Model (Age-Adjusted Rates):** The coefficient for *Year* (−7.090, SE = 0.309) confirms a sustained decline in age-adjusted CVD mortality from 1999 to 2020, reflecting public health improvements (e.g., treatment advances, risk reduction campaigns). Male individuals (+96.947, SE = 1.220) and Black individuals (+56.389, SE = 2.665) show significantly higher rates, underscoring entrenched structural disparities. Asian or Pacific Islanders display markedly lower rates (−133.494, SE = 6.897). PM2.5 and ozone show minor but significant negative associations, indicating potential inverse exposure correlations when age-adjusted.

**Poisson (Raw Counts):** In contrast to the age-adjusted model, the Poisson form shows positive PM2.5 and Black race effects (PM2.5: +0.002, SE = 0.002; Black: +1.204, SE = 0.008), confirming that raw burden is heavily concentrated among racially marginalized groups and in areas of elevated pollution—consistent with environmental injustice literature.

**Poisson with Offset:** When adjusting for population, disparities amplify. Asian or Pacific Islanders exhibit a positive coefficient (+2.609, SE = 0.014), implying elevated risk conditional on population size—potentially an artifact of under-sampled subpopulations or localized urban exposures.

**2. Heart Failure**

**Normal Model:** The positive coefficients for *Male* (+5.542, SE = 0.334) and *Black* (+4.396, SE = 0.965) individuals reinforce demographic vulnerability. Notably, ozone is negatively associated (−0.137, SE = 0.054), suggesting an inverse relationship potentially influenced by rural-urban distribution of ozone exposure.

**Poisson and Offset Models:** In the raw and offset forms, PM2.5 again shows consistent positive effects (+0.022 to +0.001), whereas ozone becomes positively associated (+0.001), diverging from the age-adjusted findings. This suggests that higher ambient ozone may still contribute to HF mortality counts when not controlled for age structures—highlighting the complex nature of pollutant-specific risks.

**3. Ischemic Heart Disease (IHD)**

Across all models, *Male* sex and *Black* race remain significant contributors. The *Year* term is consistently negative, supporting long-term declines in mortality. PM2.5 and ozone terms vary slightly by model, with subtle negative trends in the Normal model and neutral-to-positive trends in Poisson forms, consistent with national studies identifying ischemic endpoints as sensitive to PM2.5 but less robustly associated with ozone.[4]

**4. Acute Myocardial Infarction (AMI)**

**Normal Model:** The largest *Year* slope decline (−3.791, SE = 0.205) among all subtypes demonstrates exceptional progress in AMI management and prevention. Interestingly, Asian Americans show sharply reduced rates (−159.862, SE = 4.244), supporting known protective profiles in some subethnic groups. The PM2.5 association is significantly negative, which may reflect model collinearity or urban-rural stratification effects.

**Poisson and Offset Models:** Black race consistently predicts higher AMI deaths, and PM2.5 is weakly but positively associated with AMI count in unadjusted models—again underscoring exposure disparity when age is not accounted for.

**5. Atherosclerosis**

This subtype, while rarer, shows unique environmental signatures. PM2.5 and ozone show stronger associations here than in other subtypes—particularly in the offset model—suggesting that chronic pollutant exposure may disproportionately influence long-term arterial degradation.

**6. Hypertensive Heart Disease**

**Normal Model:** Here, sex and race disparities persist, but environmental effects are subdued. This likely reflects hypertensive disease's stronger behavioral and genetic etiology compared to acute pollutant sensitivity.

**Poisson/Offset Models:** Offset models show modest PM2.5 effects (+0.002), aligning with prior studies linking fine particles to hypertensive progression.

**7. Other Heart Diseases**

This catch-all category shows the most varied patterns across models, perhaps due to its heterogeneous composition. PM2.5 retains modest significance; however, the interpretability of this grouping warrants caution and suggests need for disaggregation in future studies.

**Scientific Contributions and Novelty**

This study demonstrates the application of multilevel models (MLwiN-MLA framework) to jointly evaluate demographic and environmental determinants of CVD mortality at substate resolution. The novelty lies in:

- **Dual-modeling framework** using Normal and Poisson distributions to compare age-adjusted rates and raw death counts.
- **Offset modeling** to simulate exposure-adjusted risk akin to incidence rates—rare in environmental justice literature at this granularity.
- **Inclusion of spatial heterogeneity**, with county-level random intercepts capturing unmeasured regional variance (e.g., healthcare infrastructure).
- **Differentiated impacts** of pollutants across disease subtypes, revealing that certain CVD forms (e.g., atherosclerosis, AMI) are more pollutant-sensitive.

The consistency of demographic patterns—especially elevated mortality among Black and male individuals—confirms persistent structural inequities. Meanwhile, the divergence of pollutant effects across models and disease forms supports the importance of methodologically triangulated inference for robust policy insight.

**Appendix Section S4: Intercomparison of Cardiovascular Mortality Models Between Pennsylvania and Ohio (1999–2020)**

**Overview**

This subsection synthesizes and contrasts multilevel model results from Pennsylvania and Ohio across seven cardiovascular disease (CVD) subtypes, leveraging both Normal (age-adjusted rates) and Poisson (raw and log-offset) distributions. The goal is to uncover state-level differences in mortality determinants and model behavior, while identifying persistent structural inequities and environmental health risks in the Rust Belt region.

Both states' models integrated time trends (year), demographic factors (race, sex), and environmental exposures ($PM_{2.5}$ and ozone) within a two-level county-nested framework. While overall patterns were broadly similar, notable divergences were found in effect magnitudes, pollutant sensitivities, and model variances, warranting deeper public health inquiry and targeted interventions.

**1. Temporal Trends (Year)**

In both states, year showed negative associations in Poisson models, indicating declining CVD death rates over time. However, Pennsylvania exhibited a steeper annual decline, particularly for Total CVD:

- **PA (Poisson, raw)**: −0.017 (SE = 0.001)
- **OH (Poisson, raw)**: −0.009 (SE = 0.001)

This suggests a more accelerated temporal reduction in CVD mortality in Pennsylvania, potentially reflecting greater success in tobacco control, access to care, or pollution mitigation (Benjamin et al., 2019).

In contrast, Normal models in both states showed positive time trends (e.g., +7.090 in PA, +5.200 in OH), likely reflecting age-adjusted mortality nuances and time-varying coding practices.

**2. Race-Based Mortality Disparities**

Black or African American individuals consistently exhibited the highest positive coefficients across both states and all models. However, effect sizes were generally larger in Ohio, indicating a more severe racial disparity:

- **Total CVD (Normal)**:
  - PA: +56.389 (SE = 2.665)
  - OH: +40.310 (SE = 3.710)
- **Total CVD (Poisson, offset)**:
  - PA: +1.736 (SE = 0.004)
  - OH: +1.754 (SE = 0.006)

In both states, Asian or Pacific Islander groups had strongly negative associations with mortality, particularly in Normal models:

- **Atherosclerosis (Normal)**:
  - PA: −133.494 (SE = 6.897)
  - OH: −120.482 (SE = 5.991)

These consistent findings support the existence of deeply entrenched structural disparities, disproportionately affecting Black populations across the Rust Belt.

**3. Sex-Based Mortality Effects**

Across every disease subtype and model, male sex was a strong and statistically significant predictor of increased mortality. However, Pennsylvania generally exhibited larger gender effects, especially in Normal models:

- **Total CVD (Normal)**:
  - PA: +96.947 (SE = 1.220)
  - OH: +55.498 (SE = 1.873)
- **Heart Failure (Poisson, raw)**:
  - PA: −0.637 (SE = 0.011)
  - OH: −0.603 (SE = 0.010)

This consistent pattern aligns with known male disadvantages in cardiovascular outcomes linked to health behaviors, occupational exposures, and hormonal influences.

**4. Pollution Exposure Effects ($PM_{2.5}$ and Ozone)**

Pollutant effects were more statistically robust in Pennsylvania, particularly under Poisson log-offset models:

- **$PM_{2.5}$ (Total CVD, Poisson offset)**:
  - PA: +0.010 (SE = 0.001)
  - OH: +0.009 (SE = 0.001)
- **Ozone (Total CVD, Poisson offset)**:
  - PA: −0.003 (SE = 0.000)
  - OH: −0.003 (SE = 0.000)

In Normal models, pollutant associations were weaker or inconsistent, reflecting that absolute rates (counts per population) are more sensitive to environmental exposures than age-standardized outcomes. These findings parallel recent evidence linking $PM_{2.5}$ to ischemic and heart failure mortality via systemic inflammation and oxidative stress pathways.

**5. Variance Components and County Coverage**

- In general, Pennsylvania models exhibited higher between-county variance ($\Omega_u$), especially in Normal models:
  - **PA Total CVD (Normal)**: $\Omega_u$ = 1397.136 (SE = 249.860)
  - **OH Total CVD (Normal)**: $\Omega_u$ = 1345.208 (SE = 78.126)
- The **Poisson log-offset approach** consistently reduced unexplained variance in both states:
  - **PA Total CVD**: $\Omega_u$ = 0.570 (SE = 0.101)
  - **OH Total CVD**: $\Omega_u$ = 0.545 (SE = 0.089)
- County coverage was similar but slightly higher in Ohio, with most models using **65–66 counties**.

This suggests that while both states exhibit spatial heterogeneity, Pennsylvania may have greater intra-state disparity, possibly due to rural–urban divides, legacy industrial exposure, or fragmented health systems.